\pdfoutput=1

\documentclass[]{emulateapj}
\usepackage[]{hyperref}
\usepackage{float}
\usepackage{color}
\usepackage{graphicx}
\usepackage{natbib}
\newcommand{\fluxunits}{erg cm$^{-2}$ s$^{-1}$}
\newcommand{\cgs}{erg cm$^{-2}$ s$^{-1}$ s.r.$^{-1}$ \AA$^{-1}$}
\newcommand{\feone}{Fe \textsc{ii} $\lambda2814.45$}
\newcommand{\fetwo}{Fe \textsc{ii} $\lambda2832.39$}
\newcommand{\kms}{km s$^{-1}$}

\newcommand{\lamr}{$\lambda_{\rm{rest}}$}
\newcommand{\marflare}{SOL2014-03-29T17:48}

\begin{document}

\title{The Atmospheric Response to High Nonthermal Electron Beam Fluxes in Solar Flares I:  Modeling the Brightest NUV Footpoints in the X1 Solar Flare of 2014 March 29}
\author{Adam~F.~Kowalski}
\affil{Department of Astrophysical and Planetary Sciences, University of Colorado Boulder, 2000 Colorado Ave, Boulder, CO 80305, USA. }
\affil{National Solar Observatory, University of Colorado Boulder,  3665 Discovery Drive, Boulder, CO 80303, USA.}
\affil{Department of Astronomy, University of Maryland, College Park, MD 20742, USA.}
\affil{NASA/Goddard Space Flight Center, Code 671, Greenbelt, MD 20771.}
\email{Adam.Kowalski@lasp.colorado.edu\\} 
\author{Joel C. Allred}
\affil{NASA/Goddard Space Flight Center, Code 671, Greenbelt, MD 20771.}
\author{Adrian Daw}
\affil{NASA/Goddard Space Flight Center, Code 671, Greenbelt, MD 20771.}
\author{Gianna Cauzzi}
\affil{INAF-Osservatorio Astrofisico di Arcetri, I-50125 Firenze, Italy.   }
\and
\author{Mats Carlsson}
\affil{Institute of Theoretical Astrophysics, University of Oslo, PO Box 1029 Blindern, 0315 Oslo, Norway. }
\begin{abstract}

The 2014 March 29 X1 solar flare (SOL20140329T17:48) produced bright continuum emission in the far- and 
near-ultraviolet (NUV) and highly asymmetric chromospheric emission lines, providing long-sought constraints on the 
heating mechanisms of the lower atmosphere in solar flares.  We analyze the continuum and emission line data from the Interface Region Imaging Spectrograph (IRIS) of the brightest 
flaring magnetic footpoints in this flare.  We
compare the NUV spectra of the brightest pixels to new radiative-hydrodynamic predictions calculated with the RADYN code using
constraints on a nonthermal electron beam inferred from the collisional thick-target modeling of hard X-ray data from RHESSI.
We show that the atmospheric response to a high beam flux density satisfactorily achieves the observed continuum brightness in the 
NUV.  The NUV continuum emission in this flare 
is consistent with hydrogen (Balmer) recombination radiation that originates from low optical depth in 
a dense chromospheric condensation and from the stationary beam-heated layers just below the condensation. 
A model producing two flaring regions (a condensation and stationary layers) in the lower atmosphere is also
  consistent with the asymmetric Fe \textsc{ii} chromospheric emission line profiles observed in the impulsive phase.

\end{abstract}

\keywords{}

\section{Introduction}

The spectral energy distribution of the ultraviolet, optical, and infrared continuum (white-light) emission contains important information on the atmospheric response 
at the highest densities in the flare atmosphere but has remained largely unconstrained due to a lack of broad-wavelength spectral
observations \citep{Fletcher2007}.
The white-light emission provides important constraints on the strength and depth of flare heating resulting from magnetic energy release in the corona.  White-light emission is thought to be produced by the energy deposition by nonthermal electrons due to the close spatial 
and temporal coincidence with hard X-ray emission \citep[e.g.,][]{RustHegwer1975, Hudson1992, Metcalf2003, Martinez2012}.
The source of these nonthermal electrons is controversial;  in the standard flare model, they are accelerated in the high corona as ``beams'' \citep[the collisional thick-target
model;][]{Brown1971, Emslie1978}, but recently it has been suggested that limits on electron numbers and propagation effects (e.g., beam instabilities) require
an alternative mode of energy transport, such as 
acceleration of particles in the lower corona or chromosphere by Alfv\'en waves \citep{Fletcher2008}.  
 Continuum measurements from spectra are necessary
to test heating models by
 providing constraints on the optical depths and electron densities that are attained
in the deepest layers of the flare atmosphere.  

Spatially resolved flare spectra have been obtained at NUV/blue wavelengths from the ground
and have shown a range of spectral properties around the expected location of the Balmer jump \citep{Neidig1983, Kowalski2015HSG}.
The continuum emission from these spectra has been interpreted as optically thin hydrogen recombination radiation
\citep{NeidigWiborg1984}, although an emission component from increased H$^{-}$
emission in the upper photosphere has also been suggested to explain some of the observed variation 
 \citep{Hiei1982, Boyer1985}.  However, the brightest regions of solar flares were rarely and poorly sampled in the past:  the locations 
of the spectrographic slit relative to the small white-light footpoints were not precisely known \citep{NeidigWiborg1984, Neidig1989}, due to the
rapid spatial and temporal evolution of the emission and also to variable and poor seeing during flares observed from the ground prior to the advent
of adaptive optics \citep{Donati1984}.  Therefore, the most extreme atmospheric conditions in the lower atmosphere during solar flares are not well constrained.

Recently, Sun-as-a-star observations from SOHO/VIRGO's Sun PhotoMeter (SPM) have indicated the presence
of a hotter blackbody emission component in the optical with a color temperature of $T\sim9000$ K \citep{Kretzschmar2011} which implies very large heating at high densities. High spatial resolution observations from Hinode during two X-class flares have shown a much lower color temperature in the optical of only $T\sim5000-6000$ K \citep{Watanabe2013, Kerr2014}, which is consistent with photospheric heating by several hundred K or optically thin hydrogen recombination radiation from heating of the mid-chromosphere.   
 Blackbody fitting of NUV and optical spectra and broadband photometry of magnetically active M dwarf stars in the gradual and impulsive 
phases of flares results in larger color temperatures, $T\gtrsim 8000$ K but typically $T\sim9000-12,000$ K \citep{Hawley1992, Hawley2003, Zhilyaev2007, Fuhrmeister2008, Kowalski2010, Kowalski2013}.
  Clearly, a more thorough spatially resolved characterization of the brightest
 flare footpoints is necessary 
to determine the prevalence of hot blackbody-like emission in solar flares, and if any proposed heating model can
self-consistently explain the implied heating requirements.  

 Previous spectral observations of the NUV and optical have been interpreted using static isothermal slab
models \citep[e.g.,][]{Donati1985} or semi-empirical static models \citep{Machado1980, Mauas1990, Kleint2016}, but
the flare atmosphere is known to be highly dynamic and stratified \citep{Cauzzi1996, Falchi2002}.
It has been proposed that the continuum emission may originate in impulsively-generated downflows in the upper chromosphere \citep{Livshits1981, Gan1992}, or chromospheric ``condensations'' \citep[hereafter CC;][]{Fisher1985b, Fisher1989}, which are also
attributed to the formation of H$\alpha$ red-wing emission components that are often observed in solar flares \citep{Ichimoto1984, Canfield1987, Canfield1990}.  \cite{Kowalski2015} recently
found that an extremely large electron beam flux of $10^{13}$ \fluxunits\ could produce hot $T\sim10,000$ K blackbody-like emission in very dense CCs.  
With current computational facilities and constraints on electron beam fluxes from the Ramaty High Energy Solar Spectroscopic Imager \citep[RHESSI;][]{Lin2002} and 
high spatial resolution imagery, it is timely to critically examine the hydrodynamic and time-dependent
 radiative response of the models and compare to new spectral observations of solar flares.  

  We have begun a large campaign to characterize the emission properties of the brightest flaring magnetic footpoints during Cycle 24 flares using
new NUV and far-ultraviolet (FUV) spectra from the Interface Region Imaging Spectrograph \citep[IRIS;][]{DePontieu2014}. 
The high spatial resolution of IRIS allows improved intensity measurements of the continuum emission, which 
is observed in compact sources, or kernels, as small as 0.\arcsec3 \citep{Jess2008}. 
 In this paper,  we present radiative-hydrodynamic (RHD) modeling of the brightest continuum flaring pixels in
 the X1 flare of 2014-Mar-29, which has been extensively studied by \cite{Judge2014, Young2015, Liu2015, Battaglia2015, Kleint2015, Matthews2015, Kleint2016, Fatima2016}.
In \cite{Heinzel2014} and \cite{Kleint2016}, the bright NUV continuum emission from this flare was identified and compared to static beam heated model atmospheres from 
\cite{Ricchiazzi1983} and to static phenomenological models with the RH code. \cite{Heinzel2014} concluded that
the NUV continuum intensity was consistent with optically thin Balmer continuum emission.  However, the time-dependent radiative transfer and the hydrodynamics, which can affect beam propagation through
evaporation and condensation, have not yet been compared in detail to the continuum observations; \cite{Heinzel2015IAUS} and \cite{Fatima2016} present new
RHD simulations with the \emph{Flarix} and RADYN codes, respectively, for relatively low beam fluxes compared to the flux inferred from imaging spectroscopy of the brightest source in the flare \citep{Kleint2016}.  In this paper, we use the state-of-the-art Fokker-Planck treatment of energy deposition \citep{Mauas1997, Battaglia2012, Liu2009, Allred2015} from a high-flux electron beam, in order
to understand the time evolution of the atmospheric stratification that self-consistently explains both the NUV continuum emission and chromospheric line profiles 
in the brightest flaring footpoints.

By rigorously testing new RHD models guided by the combined information from RHESSI and new data of the white-light continuum and chromospheric lines, we seek 
answers to the following questions:

\begin{itemize}
\item Using electron beam parameters inferred from standard thick target modeling of RHESSI X-ray data, do electron beams
produce an atmospheric and radiative response that is consistent with the high spatial and spectral line and continuum constraints from IRIS?

\item Does a CC form that is hot and dense enough to explain the observed IRIS line and continuum emission?  

\item Does the hydrodynamic response of the atmosphere to beam heating result in flare continuum emission in the IRIS NUV channel that is predominantly optically thin hydrogen recombination radiation? Is there evidence for hot ($T\gtrsim9000$ K) blackbody-like radiation from photospheric densities?  

\end{itemize}

The paper is organized as follows.  In Sections \ref{sec:calibration}-\ref{sec:area}, we describe the calibration of the IRIS observations and the high spatial resolution flare footpoint development;   in Section \ref{sec:wl}, we discuss the constraints on the continuum intensity and emission line profiles in the spectra of the brightest flaring footpoints; in Section \ref{sec:radyn} we describe the radiative-hydrodynamic modeling of these spectra and the formation of the NUV continuum and Fe \textsc{ii} emission lines; in Section \ref{sec:sjibright}, we compare the modeling results to the brightest sources in the slit jaw images; in Section \ref{sec:limitations} we discuss several limitations of the modeling and future work; in Sections \ref{sec:summary} - \ref{sec:conclusions} we present our conclusions.  Appendices A and B discuss broader wavelength model predictions for the optical continuum emission.

\section{Intensity Calibration of the IRIS Observations} \label{sec:calibration}

IRIS is a rastering 
spectrograph with simultaneous coverage in the near-ultraviolet (NUV, 2782.7-2835.1\AA) and far-ultraviolet
 (FUV1, 1331.7-1358.4\AA; FUV2, 1389.0-1407.0\AA) including both lines and continua.  
IRIS has a spatial resolution of 0\arcsec.33 in the FUV and 0\arcsec.4 in the NUV. The IRIS observations covering the GOES X1 flare SOL2014-03-29T17:48 
were obtained with an 8-step raster across AR 12017 (NOAA 12017, $\mu=0.82$).  We discuss the flare footpoint development of this flare in Section \ref{sec:fpevol} 

The 8-step raster was obtained with an 8-second exposure
time per step, resulting in a 75~s raster cycle.  During the flare, the automatic exposure control (AEC) decreased
the NUV spectral exposure times to 2.4~s at UT 17:46:13.
The spectra were binned in the dispersion direction
in the FUV only, resulting in a spectral pixel width of $\sim26$ m\AA\ in both channels.  
We employ the 
2830 \AA\ slit-jaw (SJI 2832) images which remained unsaturated for the duration of the flare and have the same
spatial scale (0\arcsec.166 pixel$^{-1}$) as the spectra.  The SJI 2832 images
(\#0 to \#179) are obtained once per raster at the third slit position.   The bandpass of SJI 2832
has a FWHM of 4 \AA\ centered on $\lambda=2830$ \AA, which is far into the red absorption wing
of Mg II $h$.

IRIS level 3 data cubes were created by shifting the NUV spectra by $-$2 pixels and the FUV spectra by $+1$ pixels
in order to align with the fiducial marks of SJI 2832.  The NUV wavelength calibration was adjusted by $+0.025$ \AA\ based on the 
alignment of the Ni \textsc{i} 2799.474 \AA\ line.  We applied an intensity calibration to the spectra 
\citep[IRIS Technical Note 24;][]{Kleint2016} and SJI 2832 using the post-flight effective area curves ($A_{\rm{eff}}(\lambda)$)
from the routine iris\_get\_response.pro developed by J. P. Wuesler.  Specifically, the conversion of the spectra with calibrated units of DN s$^{-1}$ pixel$^{-1}$ to units of \cgs\ was performed by multiplying by the following factor:  

\begin{equation} \label{eq:conv}
C =  \frac{\mathrm{gain}}{A_{\rm{eff}}(\lambda)\Delta \lambda} \times S \times E
\end{equation}

\noindent where the gain is 18 photons DN$^{-1}$ for the NUV spectra, 4 photons DN$^{-1}$ for the FUV spectra, $\Delta \lambda$ is the dispersion, $S$ is the solid angle conversion from pixel$^{-1}$ to steradian$^{-1}$ for a slit width of 0\arcsec.33, and
$E$ is the energy per photon.  We convert the SJI 2832 calibrated count rate (DN s$^{-1}$ pixel$^{-1}$) to an equivalent (constant) continuum intensity over the SJI bandpass by using equation \ref{eq:conv} with $A_{\rm{eff}}(\lambda)\Delta \lambda$ replaced by $\int A_{\rm{eff}}(\lambda) d\lambda= 0.02$ cm$^2$ \AA\ as a proxy for the line and continuum emission
brightness in the SJI 2832 images during the flare.

\section{IRIS Slit Jaw 2832 Image Analysis} \label{sec:area}
Compact flaring magnetic footpoints are readily identified using the high spatial resolution 
 SJI 2832 data. 
  The flare intensity
in SJI 2832 is also used to calculate a flare footpoint area, which we compare to the area obtained from 
 partially unresolved RHESSI hard X-ray observations as described in \cite{Kleint2016}.  
The flare footpoint area is critical for inferring a nonthermal electron beam energy flux, which is an input for 1D RHD modeling (Section \ref{sec:radyn}).  

We calculate the excess intensity in SJI 2832 at each time during the flare by subtracting a pre-flare
image (SJI 2832 \#171).  Because the observing cadence (75~s) is relatively low compared to the timescale ($\sim$120~s) of the hard X-ray impulsive
phase of the flare, we use a pre-flare image instead of a running difference to define the flare area.  It should be kept in mind that
the flare area consists of both newly heated flare regions and decaying flare regions that were previously heated, and we do not know 
if the bright pixels in this flare are observed during their rising or decaying evolution.

\begin{figure}[H]
\begin{center}
\includegraphics[scale=0.55]{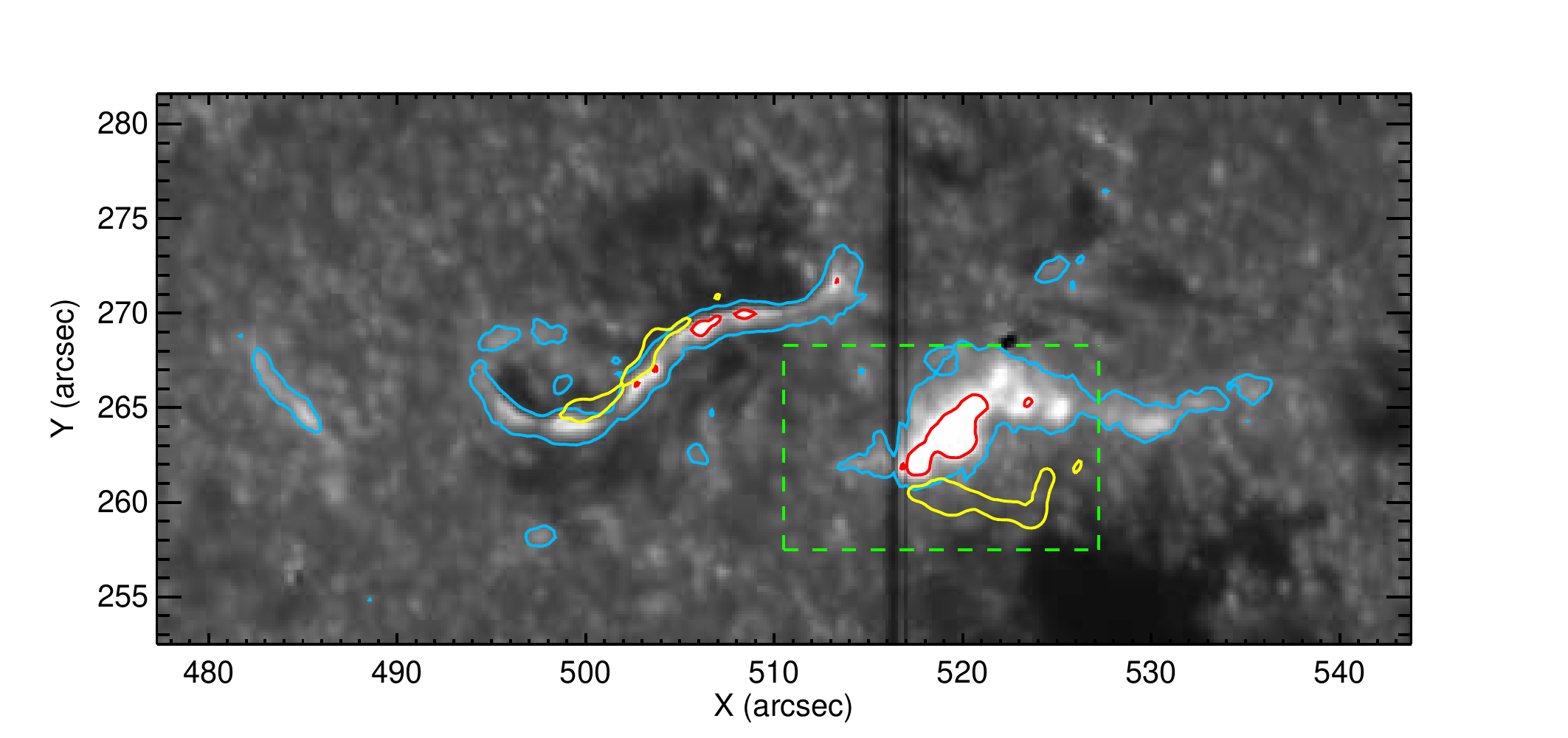} \\
\includegraphics[scale=0.6]{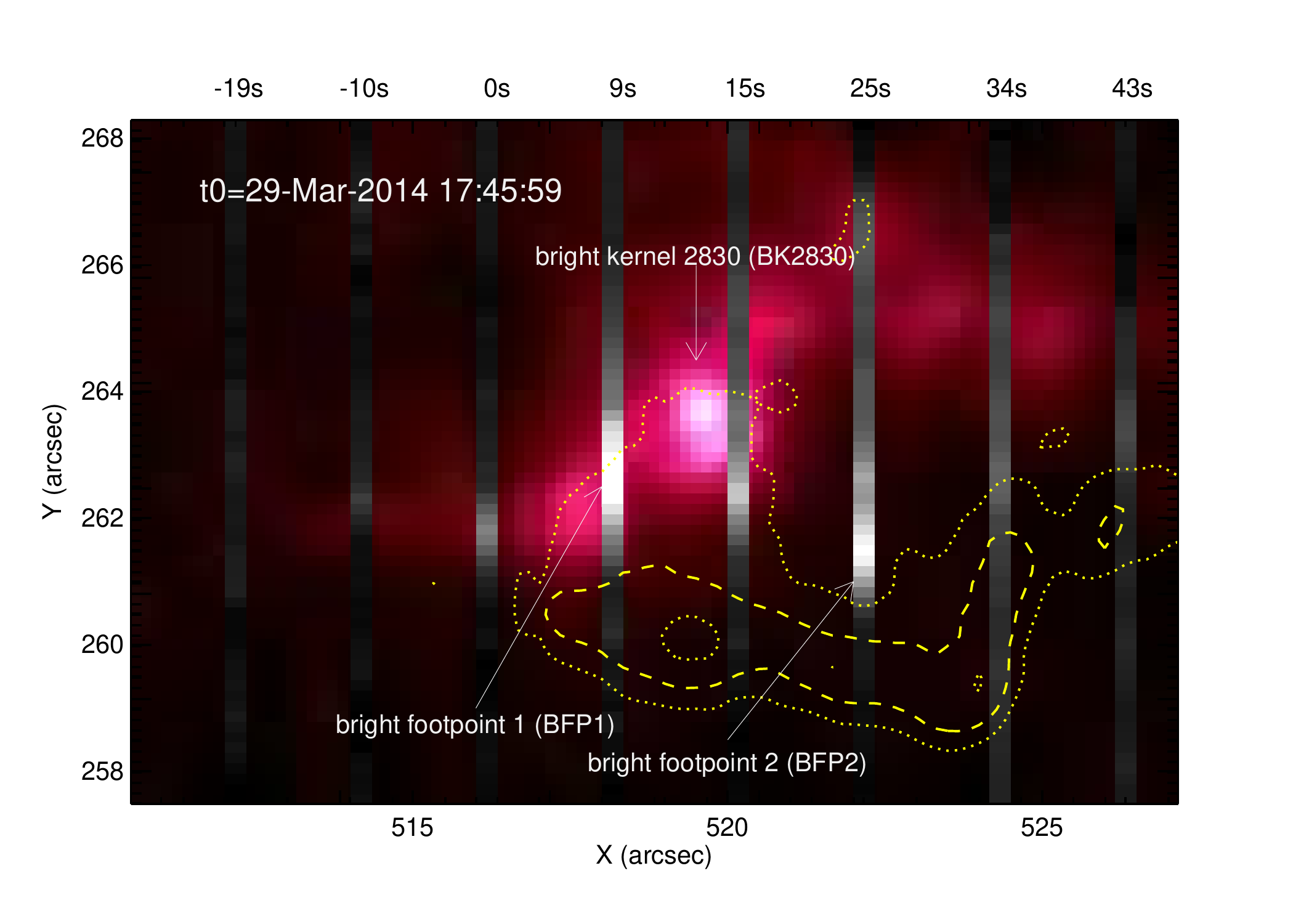}
\caption{
\small
(Top) Total intensity in SJI 2832 \#173.  The areas corresponding to the high
  thresh (red) and low thresh (light blue) values for the excess intensity SJI 2832 \#173 at the time of 29-Mar-2014 17:45:59 are overlayed.  The high thresh area for SJI
  2832 \# 174 is shown in yellow contours.
   (Bottom) An expanded view of the green dashed box in the top panel, showing the excess intensity of SJI 2832 \#173.
  The spectroscopic raster \#173 during the hard X-ray impulsive phase is shown for
  the excess C2826 (NUV continuum) in gray scale, and the times for each spectrum relative to $t0$
  are indicated on the top x-axis.   The high thresh (3$\times10^6$ \cgs) area for SJI 2832 \#174 is indicated by a yellow dashed contour.
  The yellow dotted contours show
   the area with excess intensity above 1.5$\times10^6$ \cgs\ and 6$\times10^6$ \cgs\
   (half of high thresh value and twice the high thresh value, respectively). 
   BK2830 is located at $(x,y)=(519.5,263.8)$.  The ribbon progresses downward over time
   and the IRIS slit crossed two locations of bright NUV flare continuum emission as indicated by BFP1 and BFP2 with arrows.
   \normalsize
     }\label{fig:ref2832}
   \end{center}
\end{figure}

In Figure \ref{fig:ref2832} (top panel) we show the SJI 2832 \#173 image which exhibits brightness variations outside the flare ribbons due to
the sunspots and granulation.  As a result of the evolution of granulation
 and a drift in the instrument pointing, temporal variations of $10-20$\% are present in the excess intensity images.  Thus we define the 
 flare area using
two significance thresholds as follows:

\begin{enumerate}
\item \textbf{Low Threshold Excess (``low thresh'')}
We subtract a pre-flare image (\#171) from each image in the observation.  For the excess images
prior to \#171, we find that the standard deviation of the 
excess count rate approaches a value of $\approx$15 DN s$^{-1}$ pixel$^{-1}$ before the flare, which is due to the granulation variation.
Our low threshold excess area (``low thresh'') corresponds to the flare area with an excess count rate of 60 DN s$^{-1}$ pixel$^{-1}$, which is $\approx0.6\times10^6$ \cgs\  or 4$\sigma$.

\item \textbf{High Threshold Excess (``high thresh'')}
We seek to determine the physical processes that produce the continuum intensity in the brightest footpoints in the flare determined
by the SJI 2832 sources.
We define a high threshold excess that corresponds to an excess
count rate of $\sim$290 DN s$^{-1}$ bin$^{-1}$ or $3\times10^6$ \cgs, which is a factor of five larger than the low-thresh value.  
 The high-thresh level corresponds to $\sim$30\% of the maximum excess pixel value in 
SJI 2832 \#173.  The low- and high-thresh areas are shown as contours on the total intensity image SJI 2832 \#173 in Figure \ref{fig:ref2832} (top panel).  

An approximate comparison between temporally contiguous and spatially
adjacent measurements of the continuum intensity from the spectra and in SJI 2832 justifies the high-thresh value. 
The spectral region at $\lambda=2825.6-2825.9$ \AA\ (hereafter, C2826) shows bona-fide continuum
emission (Section \ref{sec:wl}). At the same spatial ($y$) location in
the excess spectra and in the excess SJI 2832 image,
the spatial extent in the $y$-direction of the excess spectral continuum 
($x=$518\arcsec.2 at the raster position corresponding to $t0+9$~s in Figure \ref{fig:ref2832} (bottom)) has a FWHM of $\approx11$ pixels, or 1\arcsec.8
with a half-maximum excess intensity value of $1.1\times10^6$ \cgs.  
At one raster step earlier in SJI 2832 \#173 (corresponding to $t0$ in Figure \ref{fig:ref2832}), we find that a
N-S spatial extent of 11 pixels (thus, the same extent as the FWHM of
the C2826 continuum emitting area in the spectra) 
corresponds to an excess slit jaw intensity of approximately $3\times10^6$
\cgs.  Thus, a high-thresh value of $3\times10^6$ \cgs\
from SJI 2832 reasonably represents 
  an area corresponding to bright continuum emission, with an excess C2826
  value $I_{\lambda,\rm{excess}}>10^6$ \cgs\ (assuming that the emission has not significantly
  decayed over the time from slit jaw image to the spectral observation).  The high-thresh value is 70\% of the maximum over the $y$-direction
profile at the time of SJI 2832 \#173, and suggests that excess flare continuum
intensity constitutes $\approx50$\% (2.2$\times10^6$ \cgs / 4.3$\times10^6$ \cgs)   of the excess intensity in the
SJI 2832 images. 
 
   \end{enumerate}

\subsection{Comparison to Hard X-ray Emission}

RHESSI data provide critical information on the flare impulsive-phase hard X-ray emission, which is used to infer a single or double power-law distribution of an electron beam using the collisional thick-target model
\citep[e.g.,][]{Holman2003, Milligan2014}.  The collisional thick-target modeling of the hard X-ray RHESSI data and the hard X-ray imaging for this flare has been
performed by \cite{Battaglia2015}, \cite{Kleint2016}, and \cite{Fatima2016}, and we use representative parameters from these fits for RHD modeling (Section \ref{sec:radyn})
of the early impulsive phase before the bright thermal component becomes very bright in the X-ray spectrum.

\begin{figure}[H]
\begin{center}
\includegraphics[scale=0.6]{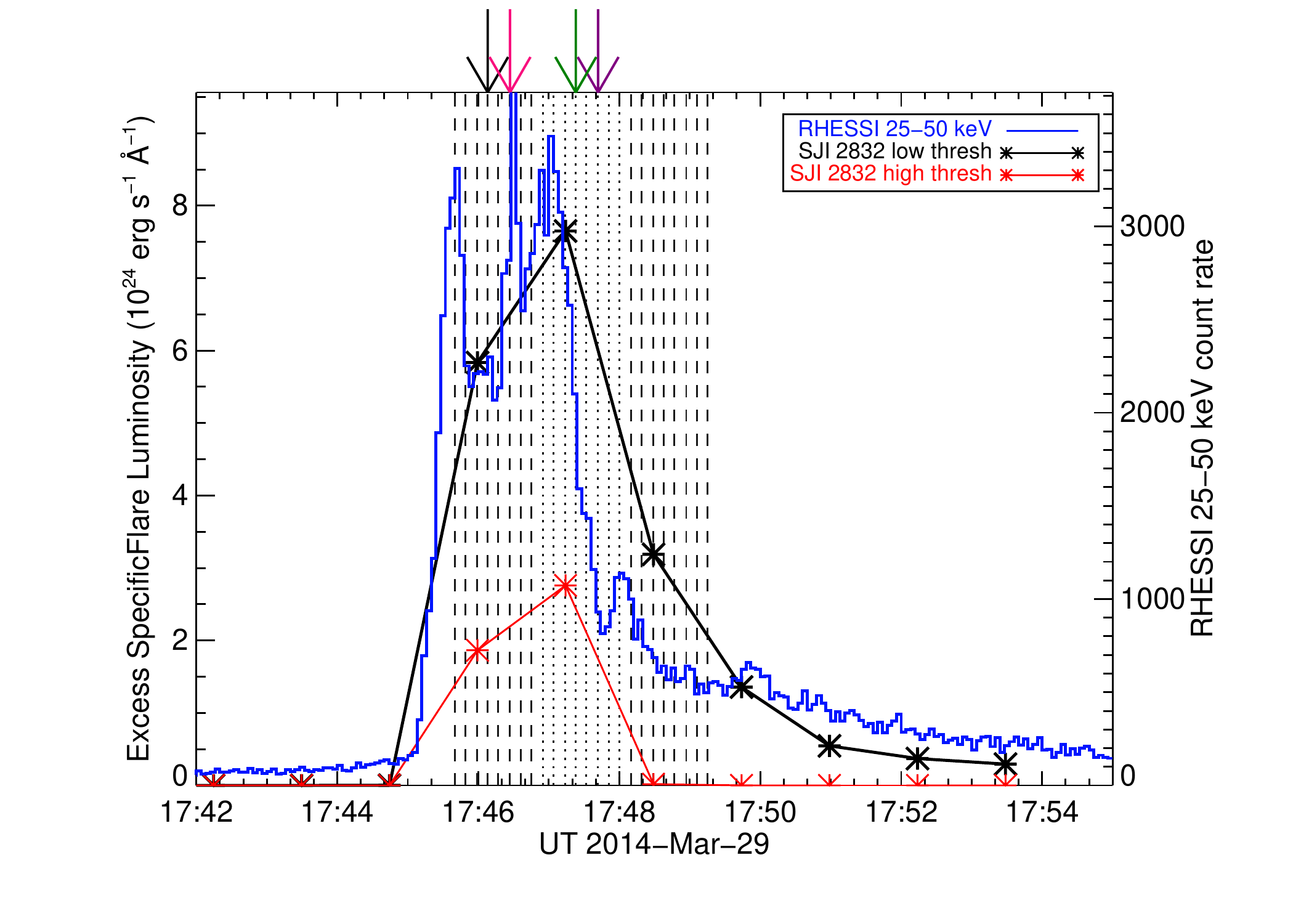}
\caption{Excess specific flare luminosity in the SJI 2832 integrated over the low thresh and high thresh areas
(left axis) compared to the corrected count rate RHESSI hard X-ray $E=25-50$
keV light curve (summed for the detectors 1F, 2F, 3F, 4F, 5F, 6F, 7F,
and 9F; right axis). 
 The vertical  dashed and dotted lines indicate the times of the
 spectral raster observations (the leftmost grouping of vertical dashed lines corresponds to raster \#173, the vertical dotted lines to raster \#174). The arrows at the top indicate the times of the four flare spectra shown
in Figure \ref{fig:brightest}, and are color-coded to the spectra. The spectral observations of bright NUV continuum emission in BFP1 at 17:46:08 (black arrow) and BFP2 at 17:46:24 (pink arrow)
correspond to the middle of the impulsive phase of the hard X-ray light curve.  } \label{fig:rhessi1}
\end{center}
\end{figure}

The hard X-ray $E=25-50$ keV RHESSI light curve (after applying the approximate adjustment for attenuator state changes)
 is shown in Figure \ref{fig:rhessi1} with 
the times of the IRIS spectral observations are indicated by vertical dashed and dotted lines.  The $E=25-50$ keV light curve
defines the impulsive phase of the flare (120 s FWHM), and consists of several peaks with varying durations of $8-30$~s 
superimposed on a gradually varying emission component.
  The eight slit positions in Figure \ref{fig:ref2832} (bottom) 
  correspond to raster \#173, and the times of these spectral observations are indicated by the leftmost eight vertical dashed lines in Figure \ref{fig:rhessi1}. 
   We refer to SJI 2832 \#173 (17:45:59) as the ``mid peak phase'' and SJI 2832 \#174 (17:47:14) as the ``early fast decay phase'' according to
  the phase of the hard X-ray light curve.  Notably, SJI 2832 \#173 occurs just after the 
hard X-ray event peaking between 17:45:36\,--\,17:45:48 that is evident in the high energy bands $E>25$ keV in Figure \ref{fig:rhessi1}.

The excess specific luminosity from 
 the low- and high-thresh flare areas in SJI 2832 are shown in
Figure \ref{fig:rhessi1};  the time profiles are generally similar to the
coarse evolution of the $E=25-50$ keV X-ray light curve.  The areal evolution of the low-thresh (not shown) is similar to the specific luminosity but exhibits 
a faster decay relative to the peak. During the impulsive phase (SJI 2832 \#173 and 174), the 
low thresh area is approximately a factor of 10 larger than the high thresh area, but only 30\% of the excess specific flare luminosity originates from the high-thresh area. In the gradual phase of the hard X-ray light curve, the high thresh area is just six pixels, and thus this threshold
effectively separates the impulsive and gradual phases of the hard X-ray emission in this flare.

\subsection{Flare Footpoint Evolution in SJI 2832} \label{sec:fpevol}
In Figure \ref{fig:ref2832} (top panel), we show the spatial evolution in SJI 2832 of the
low-thresh and high-thresh flare areas. 
The high-thresh areas are shown as red (SJI 2832 \#173) and yellow (SJI 2832 \#174) contours, and
the low-thresh (SJI 2832 \#173) are the light blue contours. Similar to recent high spatial resolution 
data of other C, M, and X-class two-ribbon flares \citep{Sharykin2014, Krucker2011, Kowalski2015HSG},
one relatively narrow (NE) ribbon develops across the umbra and another (SW) ribbon that is more spatially diffuse develops in the plage.
From the mid peak phase (SJI 2832 \#173) to early fast decay phase (SJI 2832 \#174), the location of the high thresh flare area changes relatively rapidly, following
motions that are both perpendicular and parallel to the magnetic polarity inversion line, which runs diagonally from SE to NW through the spots at (508\arcsec, 270\arcsec) \citep{Kleint2015}. 
 From the mid peak  to the early fast decay phases,
the (apparent) motion of the high-thresh footpoints in the SW (plage) ribbon is predominantly perpendicular to the polarity inversion line (nearly parallel to the slit in the image y-direction), as can be seen by comparing the high-thresh locations at the two times in Figure \ref{fig:ref2832}.  However, the brightest regions of the NE (umbral) ribbon rapidly move from the western to the eastern side of the ribbon (in the direction away from the IRIS slit).  
The SW ribbon moves rapidly through the plage towards the large sunspot of negative polarity \citep{Kleint2015} at (530\arcsec, 255\arcsec);  the average apparent
speed of this ribbon in the image y-direction is $30-40$ km s$^{-1}$ and is as high as 60 \kms\ at some locations.  The NE ribbon
moves apparently slower in the direction perpendicular to the polarity inversion line, since it develops over the umbrae of the spots where its spatial development may be limited by the umbral magnetic field.

In Figure \ref{fig:ref2832} (bottom panel) we show a zoomed region of the excess intensity in the SW (plage) ribbon at SJI 2832 \#173 ($t0$). 
This ribbon contains the brightest kernel (BK2830) in the flare at $(x,y)=(519\arcsec.5, 263\arcsec.8)$. 
The high-thresh contours of the excess intensity are shown for SJI 2832 image \#174, which illustrates that a 
 trail of emission extends from BK2830 in SJI 2832 \#173 to the brightest regions in the ribbon in SJI 2832 \#174 as the ribbon has apparently moved toward the 
 SW umbra.  The spectroscopic slit positions and relative timing are indicated, showing that the slit
does not cross the brightest regions of BK2830; the brightest points in BK2830 move diagonally from NW to SE as the raster ``hops'' over it from $t0+9$~s to $t0+15$~s.
Imaging spectroscopy of RHESSI data shows that the brightest hard X-ray source corresponds approximately to this location \citep{Kleint2016, Battaglia2015}. 
After SJI 2832 \#174 (in the hard X-ray gradual phase), the ribbons become fainter and they continue to separate but at significantly slower apparent speed.  

\section{Spectral Analysis of the Brightest Footpoint Emission} \label{sec:wl}
In this section we calculate the continuum intensity and chromospheric line bisector velocities for the 
 two brightest NUV pixels, labeled ``bright footpoint 1'' and ``bright footpoint 2'' in Figure \ref{fig:ref2832} (bottom).    The quantities from these spectra will be compared to the RHD model outputs in Section \ref{sec:radyn}. 
 
\subsection{Continuum Intensity}
\begin{figure}[H]
\begin{center}
\includegraphics[scale=0.6]{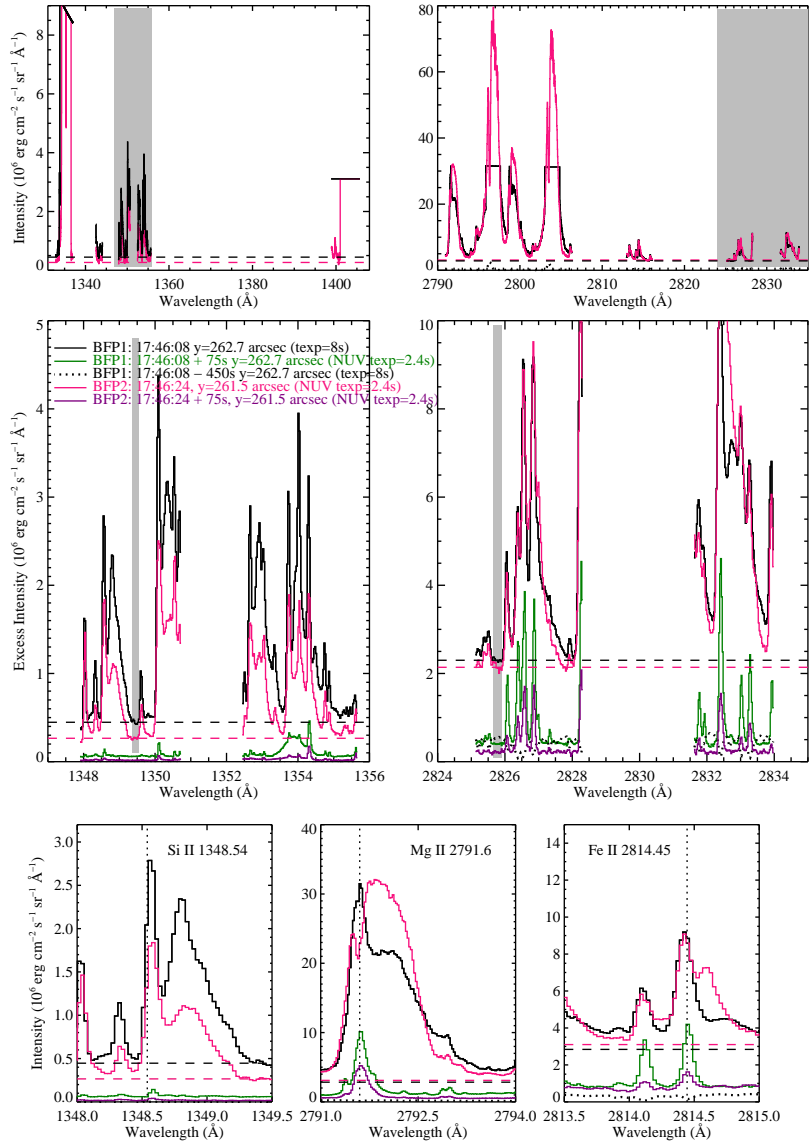}
\vspace{-5mm}
\caption{(Top) FUV and NUV spectra extracted from the brightest flaring pixels in Figure \ref{fig:ref2832} at $t0+9$~s (BFP1, black) and at $t0+25$~s (BFP2, pink) over the wavelength range of IRIS.  Note that only select wavelength regions of the full possible spectral coverage
  were recorded for the observations of this flare.    (Middle)
   Enlarged view of the wavelength ranges indicated by the shaded gray regions in the top panel.
   The wavelength ranges of the continuum regions C1349 and C2826 are indicated
   by shaded vertical bars. The spectra at the same spatial positions as BFP1 and BFP2 are shown one raster later (\# 174) as dark green and purple
   spectra, respectively.  A pre-flare spectrum is shown for the NUV as a dotted spectrum.  (Bottom) Selected chromospheric spectral lines for the same locations
   and times as the spectra in the middle panels;  note that the pre-flare spectrum has not been subtracted here.  Each singly ionized species shows a redshifted emission component and a component
   centered near the rest wavelength.  The redshifted emission component is much less prominent as the continuum level decreases in the hard X-ray fast decay phase.  In all panels, the horizontal dashed lines 
  are the values of C1349 and C2826 from BFP1 at 17:46:08 and BFP2 at 17:46:24 extrapolated to all wavelengths.   } \label{fig:brightest}
\end{center}
\end{figure}

The continuum intensity at the brightest footpoints provides rigorous constraints 
for flare heating model predictions. We search all spectra for bright Mg \textsc{ii} $\lambda$2791.6 and excess C2826 emission to identify the brightest flare sources that cross the IRIS slit.  
We calculate the excess continuum intensity in the FUV (1349.35-1349.53\AA; hereafter C1349) and in the NUV (2825.64-2825.90\AA; C2826).  The excess values are obtained 
by subracting the pre-flare continuum intensity at the same spatial location.  For the emission line analysis, the 
pre-flare is not subtracted because this can affect the line profiles\footnote{The continuum shape can also
be affected by subtracting the pre-flare \citep{Kleint2016}.  Whereas a physical quantity (velocity) is inferred from the line bisector, the observed excess is used only as a direct comparison to the model excess.}.  
The excess C2826 raster image in Figure \ref{fig:ref2832} (bottom) shows the 
 two brightest flaring locations in Mg \textsc{ii} and in the C2826 continuum, which are labeled ``bright footpoint \#1'' (hereafter, BFP1) at the 4th raster position and ``bright footpoint \#2'' (hereafter, BFP2) at the 6th raster position in raster \#173.  
Most of the bright regions in the excess C2826 raster image
aligns with the excess intensity in the SJI 2832 images, but the location of BFP2 in the 6th raster position appears
between the high-thresh contours of SJI 2832 \#173 and \#174; this can be attributed  
 to the rapid spatial development of the flare ribbon towards the SW umbra over the time of the spectral raster (Figure \ref{fig:ref2832}).
The spectra from BFP1 (black) and BFP2 (pink) are shown in the top panel of Figure \ref{fig:brightest}.  The BFP1 spectra (black) were obtained
with exposure times of 8~s in the FUV and NUV, and the BFP2 spectra (pink) were obtained with
an exposure time of 8~s in the FUV and 2.4~s in the NUV.  In the FUV, the major emission
lines (e.g., C \textsc{ii} and Si \textsc{iv}) are saturated for both spectra, but the Mg II h$+$k lines in the NUV
are not saturated for the spectrum of BFP2.  The Mg II triplet $\lambda2791.6$ is not saturated in either spectra.

The excess NUV and FUV spectra for BFP1 (black) and BFP2 (pink) are shown in the middle panel of 
Figure \ref{fig:brightest} for the wavelength regions indicated by gray bars in the top panel of Figure \ref{fig:brightest}.
The wavelength regions of C1349 and C2826 are shown as gray bars in the middle panel of Figure \ref{fig:brightest}. 
 These wavelength regions are
continuum regions outside of major and minor flare emission lines \citep[the
 NUV continuum wavelength range is the region considered by][]{Heinzel2014, Kleint2016}.  
We note that although the gray bar in Figure \ref{fig:brightest} corresponding to C1349 does not include any prominent emission lines, 
this wavelength region includes the rest wavelength of Fe XII $\lambda$1349.4.  This wavelength region is nonetheless the most optimal
estimate of the continuum among the limited wavelength regions that were read out for these observations. 
 The times of BFP1 and BFP2 
compared to the RHESSI impulsive phase are indicated by the two leftmost arrows at the top of Figure \ref{fig:rhessi1}.  BFP1 occurs 
after the hard X-ray peak from 17:45:36\,--\,17:45:48, and there is a short hard X-ray event near the time of BFP2, indicated by the pink arrow in Figure \ref{fig:rhessi1}.
Extrapolations of C2826 and C1349 over the NUV and FUV spectral ranges are 
 shown as horizontal dashed lines, which 
indicate that these continua are the lower pedestal of the emission compared to the rest of the spectra.  
A preflare spectrum and a spectrum corresponding to the early fast decay phase (raster \#174) at the locations of BFP1 and BFP2 are also shown in Figure
\ref{fig:brightest} to illustrate the evolution of the excess continuum emission.
Within one raster (75~s), the excess NUV continuum values have decreased by a factor of $>4$, which is consistent with the rapid motion of the ribbon front towards the sunspot (Figure \ref{fig:ref2832}).

The times, locations, and measured intensity values of C2826 and C1349 are given for BFP1 and BFP2 in Table \ref{table:bfpvals}.  
The values of the excess continuum intensity of BFP1 and BFP2 are
$\sim2.1-2.2\times10^6$ \cgs\ and $\sim0.27-0.45\times10^6$ \cgs\ for
the C2826 and C1349 continuum regions, respectively.  The ratio of excess NUV to excess FUV continuum intensity for BFP1
is $\sim5$.  For BFP2, the exposure times are not the same for the NUV and FUV due to the IRIS AEC adjustment.  
The intensity contrast in the NUV for BFP1 and BFP2 are $200-400$\%
\citep{Kleint2016},  which is 
a large range due to a significant spatial variation in the NUV continuum in the pre-flare spectrum (which typically varies from pixel to pixel by $10^5$ \cgs\ or $10-20$\%).
A third bright footpoint (BFP3) in the 5th raster position at ($x,y)\sim(520$\arcsec,$262$\arcsec) in Figure \ref{fig:ref2832} exhibits an excess C2826 of $\sim1.5\times10^6$ \cgs; but this footpoint is not 
analyzed further in this paper.

\subsection{Bisector Velocity of the Asymmetric Chromospheric Line Profiles} \label{sec:obsRWA}

In Figure \ref{fig:brightest} (bottom panels), we show the emission line profiles of Si \textsc{ii} $\lambda1348.54$, the Mg \textsc{ii} triplet $\lambda2791.6$, and Fe \textsc{ii} $\lambda2814.45$ for BFP1 and BFP2.   
The rest wavelengths are indicated as vertical dotted lines.
A redshifted, broadened emission component is present in each of
these line profiles, as discussed by \cite{Liu2015} for this flare for the Mg \textsc{ii} $h+k$ lines. We refer to this redshifted emission component
as a ``red wing asymmetry'' (RWA), since this redshifted component is qualitatively similar to the 
red wing asymmetry observed in H$\alpha$ profiles from \cite{Ichimoto1984}.
 A spectrally resolved RWA component is no longer present by the next raster 75~s
 later (dark green and purple spectra), while the emission centered near the rest
wavelength remains relatively bright.  We interpret the RWA as
evidence of downflowing, heated chromospheric material as discussed
for the Mg II lines in
\cite{Graham2015} for the X-class flare SOL2014-09-10T17:45.  The rapid disappearance of the 
RWA component over the raster cycle of 75~s is consistent with the findings of \cite{Graham2015} showing that the 
condensation lifetime is $\sim30-60$~s.  
 
Following \cite{Graham2015}, we measure the 30\% bisector of the Mg
 II triplet line to infer a
 chromospheric velocity\footnote{Note, we do not subtract a pre-flare
   spectrum before measuring the bisector.}.  
 The bisector velocities for the BFP1 and BFP2 spectra
 in Figure \ref{fig:brightest} are $\sim40$ km s$^{-1}$, but the detailed line profiles 
differ considerably between the two spectra.

\subsubsection{NUV Fe II Emission Line Analysis} \label{sec:feii}
Many Fe \textsc{ii} emission lines become bright in the NUV during the flare and exhibit complex line profiles
that can be used as a diagnostic of the flare chromosphere \citep{Walkowicz2008}.  These lines are never saturated in IRIS flare spectra and are thus a useful parameter to compare among flares.    Moreover, they are much more optically thin than other, brighter chromospheric flare lines such as the Mg II lines and may provide a favorable alternative as a diagnostic of the flare velocity field through the white-light continuum emitting layers. The Fe II lines have not yet been characterized in IRIS flare spectra. In this section we describe the properties of the line profiles and in Section \ref{sec:feii_model} we compare to model predictions.

The Fe \textsc{ii} flare lines in the NUV
typically have upper levels with excitation energies of $E/hc\sim61,000$ cm$^{-1}$, and we focus our
analysis on the \feone\ and \fetwo\ lines.  
The line profiles for BFP1 and BFP2 are shown in the top panels of Figures \ref{fig:feii_2814} and 
\ref{fig:feii_2832} for \feone\ and \fetwo, respectively.  
The \feone\ and \fetwo\ lines exhibit an emission component with a peak that is within 1 pixel of the rest wavelength
and a RWA emission component that peaks at least 5 pixels to the red of the of the rest wavelength.

 The \feone\ line is one of the least blended with other emission
lines, thus allowing a clean characterization of the RWA emission component.  The peaks of the RWA
emission components for this line in BFP1 and BFP2 are indicated by ``RWA'' in Figure \ref{fig:feii_2814}.
 In the BFP1 spectrum (black) of \feone, the RWA component 
is much less intense than the peak of the line at rest-wavelength (hereafter, indicated as $\lambda_{\rm{rest}}$) and is shifted to the red to $\lambda \sim$2814.75 \AA.  The peaks of the RWA components of \feone\ are redshifted to $\lambda-$\lamr$=16$ \kms\ (BFP2) and $\lambda-$\lamr$=32$ km s$^{-1}$ (BFP1), which are indicated by vertical lines with \lamr\ in Figure \ref{fig:feii_2814}. 
 Notably, the emission line components at \lamr\ and the extrapolated excess C2826 continuum values are equal in intensity for BFP1 and BFP2.
In the BFP2 spectrum (pink), the RWA component is much more prominent compared to the peak of the line at $\lambda_{\rm{rest}}$
and is shifted less to the red.  For BFP2, we subtract the extrapolation of C2826 and measure the 30\% bisector velocity as $+9$ \kms, compared to a redshift of $+16$ \kms\ for the peak of the RWA emission.  The Mg \textsc{ii} $\lambda2791$ bisector velocity
is a factor of 4 larger than the \feone\ bisector velocity, which is due to the much brighter and broader RWA component in the Mg \textsc{ii} line. 

Figure \ref{fig:feii_2832} shows the observations for BFP1 and BFP2 for the \fetwo\ line, which is a stronger line than \feone\ in the flare, but has several other lines (Fe I and Ti II) to the red of \lamr.
In Figure \ref{fig:feii_2832} we also indicate the wavelengths that correspond to velocities of $\lambda-$\lamr$=16$ \kms\ and $\lambda-$\lamr$=32$ \kms\ by vertical lines, which 
identifies similar wavelength positions of the RWA peaks as for \feone.  

We define the quantity $I_{\rm{RWA}} / I_{\lambda_{\rm{rest}}}$ as the ratio of the peak intensity of the RWA component to the peak of the \lamr\ component \citep[see also][]{Liu2015}.
For the BFP2 spectrum (pink spectrum), $I_{\rm{RWA}} / I_{\lambda_{\rm{rest}}} \sim 2/3$ for the \feone\ line, whereas
for the \fetwo\ line, $I_{\rm{RWA}} / I_{\lambda_{\rm{rest}}} \sim 1$, and the value of 
$I_{\rm{RWA}} / I_{\lambda_{\rm{rest}}}\sim1.7$ for the Mg II triplet line (Figure \ref{fig:brightest}). 
For the BFP1 spectrum, the $I_{\rm{RWA}} / I_{\lambda_{\rm{rest}}} << 1$ for \feone\ whereas
for the \fetwo\ line, $I_{\rm{RWA}} / I_{\lambda_{\rm{res}}} \sim 0.5$, and for the Mg II triplet line $I_{\rm{RWA}} / I_{\lambda_{\rm{res}}} \sim 0.6$.  
The RWA component is more intense compared to the \lamr\ component for lines that are more
optically thick, and when the RWA peak is shifted further to the red it has a lower intensity.  
In Section \ref{sec:feii_model} we test our model predictions for the appearance of two emission
line components (RWA and \lamr) and the variation in relative brightness of these line components between \feone\ and \fetwo.

\section{Radiative-Hydrodynamic Flare Modeling}\label{sec:radyn}

\subsection{RADYN Flare Model Setup}
The high spatial resolution measurements of continuum intensity and chromospheric line profiles from IRIS, combined
with constraints from RHESSI on the parameters of nonthermal electron beams, motivate new 
RHD flare models.  
We perform 1D plane-parallel, RHD flare modeling 
with the RADYN code \citep{CarlssonStein1992, Carlsson1994,
  Carlsson1995, Carlsson1997, Carlsson2002} using the updated flare version described
in \cite{Allred2015} to simulate flare heating from a nonthermal
electron beam, in order to determine if the models give 
consistent results with the new IRIS data.  We refer the reader to \cite{Allred2015}
for a detailed description of the flare modeling method.

As our starting model atmosphere, we use a semi-circular loop of half-length 10 Mm that corresponds
to the QS.SL.HT model from \cite{Allred2015},
which has an apex electron density and temperature of $8\times10^9$
cm$^{-3}$ and 3.2 MK, respectively.  This starting atmosphere is
closest to the plage environment where the majority of the high-thresh
emission is observed in this flare (Figure \ref{fig:ref2832}), and the model loop length is consistent with the RHESSI hard X-ray footpoint separation
in the early impulsive phase \citep{Battaglia2015}.  In this pre-flare atmosphere, the transition region occurs at $z\sim1150$ km, where $z=0$ is defined as the height where 
$\tau_{5000}=1$ at $t=0$.  \cite{Carlsson2015} modeled Mg II lines from IRIS and concluded that plage regions have deeper transition regions due to the conductive flux from a hot, dense corona.  

The equations of conservation of mass, momentum, energy, charge are solved together with the rate equations and the equation of radiative transfer on an adaptive grid scheme from \cite{Dorfi1987}.  The non-LTE (NLTE) problem is solved in RADYN using the technique of
\cite{Scharmer1981} and \cite{Scharmer1985}
for three model atoms:  a hydrogen atom (six levels including H \textsc{ii}), a helium atom (five levels for He \textsc{i}, three levels for He \textsc{ii}, and He \textsc{iii}), and a singly ionized
calcium ion (six levels including the Ca \textsc{iii} ground state),
giving a total of 22 bound-bound (b-b) and 19 bound-free (b-f)
transitions calculated in detail.  The optically thin loss function
from \cite{Kowalski2015} accounts for other b-b transitions not treated in detail, but excludes several ions that 
are likely optically thick at low temperature near $T\sim10,000$ K (see Section \ref{sec:limitations}).  Everything else is kept the same as in
\cite{Allred2015} except that we have excluded the Mg \textsc{ii}
ion from the detailed radiative transfer and cooling rates.
  The proper calculation of the Mg \textsc{ii} $h$ and $k$ lines
during flares requires a prescription for several important effects that are not yet included in the RADYN code, such as
  overlapping b-b transitions and partial frequency redistribution, which may be affected by elastic collisions with the electron beam \citep{Hawley2007}.  The cooling rate from the Mg \textsc{ii} ion can be important for flare
atmospheres \citep{Avrett1986} but modifying RADYN to include overlapping transitions and partial redistribution is outside the scope of this work.

Wavelengths where the continuum is calculated in detail have been added at $\lambda=$2826 \AA\ in the NUV and $\lambda = 1332, 1358, 1389, 1407$ \AA\ in the FUV in order
to compare to observed intensity measurements from IRIS.  The continuum emission in the NUV calculated in RADYN includes all possible continua 
except for the opacity from the absorption wings of Mg \textsc{ii}, which extend through the entire IRIS NUV wavelength range. 
We use the RH code \citep{Uitenbroek2001} to include the Mg II $h+k$ b-b opacities in the far wing at several times in the dynamic simulation at $\lambda=2826$ \AA\ using a method discussed in Section \ref{sec:continuum}.
Five $\mu$ values are calculated in the models ($\mu=0.05, 0.23,0.5,0.77$, and $0.95$); 
we compare the model values at $\mu=0.77$ which is closest to the observations of the flare ($\mu=0.82$).

\subsection{Flare Heating Inputs}
The brightest hard X-ray RHESSI footpoints \citep{Battaglia2015, Kleint2016} are cospatial with the brightest continuum emitting 
footpoints in the \marflare\ flare in Figure \ref{fig:ref2832}.  To model the flare heating at these locations, we use a state-of-the-art prescription for 
energy deposition from a nonthermal electron beam which has been included in the 
 RADYN flare code \citep{Allred2015}.  The energy deposition rate is determined from the solution to the steady-state
Fokker-Planck equation given a
power-law of nonthermal electrons injected at the top of the model atmosphere.  The Fokker-Planck solver has been adapted from 
\cite{McTiernan1990} and is available at \url{http://hesperia.gsfc.nasa.gov/hessi/modelware.htm}.  
The power-law parameters at the
top of the model loop are chosen to be $E_c=25$ keV and $\delta=4.2$, 
which are consistent within the uncertainty range of the thick-target modeling presented in \cite{Kleint2016}.  

An important input for the RHD models is the energy flux density in nonthermal electrons, which
cannot unambiguously be obtained from RHESSI data alone because the width of flare ribbons is known to 
be smaller than the resolution of RHESSI \citep{Krucker2011, Sharykin2014, Jing2016}.  We use the high spatial resolution of SJI 2832 to estimate
a flare footpoint area for the calculation of the nonthermal energy flux.  
\cite{Kleint2016} derived an energy flux of $3.5\times10^{11}$ \fluxunits\ for the bright hard X-ray source at 17:46:15-17:46:25 UT in the SW plage ribbon using an unresolved width of 1\arcsec, which gives an area of $2.4\times10^{16}$ cm$^{2}$ for the source \citep[see also][]{Judge2014}. The high-thresh area\footnote{Areas are de-projected using the $\mu$ value of 0.82 for these observations.} in the SW plage ribbon in SJI 2832 corresponds to 
$\lesssim 5\times10^{16}$ cm$^{2}$, which is an upper limit because the cadence of the observations is not fast enough to determine which areas are newly
formed flare areas and which areas correspond to decaying emission.   In the spatial profile of the excess C2826 (Figure \ref{fig:spatial}, right panel), there are several
  peaks in the wake of the brightest part of the ribbon.  A Gaussian with FWHM of 1\arcsec.2 can account for the ``leading edge'' \citep{Isobe2007, Krucker2011} of the flare ribbon.
  Using the high-thresh area gives an acceptable 
  lower limit on the flux of $1.5 \times10^{11}$ \fluxunits, which is likely a factor of $\sim2$ too low because the leading edge is the relevant area to divide into the hard X-ray power.

  If we use the area of BK2830 ($A\sim1.5\times10^{16}$
cm$^2$ at FWHM intensity) at $(x,y)=(519\arcsec.5,
263\arcsec.8)$ in Figure \ref{fig:ref2832} (bottom), the flux would be 
$\ge 5\times10^{11}$ \fluxunits.  As discussed by \cite{Battaglia2015}, the RHESSI $35-100$ keV images at
17:46 (SJI 2832 \#173) show
two sources corresponding to the umbral and plage 
ribbons, and thus it is reasonable to assume that the spatially integrated nonthermal power 
 originates from some flaring area that is associated with both ribbons and the RHESSI emission from the SW ribbon may include more flare area than BK2830. 
The high-thresh area for the mid peak phase
 includes area from both ribbons (see Figure
 \ref{fig:ref2832}, top panel) while the area of high thresh and hard X-ray emission are comparable (within a factor of 2),
suggesting that the high thresh is a reasonable proxy for the 
hard X-ray RHESSI emission in this flare.

Thus, we consider a range of nonthermal energy flux for our modeling to bracket the (spatially
averaged) values in the plage flare ribbon from \cite{Kleint2016}:
a model with a flux of $F_{\rm{NT}}=10^{11}$ \fluxunits\ (F11; discussed in \cite{Kuridze2015}) and a high beam flux
model with $F_{\rm{NT}}=5\times10^{11}$ \fluxunits\ (5F11).  We note that lower values of the flux of $2-5\times10^{10}$ \fluxunits\ 
have been inferred for this flare \citep[e.g.,][]{Battaglia2015};  see \cite{Heinzel2015IAUS} for RHD modeling
of these flux levels using the \emph{Flarix} code.

The nonthermal beam energy deposition duration is chosen to be 20~s for the F11 model.  For the 5F11
model, we choose a short duration of flare heating for 4~s and an extended duration of heating for 15~s.  
\cite{Fatima2016} used the derivative of the GOES $1-8$ \AA\ light curve to infer heating timescales of $8-20$~s at times around 17:46.
However, we note that there are hard X-ray variations in the RHESSI light curve in Figure \ref{fig:rhessi1} that occur on shorter timescales, and 
\cite{Penn2016} show demodulated RHESSI light curves (1~s integration times) for another flare that 
exhibit short duration hard X-ray events with FWHM durations of only $4-5$~s.
 Thus, we explore a range of heating durations ($4-15$~s) for the 5F11 simulation.   
 We use a constant heating profile and power-law index
 over the heating durations.   
 
 In the 5F11 model, shocks develop in the chromosphere (Section \ref{sec:dynamics}) and the time-steps
 become unmanageably small due to a radiative instability near the upper, lower density shock \citep[see][for a detailed analysis of this atmospheric region]{Kennedy2015}.  At $t=3$~s we adjust the 
 second derivative of the adaptive grid weights to weight toward the higher density shock, and we decrease the accuracy of the minor level populations \citep[see][]{Kowalski2015}. These adjustments
 allow the 5F11 simulation to progress with larger time-steps.
 After 4~s of heating in the 5F11 model and 20~s of heating in the F11 model, the atmosphere 
is allowed to relax for 9~s and 60~s, respectively, but the gradual phase evolution is not analyzed here. 
 
We also analyze a coronal heating simulation without energy deposition from an electron beam.  We simulate
the atmospheric response to an energy flux of $10^{11}$ \fluxunits\ deposited uniformly ($Q=125$ erg cm$^{-3}$ s$^{-1}$) over the upper
7.5 Mm of the model corona.  The heating duration for this model is 5~s, and produces a corona with a temperature of $T\sim$30 MK.
This simulation is used for a comparison of the NUV continuum emission and line profiles that are produced from a large conductive flux into 
the chromosphere, as done by other authors \citep[e.g.][]{ReepHolman2016}.  A flux of $10^{11}$ \fluxunits\ is 50 times larger than the value of the conductive flux from the heated corona in this flare found at the locations
without detectable RHESSI hard X-ray emission \citep{Battaglia2015}.

The modeling analysis is divided into the following sub-sections:  in Section
\ref{sec:observables}, we calculate the excess NUV continuum
intensity and 30\% bisector velocity for the H$\alpha$ line.  The continuum quantities can be directly compared
to the observations, and the H$\alpha$ bisector is used as a proxy to what we expect for an optically thick
line like the Mg \textsc{ii} lines in the NUV.  In Section \ref{sec:dynamics} we summarize the hydrodynamics in the 5F11 simulation, which most
adequately explains the observed properties of BFP1 and BFP2.
 In Section \ref{sec:continuum}, we discuss the
origin of the NUV continuum emission in the 5F11 model.  In Appendix A, we discuss the upper photospheric heating and optical continuum emission
in the 5F11 model.
In Sections \ref{sec:feii_model} and \ref{sec:feii_model2}, we analyze the physical processes that reproduce the Fe \textsc{ii} line profiles from Section \ref{sec:feii}. 
We analyze two time steps in the 5F11 model in detail at $t=1.8$~s and 3.97~s, and we also consider the 
differences in the model predictions at $t>4$~s in the short heating run (4~s heating duration) and the extended heating run (15~s heating duration).

\subsection{Model Observables} \label{sec:observables}
For the model runs, we calculate an excess C2826 continuum intensity and 30\% line bisector for comparison to the 
observations of BFP1 and BFP2 (Section \ref{sec:wl}).  
 The excess continuum values are calculated by subtracting the pre-flare model spectrum.  
 
 The model observables are summarized in Table \ref{table:model}.
 For the excess C2826 continuum intensity, 
we first discuss the excess from the RADYN calculation (column 2), which does not include the Mg II wing opacity.   In Section \ref{sec:continuum}, 
we use the RH code to include the Mg II level populations for a refined calculation of the excess C2826 at select time steps (column 3).  The
excess C2826 values with Mg II wing opacities are comparable to or larger than the excess C2826 values obtained directly from RADYN;  these differences are discussed in Section \ref{sec:continuum}.

\subsubsection{Excess Continuum Intensity} \label{sec:radyn_excess}

The time-evolution of the excess C2826 from the RADYN simulation is shown in Figure \ref{fig:lc2826}
for the the 5F11 model (for the extended 15~s and short 4~s beam heating durations).
The C2826 attains a value of 2.2$\times10^{6}$ \cgs\ after 1.8~s, and thus the excess NUV continuum intensity
of the brightest flare footpoints BFP1 and BFP2 (Section \ref{sec:wl}, Table \ref{table:bfpvals}) is achieved 
 in the 5F11 simulation.
 The NUV continuum continues to brighten to 5$\times10^{6}$ \cgs\ at 4~s in the 5F11.
After 5~s, the
NUV continuum decreases slowly for the extended heating model and decreases
rapidly in the short heating model.
 The F11 produces an excess C2826 that is at least a factor of three
 lower at all times.   The coronal heating model produces an excess C2826 of $\lesssim4\times10^5$ \cgs, which is nearly a factor of two lower than the value that the F11 attains 
 at comparable times.  The relationship between the values of the excess C2826 and the atmospheric response are discussed in detail in Section \ref{sec:continuum}.

     The IRIS integration times of the NUV spectra range from 2.4-8~s, which are long compared to the continuum brightness time-evolution in the 5F11 model.
   Accounting for an IRIS integration time by averaging over the first 8~s 
  gives excess C2826 values of $\sim4\times10^6$ \cgs\ for the extended 5F11 heating run and 1.2$\times10^6$ \cgs\ for the short 5F11 heating run.
 An excess C2826 value of $\sim4\times10^6$ \cgs\ is notably brighter than either BFP1 and BFP2; in Section \ref{sec:sjibright}, we compare the 5F11 prediction to the brightest pixels in the SJI 2832 image \#173 of BK2830 (Figure \ref{fig:ref2832}). 

\subsubsection{Chromospheric Line Bisector}
The Mg \textsc{ii} triplet lines in the NUV are not computed in the RADYN models, and thus we cannot directly compare
the bisector velocities to the observations.  As a proxy for the optically thick chromospheric emission lines, we use the H$\alpha$ line which is
calculated in RADYN. The line profiles for H$\alpha$ at $t=1.8, 3.97$~s are shown in Figure 
\ref{fig:Halpha}, and the 30\% bisector velocity values are given in Table \ref{table:model} (column 4).   In the 5F11, a $\sim30$ km s$^{-1}$ bisector velocity obtained 
from H$\alpha$ is near the values obtained from Mg \textsc{ii} $\lambda2791.6$ (30\% bisector velocities of $\sim40$ km s$^{-1}$; Section \ref{sec:obsRWA}).  
A similar 30\% bisector velocity measure of H$\alpha$  is obtained in the coronal heating model.   Large bisector redshifts 
are not obtained in the F11 model in the first 20~s of beam heating \citep[see also][]{Kuridze2015}. 

We conclude that the NUV continuum excess intensity and the bisector velocity of optically thick chromospheric lines are both adequately
reproduced in the 5F11 model.  In Sections \ref{sec:continuum} - \ref{sec:feii_model2} we use the 5F11 model to
describe the physical processes that produce the excess NUV continuum (C2826) intensity and the redshifted emission (RWA) component in the chromospheric Fe \textsc{ii} lines in Figures \ref{fig:feii_2814}-\ref{fig:feii_2832}.

\subsection{Dynamics of the 5F11 Model} \label{sec:dynamics}

The atmospheric evolution of the 5F11 is similar to the 
response of an M dwarf atmosphere to an F13 electron beam described in \cite{Kowalski2015}; and we refer the reader to Section 3.2 of \cite{Kowalski2015} for a detailed description.  
Compared to the F13, the dynamical timescales of the 5F11 are longer and the magnitudes of the changes in the atmospheric parameters are generally less extreme.
The main points are as follows:   At $t=1.8$~s, explosive mass motions 
upward  and downward have already developed from the formation of a $T\sim$10 MK temperature plug
 at mid-chromospheric heights.  The downward mass motions that originate in the chromosphere are referred to as the CC (chromospheric condensation) and the upward motions as chromospheric
evaporation. The evaporation velocities 
in the lower corona ($n_e=10^{11}-10^{12}$ cm$^{-3}$) range between $150-500$ km s$^{-1}$.  These 
were produced initially by a large increase in thermal pressure below the pre-flare transition region, and this is the evaporation of transition region material.   On the lower side of this high temperature region, the
large increase in thermal pressure drives material into a high density ($\rho_{\rm{max}}=1.7\times10^{-10}$ g cm$^{-3}$, $T\approx23,000$ K) CC with a height extent of approximately 25 km. 

The CC is a heated compression of the lower atmosphere that increases in density and decreases in temperature and speed as it accrues dense material at lower heights.
 A time step of the velocity and density evolution of the 5F11 
at $t=3.97$~s (when the NUV continuum is nearly maximum) is shown in Figure \ref{fig:vz_evol}.  The evolution here can be compared directly
to Figure 2 from \cite{Kowalski2015} that describes the CC evolution in an F13 electron beam simulation (see also Appendix B.1 here). The bottom panels of Figure \ref{fig:vz_evol} show the temperature, velocity, and mass density evolution at the location of the highest mass density in the CC for the 5F11 model with extended heating.  The downflowing velocities of the CC correspond to material with temperatures ranging from $T\approx 7500 - 5$ MK throughout the simulation, and the temperature at the maximum density decreases from $T\sim10^5$ K initially to $10^4$ K at 4~s, as the CC increases in density.  
The condensation continues to cool below $T=10^4$ K, increase in density, and decrease in speed after 4~s (for the extended heating run).  The CC evolution qualitatively follows the analytic description from \cite{Fisher1989} and the model atmosphere response to a softer ($\delta = 7-8$) electron beam over a longer ($\sim100$~s) time in \cite{Kennedy2015}.
The 5F11 model is the first model to follow the development of a high flux beam-heated CC past the peak of the NUV response on short timescales.

\subsection{NUV Continuum Analysis of the 5F11 Model} \label{sec:continuum}
Figure \ref{fig:contspec} shows the evolution of the RADYN continuum spectrum in the 5F11 model.  The flare spectrum consists of bright
emission blueward of the Balmer limit at $\lambda=3646$ \AA.  In this section, we discuss the atmospheric parameters and emission mechanisms that produce the emergent C2826 flare continuum intensity at $t=1.8$~s and 3.97~s;  these times correspond to the mid-rise and near the peak, respectively (Figure \ref{fig:lc2826}).
In Appendices A and B, we discuss the properties and evolution of continuum emission at other wavelengths in the NUV and optical that are indicated in Figure \ref{fig:contspec}.

We calculate the contribution function to the emergent continuum intensity \citep{Magain1986, Carlsson1998} and the optical depth at $\lambda=2826$ \AA\ ($\mu=0.77$) for these representative times and for $t=0$~s.  Following \cite{Kowalski2015} and \cite{Leenaarts2013}, we use the following form of the contribution function: 

\begin{equation} \label{eq:didz}
C_I = \frac{dI_{\lambda}}{dz} = \frac{j_{\nu}}{\mu} e^{-\tau_{\nu}/\mu} \frac{c}{\lambda^2}
\end{equation}

\noindent where $j_{\nu}$ is the total continuum emissivity at a given height and $\tau_{\nu}$ is the optical depth at a given
height obtained by integrating the total continuum opacity ($\chi_{\nu}(z)$) over height.  The NLTE 
spontaneous b-f emissivity and b-f opacity corrected for stimulated emission \citep[Equation 7-1 of ][]{Mihalas1978}
 are calculated using a six-level hydrogen atom with the population densities of each level given by
 the non-equilibrium ionization, NLTE populations from RADYN at each time step.  The hydrogen f-f emissivity (where $B=S$) is included using the non-equilibrium ionization, NLTE proton density.
Hydrogen-like helium has recombination edges in the NUV, and we include the NLTE b-f opacity and emissivity from these continua.
  Other continuum opacities and emissivities (from higher levels of hydrogen, H$^-$ and metals\footnote{Although the population of H$^-$ is not considered in the level population equation, its population density is calculated from the actual non-equilibrium, NLTE densities of electrons and neutral hydrogen atoms and included in the opacity in the equation of radiative transfer.})
are calculated in LTE, as done internally in RADYN.  We include the Thomson and Rayleigh scattering in the opacity and in the emissivity by calculating 
 the angle-averaged intensity with a Feautrier solver from the RADYN code.  These two processes are non-negligible compared to the total emissivity during the flare only at heights where the 
NUV continuum contribution function is very small.
 
We use the RH code \citep{Uitenbroek2001} to include the Mg \textsc{ii} wing opacity in the calculation of the emergent NUV continuum intensity, for a refined comparison to the excess C2826 obtained from the spectral observations.
 RH includes partial redistribution and overlapping opacities, which are appropriate for the Mg II $h+k$ line wings.  Using a 3 level $+$ continuum model 
 Mg \textsc{ii}$+$\textsc{iii} atom, we calculate the NLTE population densities of the upper and lower levels of the $h$ and $k$ lines with statistical equilibrium.  Thus, we are able to include the equilibrium ionization, NLTE Mg II $h+k$ wing opacity and emissivity at $\lambda=2826$ \AA\ in Equation \ref{eq:didz} with the non-equilibrium ionization, NLTE calculations of hydrogen and helium from RADYN.
 
 The excess C2826 intensity values including the Mg II wing opacities are calculated by integrating Equation \ref{eq:didz}  and are given in Table \ref{table:model} (column 3, ``RH'') at representative times: at $t=0, 1.8, 3.97, 15$~s in the 5F11 model, at $t=3, 18$~s in the F11 run, and at $t=3$~s in the coronal heating model\footnote{For $t=1.8$~s in the 5F11 and $t=3$~s in the coronal heating model, we have set the velocities in the atmosphere to 0 km s$^{-1}$ to facilitate convergence.  We do not expect this to affect the calculation of continuum wavelengths in the far wing of the line.}.
The C2826 intensity for the $t=0$~s solar atmosphere is lower by 40\% in the RH calculation than in the RADYN calculation, 
which is expected when the photospheric wing absorption profile is
included.  The RH calculation at $t=0$~s falls between the range of values for the observed pre-flare intensity (Table \ref{table:bfpvals}) as expected since BFP1 occurs near a darker region and 
BFP2 occurs near brighter plage, but there is likely an error due to 3D effects and excluding Mg I in the NLTE calculation \citep{Leenaarts2013}.
 For the excess flare intensity values, we find that the excess intensity in the F11 model at $t=18$~s is $10^{6}$ \cgs\ and is higher than from the RADYN calculation by about 40\%, yet at least a factor of two lower than the 
observed C2826 values for BFP1 and BFP2.   The refined excess C2826 values with Mg \textsc{ii} $h+k$ opacities are shown in Figure \ref{fig:lc2826} (open circles) for comparison to the time-evolution of the excess C2826 from RADYN.
The excess C2826 values in the 5F11 model are $5-10$\% larger in the RH calculation than in the RADYN calculation but follow the same trend. 
The 5F11 continuum prediction is closest to the observations at $t=1.8$~s, as concluded from the RADYN model continuum excess spectra without the Mg II wing opacity.  

At $t\le0.05$~s in the 5F11, the contribution function for $\lambda=2826$ \AA\ at photospheric and sub-photospheric heights decreases at the onset of the flare heating due to continuum dimming from high nonthermal collision rates in the chromosphere \citep{Abbett1999, Allred2005}.  After $t=0.05$~s, the maximum of the contribution function in the sub-photosphere (at $z\sim-40$ km) is $\sim70$\% of the $t=0$~s value due to opacity from increased thermal rates in the heated chromosphere and upper photosphere. 
The excess continuum intensity calculations that include the Mg II $h+k$ wing opacity (``RH'' column 3 in Table \ref{table:model}) are thus larger than the 
excess continuum values without the wing opacity (``RADYN'' column 2 in Table \ref{table:model}) because there is a smaller amount of decrease of (sub-)photospheric emission when Mg II $h+k$ wing opacity is included.

The contribution functions for the emergent intensity ($I_{\mu=0.77}$) at $\lambda=2826$ \AA\ at $t=1.8$~s and 3.97~s are shown in Figures \ref{fig:jtot1} and \ref{fig:jtot2} compared to the temperature
and density structure at these two times.  Following \cite{Kowalski2015IAU}, we also calculate the normalized cumulative contribution function
$C_I^{\prime}$:

\begin{equation} \label{eq:ciprime}
C_I^{\prime} (z,\mu) = 1 - \frac{\int_{z\ge z_{\rm{lim}}}^{z=10\rm{Mm}} C_I(z,\mu) dz}{\int_{z_{\rm{lim}}}^{z=10\rm{Mm}} C_I(z,\mu) dz}
\end{equation}

where $z$ is the height variable (the height variable is defined as $z=0$ at $\tau_{5000}=1$; $z=10$ Mm corresponds to the top of the model corona), and the denominator is equal to the emergent intensity $ I_\lambda(\mu)$ if $z_{\rm{lim}}$ corresponds to the height of the lowest grid point of the model atmosphere.   The physical depth range parameter, $\Delta z$, is the height difference between 
$C_I^{\prime}=0.95$ and $C_I^{\prime}=0.05$ and thus quantifies the vertical extent of the atmosphere over which  
a majority of the emergent intensity is formed\footnote{In \cite{Kowalski2015}, the physical depth range of
  the CC was calculated as the FHWM of the
  contribution function within the CC.}. The variation of $\Delta z(\lambda)$ is a proxy for comparing the optical depth
among continuum wavelengths.  
  $C_I^{\prime} (z)$ is also useful for determining the fraction of emergent
intensity that originates from atmospheric layers above $z$.  The $C_I^{\prime} (z,\mu)$ curves are shown in Figures \ref{fig:jtot1}-\ref{fig:jtot2} and linearly scaled from 0 to 1 on the 
right axis ($z_{\rm{lim}}=-60$ km).

\subsubsection{The Origin of the NUV Continuum Intensity in the 5F11 Model}
The excess NUV continuum intensity (C2826) during the 5F11 originates over low optical depth continuum-emitting atmospheric layers at $z>500$ km, which we refer to as the flare chromosphere.  
In Figures \ref{fig:jtot1}-\ref{fig:jtot2}, the $C_I^{\prime}$ curves are flat from $z\approx150-550$ km showing that at these times
there is not any contribution to the emergent NUV intensity over this height range.   
At $z<500$ km, there is not a relatively large amount of enhanced NUV continuum emissivity during the rise phase (from $t=0-4$~s in Figure \ref{fig:lc2826}).  Only at later times in the extended
5F11 simulation does the upper photospheric temperature increase (Appendix A) to produce a net increase in the height-integrated continuum emission at C2826 at $z<500$ km. 
For calculations of $\Delta z$ in the first 10 s, we use $z_{\rm{lim}} = 500$ km in Equation \ref{eq:ciprime} since increased emissivity predominantly occurs in higher layers.
The physical depth range for C2826 at 1.8~s as $\Delta z=$390 km, from $z\sim630-1020$ km.  At 3.97~s, $\Delta z=160$ km, from $z\sim750-905$ km.  Thus the region
producing the  NUV continuum emission becomes narrower while peaking near $z\sim900$ km during the rise phase of the excess C2826 as it increases in brightness by a factor of $\sim2$.

  In the bottom panels of Figure 
\ref{fig:jtot1} and Figure \ref{fig:jtot2}, we show the emissivity for several atomic processes that are included in the contribution function for C2826.   
Where most of the excess NUV continuum intensity originates ($z>500$ km), the dominant process for the emission of NUV photons is hydrogen (Balmer, $n=2$) b-f emission. 
 Although the optical depths increase during the flare, $\tau_{2826}$ attains a value of only 0.15 at $z=500$ km at $t=1.8$~s and increases to only 0.2 at $t=3.97$~s.  These values of the optical depth are comparable  to the value obtained in \cite{Heinzel2014} using the static F11 beam model calculations from \cite{Ricchiazzi1983}.
The density increases in the CC at $t=3.97$~s to $n_{\rm{H, max}} = 5\times10^{14}$ cm$^{-3}$ 
but the optical depth at $\lambda=2826$ \AA\ in the CC remains low at $\tau_{2826}\sim0.07$. The radiation from the CC is optically thin 
because the relatively low column density for $n=2$ level of hydrogen, which predominantly determines the optical depth at this temperature, density, and wavelength,
in the CC is only $2\times10^{15}$ cm$^{-2}$.  This value is far from the value that produces optically thick emission
in the NUV \citep[$3\times10^{17}$ cm$^{-2}$;][]{Kowalski2015}.  The physical depth range does not vary largely with wavelength over 
the NUV and optical wavelength regimes in the 5F11 and the logarithm of the contribution function for C2826 traces the logarithm of the electron density profile and evolution (Figures \ref{fig:jtot1} and Figure \ref{fig:jtot2}), which are generally indicative of optically thin hydrogen recombination radiation from a flare atmosphere\footnote{Figures 2, 4, and 6 of \cite{Kowalski2015} illustrate
how the contribution function of Balmer continuum emission over large optical depth does not follow the electron density in a flare atmosphere with a much more dense CC than in the 5F11 model, and Figure 2 of \cite{Kowalski2015IAU} shows
how the physical depth range of emergent continuum intensity is affected by large optical depth.  See also Appendix B.1 here.}.
Because the observed excess NUV continuum intensity is achieved in the 5F11 model (Figure \ref{fig:lc2826}), we conclude that the NUV continuum emission from BFP1 and BFP2 can be explained by
 Balmer continuum radiation from a flaring chromosphere with a CC that has a high density but low optical depth.

The flare-enhanced NUV continuum intensity primarily originates from two beam-heated flare layers with increased electron density at $z>500$ km: downflowing CC layers and non-moving flare layers below the CC.  
In the top panels of Figures \ref{fig:jtot1} and \ref{fig:jtot2}, the 
downflowing (negative) velocities correspond to peaks in the electron density and in the C2826 contribution function (bottom panels); these are the cooler layers of the CC with $T \lesssim 25,000$ K (the CC includes material with temperatures as high as several MK over the evolution, but only the high density cooler regions give rise to the 
continuum and NUV line emission).
We define the vertical extent of the CC as the layers where the velocity is $v<-5$ km s$^{-1}$ and where $C_{I, 2826}^{\prime} > 0.05$ (i.e., those that are visible in NUV continuum light).  The CC is only about $\sim25$ km in vertical extent at these two times. There is also continuum emission that originates from the layers with nearly negligible velocity that extend from the bottom of the CC to several hundred km below the CC. There is a small upflow velocity of $<1$ km s$^{-1}$ that develops in these layers, but this is small and therefore not possible to robustly detect in the observations.  We therefore refer to these flare layers as the stationary flare layers.  Using Equation \ref{eq:ciprime}, we calculate that 
$\sim90$\% and 20\% of the emergent C2826 intensity originates from the stationary flare layers at $t=1.8$ and 3.97~s, respectively.  The increase in brightness from 1.8~s to the peak in
Figure \ref{fig:lc2826} is a result of the large increase in the CC contribution to the emergent C2826 intensity.

The relative contributions from the stationary and CC layers to the excess C2826 intensity is a combination of the evolution of the electron density and temperature
 in these two flare layers at $z>500$ km.  In Table \ref{table:params}, we 
give the representative physical parameters in the CC and in the stationary flare layers over time steps of the simulation.  The representative electron density and temperatures in each flare layer are given as the C2826 contribution function-weighted values in the two flare layers in columns 6, 8, 9, and 10.   The bottom extent of the stationary flare layers ($z_1$) 
is defined as the height where $  C_I^{\prime} = 0.95$ (this height changes as a function of continuum wavelength), and layers just below the 
$  C_I^{\prime} = 0.95$ layer for $\lambda=2826$ \AA\ are also heated but to a lesser extent such that comparable amounts of NUV continuum emission to the contributions
in the CC and stationary flare layers are not produced at $z<500$ km.  The value of $z_2$ is the division between the stationary and CC layers;  the temperature at $z_2$ is given in column 7.   
$z_3$ is defined as the top of the CC with $v<-5$ km s$^{-1}$ and where $C_{I,2826}^{\prime} = 0.05$.
Over the evolution of the simulation from 1.8~s to 3.97~s, the CC increases in electron density 
from $7\times10^{13}$ cm$^{-3}$ to nearly $4\times10^{14}$ cm$^{-3}$. At 1.8~s, the CC and stationary flare layers have
comparable electron densities ($\sim5-7\times10^{13}$ cm$^{-3}$) but at 3.97~s, the electron density in the CC is nearly an order of magnitude larger than in the stationary layers and thus the 
relative amount of continuum emission is larger.

The ambient electron density and the temperature in the two flaring layers is determined by the evolution the beam heating rate, which is a function of 
the electron beam parameters and column mass evolution.
In both flare layers, the beam energy deposition (shown in Figures \ref{fig:jtot1} and \ref{fig:jtot2}) dominates the sources that contribute to the increase in internal energy at $z>550$ km in the stationary flare layers and in the CC for temperatures $<25,000$ K.  At the height corresponding to $T\sim25,000$ K, there is a steep temperature gradient at the lower shock (Section \ref{sec:dynamics}) where conductive heating dominates.  
  At $z \lesssim 550$ km, radiative heating from the absorption of photons in the Balmer continuum wavelength range dominates the sources that increase the internal energy. 

In the last two columns, we give the fraction of beam energy deposited at heights greater than the stationary flare layers ($z>z_2$; column 11) and heights
greater than the CC ($z>z_3$; column 12) for the 5F11 model. 
Due to the increasing density in the corona (from ablation), a larger fraction of the beam energy deposition occurs at larger heights over time.
By $t=3.97$ s, only 50\% of the 5F11 beam energy flux is deposited at heights within the CC and below ($z < z_3$) and is less at later times.
 The temperature and the electron density thus drop in the stationary flare layers (e.g., at $z\sim750$ km) from 1.8~s to 3.97~s.  
From 0 to 4~s, almost all of the material in a 175 km region from $z\sim900-1075$ km has been compressed into a 25 km region (the CC) at $z \sim 900$ km.  
Due to the increasing density in the CC (from compression), an approximately constant fraction ($\sim50$\%) of the beam energy deposition occurs in the descending, narrow height range of the CC, and thus more 
energetic electrons heat at lower heights \citep{Kennedy2015}.

 Following the analysis of \cite{Kennedy2015}, 
we calculate the classical thick-target stopping depth \citep{Emslie1978} as a function of electron energy for several times in the 5F11.
First, to confirm that the classical stopping depth formula gives reasonable values, we use the evolved 5F11 atmosphere at 3.97~s and calculate the Fokker-Planck energy deposition with $\delta=10$ and lower energy cutoffs of $E_c=25,35,50,85$ keV which are nearly monoenergetic beams with $E_{\rm{ave}}\sim E_c$. 
Beam electrons with energies initially $E>50$ keV at 1.8~s and $E>$80 keV at 3.97s heat the stationary flare layers, whereas lower energy beam electrons primarily heat the CC and the upward flows in the flare corona. 
 As less of the beam energy is deposited in the stationary flare layers (column 11) and the CC becomes brighter from an order magnitude increase in density as the temperature 
  decreases to $\sim10,000$ K, most of the emergent intensity originates from the CC and the 
  physical depth range of the continuum emission decreases from 390 km to 160 km.

\subsection{Fe II $\lambda$2814.45 Modeling} \label{sec:feii_model}
With hydrodynamic modeling, we can rigorously test the model predictions of the dynamics over the continuum-emitting
layers by synthesizing flare chromospheric lines that have low optical depth and probe the temperatures ($T\sim 10,000$ K) of white-light continuum formation.
In this section, we model the Fe \textsc{ii} lines in LTE to show that the flare atmospheric evolution of the 5F11 is qualitatively consistent with the resolved spectral components at \lamr\ and the RWA emission in the NUV spectra of BFP1 and BFP2 (Figures \ref{fig:brightest}, \ref{fig:feii_2814}, \ref{fig:feii_2832}, Section \ref{sec:feii}).

We use the atomic data for the \feone\ and \fetwo\ lines from \cite{feii_ref1}, \cite{nave}, and \cite{feii_ref2} using the National Institute of Standards and Technology (NIST) database and the partition function from \cite{Halenka1984} to calculate the LTE level populations of the upper and lower levels of the transitions.
  The temperature, electron density, mass density, and velocity are obtained from each time step in the 5F11, F11, and coronal heating RADYN simulations at approximately 0.1-0.3~s intervals (thus,
the LTE Fe \textsc{ii} calculations include the non-equilibrium ionization, NLTE electron density at each time).  We assume that scattering is not important for these lines, and therefore the LTE line emissivity and opacity can be written as Equations 2.69 and 3.87, respectively,
from \cite{Rutten2003}.  The rates for (stronger) Fe II lines in the NUV are collisionally dominated for the electron density of the quiet Sun \citep{Judge1992}; 
since the flare atmospheres increase the density and temperature, LTE is a plausible approximation for the minor Fe II lines (see also Section \ref{sec:limitations}). The continuum emissivity and opacity from Section \ref{sec:continuum} are also included in the calculation (however, we exclude Mg II from these calculations).  The time-averaged emergent intensity spectra are calculated for each model over a simulated IRIS exposure time of 8~s and are convolved with the instrumental resolution (FWHM $\sim5.5$ \kms).  For the 5F11 model,
we average the short 5F11 heating run (4~s heating) over the first 8~s of the simulation, and we average the extended (15~s) 5F11 heating run
over the first 8~s of the simulation.  We also calculate average spectra over 2.4~s intervals to compare to the shorter exposure times resulting from the IRIS AEC adjustment during the flare.  

The 8~s average model intensity spectra for the \feone\ line are shown as the black and pink curves for the short and extended heating runs, respectively, in the middle panel of Figure \ref{fig:feii_2814}.
The models produce a broad RWA component and a narrower component at \lamr, as 
 in the observations (top panel of Figure \ref{fig:feii_2814}). The value of $I_{\rm{RWA}} / I_{\lambda_{\rm{rest}}}$ from the models are $\sim1/3$ and $2/3$ as for the spectra 
 of BFP1 and BFP2, respectively (Section \ref{sec:feii}).
 The extended heating model is similar to the spectrum of BFP2, but the NUV exposure time of the BFP2
was adjusted by the IRIS AEC to be 2.4~s.  We average over the short and extended 5F11 heating runs in 2.4~s durations for the first 8~s, which are shown in the top panels of Figure \ref{fig:feii_sequence}; the two emission line components appear in the shorter exposure time average as long as the average is not the first 2.4~s of the heating.

We calculate the contribution function over the wavelength range of the \feone\ line in the 5F11 model.
The emergent intensity over the \feone\ line and in the NUV (C2826) continuum originate over the same two flaring layers at $z>500$ km.
The RWA emission component is primarily due to emission from the 
CC and the component at \lamr\ originates primarily from the stationary
flare layers just below the CC.   In Figure \ref{fig:feii_ci}, we show the contribution
function at \lamr\ for the \feone\ line\footnote{We use a constant nonthermal broadening of $\xi=7$ \kms\ at $z>500$ km; see Section \ref{sec:nonthermal}.} compared to the contribution function for the nearby continuum at $\lambda=2826$ \AA\ at $t=1.8$~s and $t=3.97$~s. At both times, the optical depth at \lamr\ reaches a value of $\tau\sim1$ at $z\sim800$ km: the \feone\ line is not optically thin over the stationary flare layers at $z>500$ km.  The continuum is optically thin over the heights of \feone\ formation and is thus formed over a larger physical depth range.  The physical depth ranges are 170 km (at 1.8~s) and 145 km (at 3.97~s) for \lamr\ of \feone, and the physical depth ranges are 390 km (at 1.8~s) and 160 km (at 3.97~s) for C2826 (using $z_{\rm{lim}}=500$ km in Equation \ref{eq:ciprime}).

The relation between the \feone\ line profile and the evolution of the flare atmosphere over the first 8~s of the 5F11 heating runs is as follows (see Figure \ref{fig:feii_sequence} for reference).
The \lamr\ emission component is bright by 0.1~s due to beam heating in the stationary flare layers at $z>500$ km.  After 2.5~s, the bright RWA emission component appears at $\lambda-$\lamr$=35$ km s$^{-1}$ as the density of the CC increases and the temperature decreases
to 24,000 K (Figure \ref{fig:vz_evol}, Table \ref{table:params}).  The RWA emission reaches maximum intensity at 4~s while still centered at $\lambda-$\lamr$=35$ km s$^{-1}$ as the CC decreases to a temperature of 12,000 K;  the C2826 intensity also increases in brightness over this time due to increasing density in the CC (Figure \ref{fig:lc2826}). For the short 5F11 heating, the continuum drops to the preflare level by 5~s (the electron beam heating is turned off at 4~s in the short heating model)
and the two emission line components decrease in brightness.  Therefore, the average of the first 8~s of the 5F11 with short heating has a relatively brighter \lamr\ component ($I_{\rm{RWA}} / I_{\lambda_{\rm{rest}}} << 1$) because the RWA does not develop until about $t=2.5-3$~s. 
At $t=4-7$~s in the extended 5F11 heating run, the \lamr\ emission
component decreases in intensity while the RWA component remains bright and its peak shifts to $\lambda-$\lamr$=15 - 20$ km s$^{-1}$ by 8~s. 
The CC maintains a high temperature between $8,500-11,000$ K with a high density of $4\times10^{14}$ cm$^{-3}$ over this time as it decreases in speed while moving into the stationary flare layers below.  From Table \ref{table:params}, the value of $z_2$ (top of the stationary flare layers)
  has decreased significantly by 7~s due to the stationary flare layers being swept up by the CC;  thus, some material that was emitting \lamr\
 emission at 4~s emits redshifted emission at 7~s (similarly, some material that emits \lamr\ photons at 1.8~s emits redshifted photons at 3.97~s; see Figure \ref{fig:feii_ci}).  
 
  The broadening and location of the peak of the RWA component in the 8~s exposure average in the extended 5F11 heating run is due to the 
 velocity evolution of the CC, which decreases from $\sim45$ \kms\ at 2~s to $\sim18$ \kms\ at 8~s.  The velocity gradient within the CC at each time also contributes toward emission between \lamr\ and the peak of the RWA but is relatively minor.
 Therefore, the average of the first 8~s of the 5F11 with extended heating has a brighter RWA component ($I_{\rm{RWA}} / I_{\lambda_{\rm{rest}}} \sim 2/3$) because of the decreasing 
 intensity of the \lamr\ emission component from material being swept up by the CC while the CC produces bright red-shifted emission after 4~s when the density is increasing and the temperature remains high  near $T\sim10,000$ K.

 \subsubsection{Nonthermal Broadening}  \label{sec:nonthermal}
\begin{figure}[H]
\begin{center}
\includegraphics[scale=0.5]{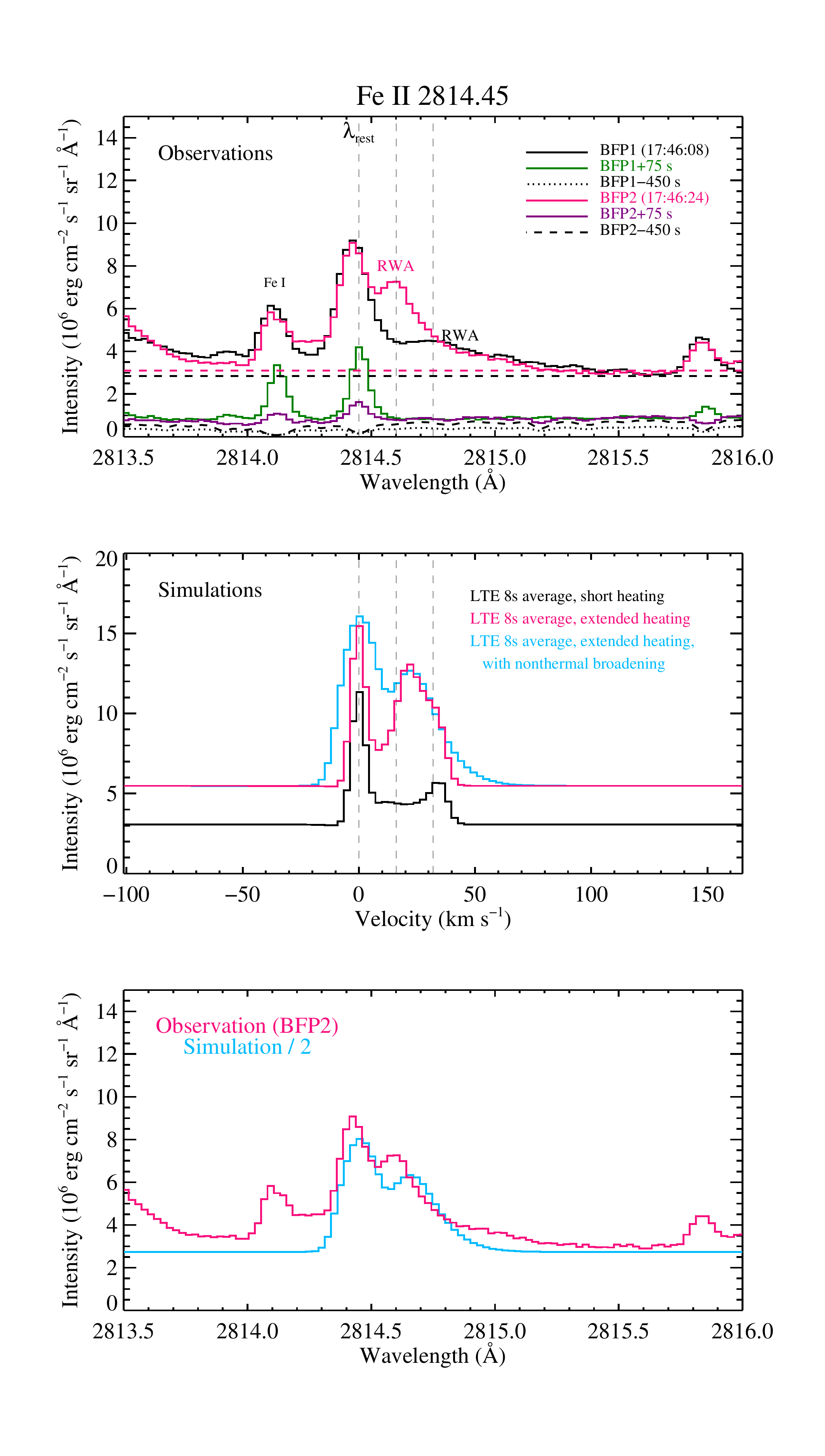}
\vspace{-5mm}
\caption{ (Top) The \feone\ profiles for the same times and locations as in Figure \ref{fig:brightest}.  
``RWA'' indicates the locations of the peaks of the red-wing asymmetry line components.  The three vertical dashed lines show the rest wavelength for \feone, 
$\lambda-$\lamr$=16$ \kms, and $\lambda-$\lamr$=32$ \kms; these velocities identify the peaks of the RWA
components for BFP2 and BFP1, respectively.  The rest wavelength component of Fe \textsc{i} $\lambda2814.11$ \citep{nave_fei} is indicated in the top panel.  The black and pink horizontal dashed
lines are the values of C2826 extrapolated to this wavelength range.  The dotted and dashed spectra indicate the pre-flare.  (Middle) LTE \feone\ profiles averaged over the first 8~s of the 5F11 model:
the short (4~s) heating model (black) and the extended (15~s) heating model (pink).  The light blue spectrum is the average of the first 8~s of the extended 5F11 heating run with a nonthermal broadening 
parameter of $\xi=7$ \kms\ included at heights $z > 500$ km, $\xi=2$ \kms\ included at $z<500$ km, and a variable value of $\xi(t)$ included in the CC (see text; Table \ref{table:xi}).  The same vertical dashed lines in the top panel are reproduced in the bottom panel. (Bottom)  Simulation of \feone\ with nonthermal broadening compared to the observation of BFP2 from the top panel.  The model intensity
has been reduced by half.   The 5F11 model averaged over the first 8~s reproduces two spectrally resolved \feone\ emission components with the same relative peak intensity as in the observations. }
\label{fig:feii_2814}
\end{center}
\end{figure}

\begin{figure}[H]
\begin{center}
\includegraphics[scale=0.6]{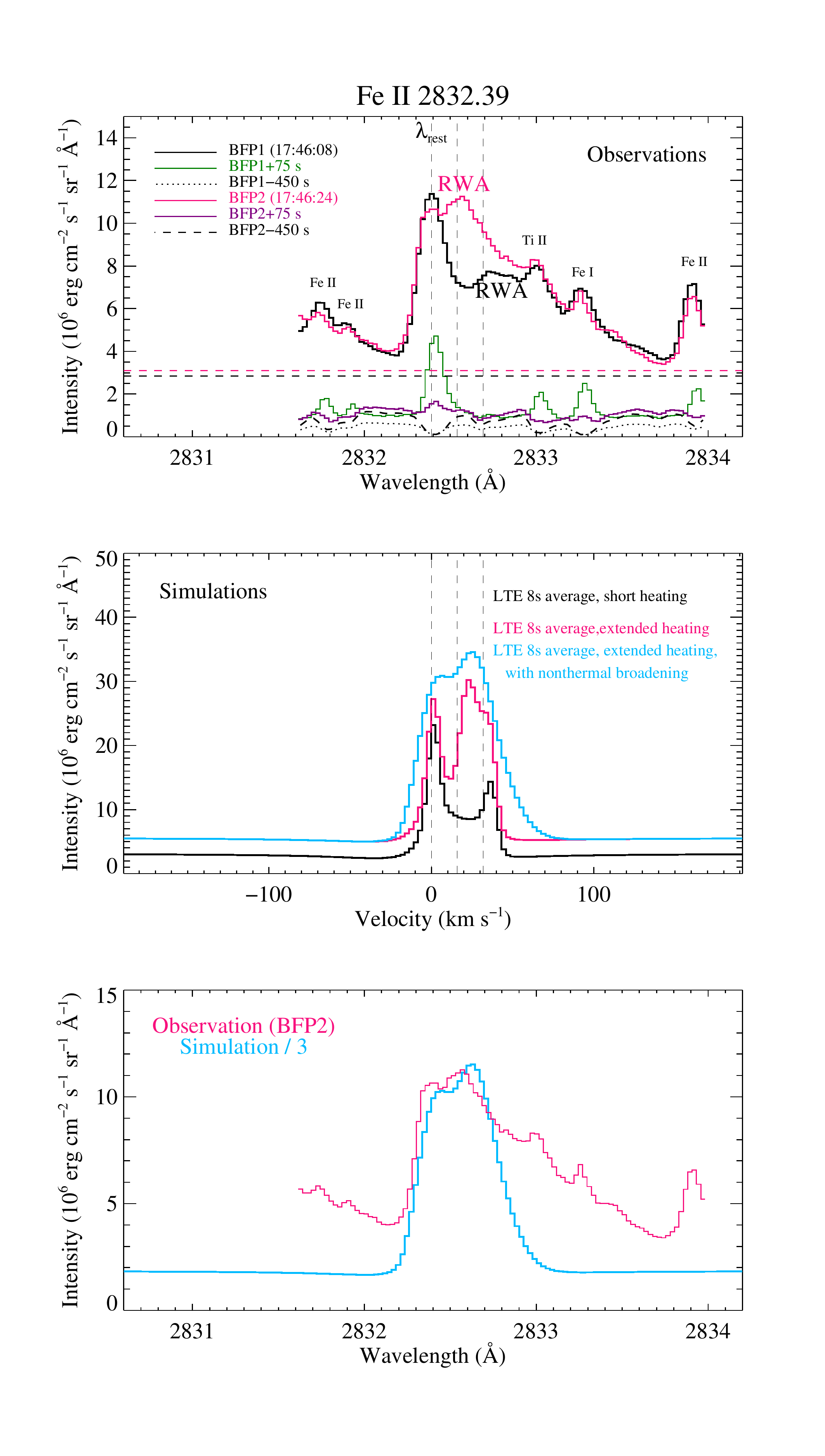}
\caption{  Same as Figure \ref{fig:feii_2814} but for \fetwo. Note the
  enlarged wavelength range.  
  Other emission lines in this range are indicated: Fe \textsc{ii} $\lambda$2833.92,
  2831.76, 2831.92 \citep{nave}, Fe \textsc{i} $\lambda$2833.27 \citep{nave_fei}, and Ti \textsc{ii} $\lambda$2832.99 (I. Zapadlik et al.\ priv.\ communication).
  The two emission components in the 5F11 model averaged over the first 8~s are adequately reproduced with the correct relative intensity but are both brighter than the observations.}
\label{fig:feii_2832}
\end{center}
\end{figure}

\begin{figure}[H]
\begin{center}
\includegraphics[scale=0.6]{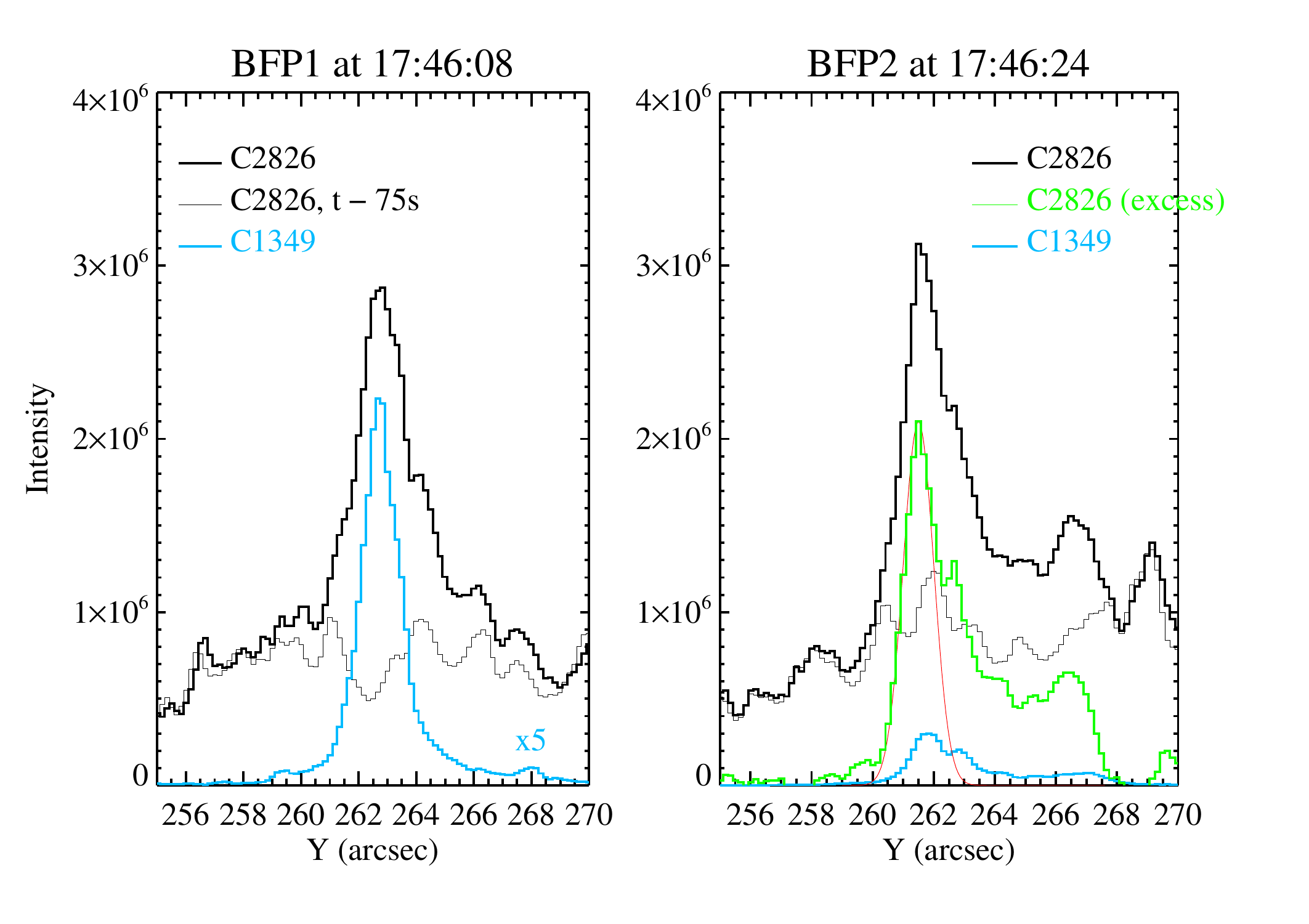}
\caption{The spatial profiles from the C2826 and C1349 continuum regions for BFP1 and BFP2 in the bottom panel of Figure \ref{fig:ref2832}.  For the NUV, the exposure time 
for the left panel is 8~s and the exposure for the right panel is 2.4~s.  The excess NUV intensity profile is shown in the right panel with a Gaussian fit (FWHM $\sim$1\arcsec.2)
to the leading edge of the profile; therefore, the leading edge of the NUV continuum emission is adequately resolved with IRIS at this location. } \label{fig:spatial}
\end{center}
\end{figure}

In the two exposure averages of the 5F11 model in Figure \ref{fig:feii_2814} (middle panel),
the \lamr\ emission components in \feone\ have comparable peak intensity values (accounting for the different continuum levels), as do the \lamr\ components in the spectra of BFP1 and BFP2.  
However, the model \lamr\ components in the Fe \textsc{ii} lines are significantly narrower 
than the width of this component in the observations (top panel of Figure \ref{fig:feii_2814}).  In these LTE calculations, we include natural damping and thermal broadening, which are small. The quadratic electron pressure broadening is applied using the prescription in the RH code, but the largest values of the quadratic electron pressure damping parameter 
in our flare atmospheres are as small as the natural damping parameter and thus do not contribute measurable broadening. 
 It is also known that theoretical quadratic electron pressure broadening 
using the adiabatic approximation underestimates the damping parameter \citep{Mihalas}, but only an extremely large discrepancy by several orders of magnitude could produce the observed broadening.  
Because the model (black) profile is too narrow in the line-center emission component, we add a nonthermal broadening (microturbulence) parameter, $\xi$.  

A nonthermal broadening parameter of $\xi=6-7$ \kms\ at $z>500$ km is adequate to reproduce the width of the \lamr\ emission, but
the observed broadening of the RWA component is still not well-reproduced.  
In Figure \ref{fig:feii_2814} (middle panel), the RWA component of the model \feone\ line is broader than the emission in any instantaneous profile due to the velocity evolution of the CC over the 8~s exposure time. The red extent of the model profile is determined by the velocity of the CC when it attains a temperature of $\sim20,000$ K. 
The extrapolation of the C2826 continuum intensity to the 
wavelength range of \feone\ in Figure \ref{fig:feii_2814} (top panel) demonstrates that the observed RWA component extends to
$\lambda-$\lamr$\sim120$ km s$^{-1}$ for BFP1; for BFP2, the tail of the RWA component decreases to the continuum level at $\lambda-$\lamr$\sim70-80$ \kms.  

 A detailed investigation of the physical origin of the nonthermal broadening is outside the scope of this paper; here, we assume the nonthermal broadening
 in the CC layers is velocity broadening (turbulent broadening with an isotropic Gaussian distribution).
To approximate time-dependent velocity broadening in the CC, a simple prescription for $\xi(t)$ is used.  We assume that $\xi(t) \propto \rho_{\rm{max}}(t)^{-1/2}$ where $\rho_{\rm{max}}$ is the maximum density of the CC given in Figure \ref{fig:vz_evol}, and $\xi(t)$ is constant as a function of height throughout the CC.  
 This relation may be reasonable if a 
 constant amount of flare energy goes into generating the kinetic energy of turbulence, such that the turbulent velocity decreases as the CC accretes mass.  The evolution
 of $\xi(t)$ is given in Table \ref{table:xi} and is normalized to 7 km s$^{-1}$ at 15~s, when the RWA emission component merges into the \lamr\ emission component.  A value of $\xi=7$ \kms\ is kept for the stationary flare layers at $z>500$ km.  
 
 The LTE, exposure-averaged profile 
 over 8~s for the 5F11 extended heating run with this prescription for $\xi(t)$ is shown as the light blue line in Figure \ref{fig:feii_2814} (middle panel) and in the bottom panel of Figure \ref{fig:feii_2814} overlayed on the observations. Although the intensity of the model is a factor of $\sim$2 larger than the observations, 
  the 30\% bisector of 13 \kms\ is in satisfactory agreement with the measurement from BFP2 (30\% bisector velocity of 9 \kms; Section \ref{sec:feii}). 
  From $t=2$~s to 8~s, the velocity of the CC decreases from $\sim45$ \kms\ to $\sim18$ \kms; notably, the gas velocity is significantly larger than the 30\% bisector velocity
of \feone.  
 The \feone\ model profile exhibits a RWA component that extends to redder wavelengths and a profile that is more consistent with the spectrum of BFP2 (top panel) than the calculation without nonthermal broadening (the variable prescription of $\xi(t)$ better accounts for the far red wing emission at $\lambda-$\lamr$\sim45-70$ \kms\ than with a time-independent value of $\xi=7$ \kms\ in the CC).  
However, more emission is required to account for the observations in the far red wing at $\lambda-$\lamr$>70$ \kms.  In the Mg \textsc{ii} $\lambda2791$ and $k$ lines, the observed RWA components
 extend to $\lambda-$\lamr$>100$ \kms\ and are much brighter (e.g., Figure \ref{fig:brightest}).  Although the \feone\ line is more optically thin and provides a better diagnostic of the velocity field over
 the continuum-emitting flare layers, calculations of the Mg II lines with partial frequency redistribution \citep[e.g.,][]{Leenaarts2012, Leenaarts2013b, Leenaarts2013} can be used to better understand the extremely red RWA emission, which we leave to future work.

\subsection{Comparison of \feone\ to \fetwo\ LTE Models} \label{sec:feii_model2}

 We repeat the LTE calculation for the \fetwo\ line profiles to compare to the observations of BFP1 and BFP2 in Figure \ref{fig:feii_2832} (top and middle panels).  In the 8~s average over
the extended 5F11 heating model, the peak of the RWA component
 is comparably bright to the peak of the \lamr\ emission component ($I_{\rm{RWA}} / I_{\lambda_{\rm{rest}}}\sim1$) and the peak wavelength is closer to \lamr, whereas in the 8~s average over the short 5F11 heating the peak of the RWA emission component is less intense than 
the peak intensity of the \lamr\ emission component ($I_{\rm{RWA}} / I_{\lambda_{\rm{rest}}}\sim1/2$) and has a larger redshift.  Similar to \feone, the short 5F11 heating model (black) spectrum qualitatively resembles 
the spectrum of BFP1, and the extended 5F11 heating model (pink) spectrum qualitatively resembles 
the spectrum of BFP2. In the middle and bottom panels of  Figure \ref{fig:feii_2832}, we show the \fetwo\ profile with the $\xi(t)$ prescription from Section \ref{sec:nonthermal}.  
Although the line intensity is significantly larger than the observations, the value of $I_{\rm{RWA}} / I_{\lambda_{\rm{rest}}}$ is well-reproduced.  
Thus, we are able to reproduce the range of observed values of  $I_{\rm{RWA}} / I_{\lambda_{\rm{rest}}}$ in the brightest pixels with the 5F11 model.

The differences in the optical depth between the \feone\ and \fetwo\ lines contribute to the differences in the values of $I_{\rm{RWA}} / I_{\lambda_{\rm{rest}}}$ for the exposure-averaged simulated profiles in Figures \ref{fig:feii_2814} and \ref{fig:feii_2832}.
   The RWA component in the \fetwo\ profile becomes comparably bright to the \lamr\ component at earlier times ($t=3$~s)
 than for the \feone\ ($t=4$~s).  The time lag to increase to comparable brightness to the emission at \lamr\ 
  is due to larger optical depth at \lamr,  and therefore a smaller physical depth range in the stationary flare layers, for \fetwo.  For example, at 3~s in the 5F11, $\tau_{\lambda,\rm{rest}}=1$ at 890 km for \fetwo\ whereas $\tau_{\lambda,\rm{rest}}=1$ at 805 km for \feone.  At $t=3.97$~s in the 5F11, the physical depth range\footnote{Using $\xi= 7$ km s$^{-1}$ at $z>500$ km.} is 145 km for \feone\ at \lamr\ and  is
 only 30 km for \fetwo\ at \lamr.  For \fetwo, the larger optical depth and smaller physical depth range at \lamr\ suppresses the \lamr\ emission component relative to the RWA emission component, producing a $I_{\rm{RWA}} / I_{\lambda_{\rm{rest}}}>1$  at times also when the 
 beam heating in the stationary flare layers is reduced due to the increasing CC and coronal density (Table \ref{table:params}).  
 The RWA component is brighter than the \lamr\ emission in \fetwo\ for a longer fraction of the exposure time average, which results in $I_{\rm{RWA}} / I_{\lambda_{\rm{rest}}} \sim 1$ for the 8~s exposure averaged profile; for \feone\ the value of $I_{\rm{RWA}} / I_{\lambda_{\rm{rest}}} \sim 2/3$ for the 8~s exposure averaged profile.
We speculate that the spectra of Mg \textsc{ii} $\lambda2791.6$ (Figure \ref{fig:brightest}), where the value of $I_{\rm{RWA}} / I_{\lambda_{\rm{rest}}}$ is larger than for the \feone\ line, can also be explained by the larger optical depth at \lamr\ for this line compared to both of the Fe \textsc{ii} lines.

 The optical depth in the RWA affects the value of $I_{\rm{RWA}} / I_{\lambda_{\rm{rest}}}$ in the opposite way:  the optical depth of the RWA wavelengths in the \feone\ line is always low, but for \fetwo\ the optical depth of the RWA wavelengths is non-negligible and this tends to decrease the value of $I_{\rm{RWA}} / I_{\lambda_{\rm{rest}}}$.
  However, the optical depth effects at \lamr\ contribute the most to the value of $I_{\rm{RWA}} / I_{\lambda_{\rm{rest}}}$ because the RWA emission originates in the CC, which has a 10x larger density than the stationary flare layers (Table \ref{table:params}), and therefore is very bright even when the physical depth range of emission at RWA wavelengths is comparably small. 

\subsection{Comparison of Heating Model Predictions for Fe II}

In Figure \ref{fig:feii_sequence}, we compare the predictions of the 5F11 \feone\ spectrum to an LTE calculation for the F11 model (lower left panel) and a coronal heating model (lower right panel) that produces a large conductive flux into the transition region and upper chromosphere.  
While the 5F11 model satisfactorily explains the two emission line components in \feone, the F11 and coronal heating model do not
produce the two line components and the observed excess NUV continuum intensity.  An RWA emission component is not produced by the F11 model due to the lack of dense, fast downflows \citep[as discussed in][for this model]{Kuridze2015}, and there is only bright 
emission at \lamr. 
The coronal heating model produces a relatively bright RWA component in \feone\ and a 30\% bisector velocity of 30 \kms\ for H$\alpha$ (Table \ref{table:model}), but the excess NUV continuum emission in the coronal heating model is faint ($\le4\times10^5$ \cgs) and is a factor of $6$ below the observed values in BFP1 and BFP2 (Table \ref{table:bfpvals}).  
In the coronal heating model, high velocity, Balmer continuum emitting layers are produced as in the 5F11 model but with much lower emergent continuum intensity.  The stationary flare layers at $z\sim750$ km are heated to $T\sim7000$ K (from $T\sim4460$ K at $t=0$~s)
 and produce the Fe \textsc{ii} emission at \lamr.  The temperature increase in the stationary flare layers is due to backwarming from the absorption of NUV photons in the range from $\lambda \approx 1500-3000$\AA\footnote{This type of chromospheric heating will be discussed in detail
 in a future modeling study of stellar flare data from the Hubble Space Telescope (Kowalski et al. 2016 in prep).}.
In the coronal heating model, the RWA for the \feone\ line develops before the line center emission, whereas the 5F11 beam model exhibits a $\sim2.5$~s delay in the brightening of the RWA after the initial brightening of the line-center component.   High time-resolution ($\Delta t < 1$~s) spectra and hard X-ray data could be useful to distinguish between these two heating scenarios \citep{Canfield1987}.

\subsection{The RWA Component in Chromospheric Emission Lines}  \label{sec:ha}
We hypothesize that the spectrally resolved red-shifted components (RWA) in the Fe \textsc{ii} line profiles of the 5F11 model  are analogous to the red-wing asymmetry often observed in the H$\alpha$ line profile which has been attributed to the phenomenon of chromospheric condensation \citep{Ichimoto1984, Canfield1987, Canfield1990}. 
In future work, we will improve on the modeling predictions of the H$\alpha$ line in Figure \ref{fig:Halpha}.  To accurately
model the H$\alpha$ line, we will include an improved prescription of the electron pressure broadening and explore the effect of the large microturbulence parameter inferred from 
the IRIS lines (the standard value of $\xi$ used in RADYN is 2 \kms). In the 5F11 model, this line is very optically thick and does not have a significant emission component at \lamr\ that originates from the stationary flare layers;
the entire line exhibits a redshift at $t=3.97$~s (Figure \ref{fig:Halpha}), which is qualitatively consistent with some previous flare observations of H$\alpha$ \citep{Canfield1990} but not the observations of \cite{Ichimoto1984}.  
However, when a redshift of the entire line is observed during flares, the maximum redshift occurs when the intensity of the line is low and this is not predicted by the 5F11 model. 

The wavelength range of the observations of H$\alpha$ for this flare \citep{Kleint2015, Fatima2016} from the DST/IBIS \citep{Cavallini2006} is indicated in Figure \ref{fig:Halpha}.  
The 5F11 prediction at $t=1.8$~s predicts that $I_{6564} \approx I_{6563} > I_{6562}$ as in the observations of the brightest regions of the flare \citep[cf. Fig 2 panel b of][]{Fatima2016}, but the intensity of the model is a factor of $5-10$ larger than these observations.

\section{The Brightest SJI 2832 Kernel in the March 29th, 2014 X1 Flare} \label{sec:sjibright}
The excess C2826 in the 5F11 model reaches a maximum of $5-5.5\times10^6$ \cgs\ between $t=4$ and $t=6$~s (Figure \ref{fig:lc2826}), which is significantly larger than
the highest values of excess C2826 obtained from the spectra of BFP1 and BFP2 ($\sim2-2.2\times10^6$ \cgs).  
The Fe II emission line profiles for the extended heating run (those that produce the strongest RWA components) also produce very bright emission line profiles that are a factor of $1.5-3$ larger than the observations (bottom panels of Figures \ref{fig:feii_2814} and \ref{fig:feii_2832}), while the short heating prediction for the \feone\ intensity is rather well-reproduced (middle panel of Figure \ref{fig:feii_2814}).  However, the locations of BFP1 and BFP2 are not the brightest locations during the impulsive phase due to the long rastering cadence of $\sim$75~s and the relatively small
spatial coverage of the slit, which notably misses the brightest regions of the ($\sim10^{16}$ cm$^2$) flare kernel BK2830 at $(x,y)=(519\arcsec.5, 263\arcsec.8)$ in SJI 2832 \#173 (Figure \ref{fig:ref2832}).
BK2830 corresponds to the location of the hard X-ray source revealed by RHESSI imaging of the plage ribbon \citep{Kleint2016}.  As noted by \cite{Kleint2016}, the 
IRIS raster intersects the half-maximum of this hard X-ray source.

Using the intensity calibration of SJI 2832 (Section \ref{sec:calibration}), we estimate the maximum excess continuum intensity in BK2830 to be $\sim10^{7}$ \cgs\ and  $\sim0.7\times 10^{7}$ \cgs\ in SJI 2832 \#173 (mid peak phase) and \#174 (early fast decay phase), respectively.  These values are $3-4.5$ times larger than the values of excess C2826 obtained from the spectra and are a factor of $\sim10$ larger than the preflare values.  The excess SJI 2832 emission includes several Fe \textsc{ii} emission lines, which may exhibit bright RWA components.  Unfortunately, the spectral region corresponding to the SJI 2832 wavelengths was not readout for these observations, and we cannot determine the relative contributions for this flare.  Using full spectral readout data during an X-class flare on 2014 Oct 25 (Kowalski et al.\ in preparation), we identify that the bright Fe \textsc{ii} emission lines within SJI 2832 wavelength range include Fe \textsc{ii} $\lambda$2826.58, $\lambda$2826.85,  $\lambda$2828.26, $\lambda$2829.46, $\lambda$2829.51 and $\lambda$2832.39, and these account for a significant fraction of the SJI 2832 bandpass intensity.

  To estimate the SJI 2832 intensity predicted by the 5F11 model, we use the LTE approximation for these Fe \textsc{ii} lines at $t=3.97$~s as done for the \feone\ and \fetwo\ lines in Section \ref{sec:feii} and include the continuum opacity and emissivity.  The intensity spectrum is converted to DN s$^{-1}$ pix$^{-1}$ by integrating over the effective area curve of SJI 2832, which is then converted to 
  a flux density as done for the data (Section \ref{sec:calibration}).
The excess SJI 2832 intensity from the 5F11 model at  $t=3.97$~s corresponds to $\sim10^{7}$ \cgs, which is a lower bound because a few emission lines in the bandpass aside from Fe \textsc{ii} are not included in the calculation.  At its brightest times, the 5F11 model is consistent with the excess intensity from the brightest pixels of BK2830 in SJI 2832 image \#173.  Furthermore, this model predicts that the excess SJI 2832 intensity values include nearly equal contributions from flare continuum emission and Fe II lines.  
In Section \ref{sec:area}, we inferred similar proportions of line and continua emission using spatially coincident but temporally contiguous observations from the slit jaw images and spectra.

\section{Model Uncertainties and Assumptions} \label{sec:limitations}

In our modeling approach, we have made several assumptions that warrant discussion in light of our findings. 

\subsection{The LTE assumption for \feone\ and \fetwo\ profiles}
Using the RH code for the 5F11 at $t=3.97$~s and a large Fe \textsc{ii} atom \citep{Walkowicz2008} that comes with the 
 RH distribution, we confirmed that the LTE profile of \feone\ satisfactorily matches the NLTE profile from RH.  
All relevant sources of broadening (including electron pressure broadening) are included in this calculation, which justifies implementing an ad-hoc
 microturbulent parameter to match the line broadening.  
 Because the LTE assumption is employed at all time steps in the simulations to produce the exposure-averaged spectra, the spectra for the \feone\ and \fetwo\ lines
(Section \ref{sec:feii_model}) are approximations. 
The LTE modeling assumption of the Fe \textsc{ii} lines within the SJI 2832 camera effective area at $t=3.97$~s provides a reasonable approximation to the amount of excess SJI 2832 emission that is due to Fe II flare lines with bright RWA components because the collisional rates are high at this time.

\subsection{Return current and beam instability effects}
  The large electron flux ($8.5\times10^{18}$ electrons s$^{-1}$ cm$^{-2}$) for the 5F11 beam model produces an electric field given that nonthermal protons 
do not neutralize the beam \citep{Emslie1980, Oord1990}.  Using the equations from \cite{Holman2012}, we estimate that an energy loss of $\sim$30 keV per beam electron would occur over the upper 8 Mm of the corona due to the return current electric field (at $t=0$~s). The return current electric field imparts a drift velocity to ambient electrons, and the energy loss from the beam over the potential drop thus increases the thermal energy of the corona from the Joule heating \citep{Holman1985, Holman2012}.  

We have used an approximate prescription for the heating from the return current ($Q_{\rm{RC}}$) in the F11 and 5F11 models \citep{Holman2012, Allred2015}.  The maximum temperature in the corona for the 5F11 model at 4~s is 30 MK and is due to comparable amounts of return current heating and beam heating; after 10~s, the maximum temperature exceeds 50 MK.  We note that a coronal temperature of $T\sim30$ MK is reasonable compared to the RHESSI observations \citep[$T\sim25$ MK;][]{Battaglia2015} but a detailed emission measure analysis of the model Fe XXI line in the IRIS range is necessary for a meaningful comparison. The mean free path of the ambient drifting electrons (at $t=0$~s) is $\sim150$ km so they are thermalized in a relatively short distance.  At all times, we assume the electrons are thermalized over a short distance so that the return current is in steady state ($E_{\rm{RC}}=\eta J_{\rm{beam}}$, where $\eta$ is the plasma resistivity).  We also assume that ambient electrons are not accelerated out of the thermal distribution, but the 5F11 is at the threshold where this assumption starts to break down.

High beam fluxes of $\gtrsim10^{12}$ \fluxunits\ are commonly inferred from high spatial resolution flare data of the brightest flaring footpoints for small and large GOES class flares alike \citep[L. Fletcher, priv. communication][]{Hudson2006, Fletcher2007, Krucker2011, Milligan2014, Gritsyk2014, Jing2016} and return currents have been attributed to the spectral 
 differences between coronal and chromospheric hard X-ray sources \citep{Battaglia2006, Battaglia2008}.  A method for determining the importance of return currents in a large sample of flares (including the \marflare\ flare) using a self-consistent analysis of the soft and hard X-ray spectra will be presented in Alaoui et al. in prep.
 The potential drop from a return current electric field is expected to flatten the electron distribution function at low energies \citep{Zharkova2006}, but the early impulsive phase of the \marflare\ flare can be fit satisfactorily by a single power law \citep[][Alaoui et al.\ in prep]{Kleint2016}.  The presence of a large potential drop of 30 keV from a high flux (5F11) beam at $t=0$~s cannot be affirmed from the hard X-ray data of this flare, but the potential drop is expected to lower as the resistivity decreases as the corona heats up from its initial temperature.  

The 5F11 beam density is 10\% of the ambient coronal (pre-flare) density, which is $8\times10^9$ cm$^{-3}$ in our model and is consistent with the upper limits of
 the solar coronal density above an active region \citep{Krucker2012}.  This beam density results in a return current drift speed of $\sim10^9$ cm s$^{-1}$
 and exceeds the electron thermal speed of the pre-flare corona.  We do not consider the effects of beam instabilities and double-layers that result from a  
 large return current drift speed \citep[e.g.,][]{Lee2008, Li2014} or from a sharp low-energy cutoff \citep{Hannah2009}.  The effects from beam instabilities (which can form on very short timescales) for $n_{\rm{beam}}/n_{\rm{bckgd}} < 1$ will likely need to be addressed in more detail in future work on high beam fluxes.
 
Lower electron beam fluxes have been considered in previous radiative hydrodynamic flare modeling work of this flare, but these simulations do not produce the bright red wing emission of the Mg II $h$ and $k$ lines although the \lamr\ intensity is closer to the observations \citep{Fatima2016}.  The F11 model also produces a continuum intensity that is inadequate to explain the brightest footpoint emission \citep[see also][]{Heinzel2015IAUS}.
A combination of several heating mechanisms, such as electron beams, proton/ion beams \citep{Allred2015}, and Alfven waves \citep{Russell2013, Reep2016} may alleviate the requirement of large nonthermal electron beam fluxes to account for both the dense, heated CCs and stationary flare layers in the impulsive phase.  However, one must still account for the observed hard X-ray bremsstrahlung emission.
 \cite{Fatima2015} presented a novel modeling method  
using an initial electron distribution from stochastic acceleration theory \citep{Petrosian2004}, which could produce downflows for low beam fluxes thus mitigating return current effects \citep[as also suggested by][for this flare]{Fatima2016}.  The downflows are conductively driven from low-energy electron energy deposition in the corona.  It remains to be determined whether beam distributions from stochastic acceleration theory have enough energy in high energy electrons to heat the downflows and stationary flare layers enough to produce the observed NUV continuum intensity. 
Alternatively, it has been proposed that the bulk of the nonthermal electrons are accelerated by Alfven wave energy transported to lower heights in the atmosphere \citep{Fletcher2008}.

\subsection{Spatial Resolution}
We compare the ($\mu=0.77$) intensity from the model to the observations, which requires that the NUV continuum footpoints are sufficiently spatially resolved.  
The spatial profiles of BFP1 and BFP2 have a FWHM of $\sim$11 pixels (Section \ref{sec:area}, Figure \ref{fig:spatial}) and thus they are resolved.  We estimated the leading edge 
spatial extent to $\sim$ 7 pixels, which is also adequately resolved.  Several locations of the ribbon, such as the third bright footpoint BFP3 ($(x,y)\sim(520$\arcsec,$262$\arcsec) in Figure \ref{fig:ref2832}), do not appear
 resolved.  If the filling factor of an IRIS resolution element is significantly less than $1$, then the model continuum intensity must be degraded and a series of sequentially heated elements would be more comparable to the observations from a single IRIS resolution element \citep{Heinzel2015IAUS}.

\subsection{Other modeling approximations}
Ionization equilibrium is assumed for atoms and ions included in the optically thin loss function.
Also, several ions (Fe II, Si II, Mg II) are excluded from the optically thin loss function because these elements are likely optically thick at low temperature. The ionization fraction of neutral magnesium is not included in the calculations of the Mg II wing opacities.  The Mg I 2852 line 
wing affects the intensity in the wavelength range of SJI 2832 \citep{Leenaarts2013}, but this is likely a small effect compared to the chromospheric flare intensity produced in the 5F11 model.
  Mg I also has edges in the NUV (e.g., $\lambda=2513$ \AA), and the population may be 
reduced in the lower atmosphere during the flare from absorption of Balmer continuum photons.  We employ a 3 level$+$continuum 
Mg II ion for the RH calculations, and using a more complex model ion \citep[like that in][]{Leenaarts2013b, Fatima2016} may slightly affect the 
wing emission in the upper photosphere.

\subsection{Future Work}
Solar flares exhibit a range of inferred electron beam parameters \citep[e.g.,][]{Fletcher2007, Kuhar2016}, properties of the Balmer jump spectral region \citep[e.g.,][]{Neidig1983}, H$\alpha$ profiles
\citep[e.g.,][]{Ichimoto1984, Canfield1990, Kuridze2015}, and Mg II $h$ and $k$ profiles \citep[e.g.,][]{Kerr2015, Graham2015}.  
 A comprehensive flare model must be able to self-consistently explain the range of these properties observed in different flares. 
In future work, we will analyze IRIS observations of the brightest kernels in many flares and use the NUV continuum and Fe \textsc{ii} modeling techniques developed here
to compare to RHD simulations covering a range of electron beam parameters inferred from RHESSI.  We will also employ constraints from the FUV continuum from IRIS for a multiwavelength  characterization of the continuum distribution.  Higher cadence sit-and-stare observations will be used to improve on the constraints of time-evolution of the continuum intensity and RWA evolution (Section \ref{sec:feii_model}).  
 
 In future modeling work for the March 29th flare, we will explore high-flux ($\ge$F11) beam
models with $E_c<25$ keV to determine if these models produce a more accurate exposure-averaged continuum
intensity and Fe \textsc{ii} line profile shapes and intensity; the 5F11 with $E_c=25$ keV model achieves a continuum and line brightness that exceeds the spectroscopic constraints by a factor of $\sim$2.  We will also incorporate the NLTE predictions of the Mg \textsc{ii} line profiles \citep[as in][]{Fatima2016} with the effects of nonthermal collision rates and compare our CC models in detail to observations of H$\alpha$ with the DST/IBIS \citep{Kleint2015, Fatima2016} and to lines from IRIS that probe hotter temperatures \citep{Young2015}.  In flares, the Si IV lines also exhibit two, spectrally resolved emission components \citep[referred to as ``CB'' and ``CR''][]{Brannon2015} which will be compared to the \lamr\ and RWA components in the chromosphere flare lines.  

Beam propagation effects such as the return current will need to be addressed in future work modeling of the impulsive phase. The self-consistent treatment of energy loss from the beam and ambient heating will be included in the Fokker-Planck equation and modeled in future work (Allred 2016 in prep).  A prescription for ramping the beam flux down \citep{Kasparova2009} and other gradual phase heating mechanisms will be necessary for a comparison to observations of the gradual phase.  An abrupt turn-off of the flux in the short heating 5F11 run results in a $\sim$100\% decrease in the NUV continuum within two seconds, whereas the observations show that continuum intensity decreases by only $\sim$80-90\% in 75~s (Figure \ref{fig:brightest}).

\section{Summary} \label{sec:summary}
We modeled the 1D radiative-hydrodynamic response of the solar atmosphere to a high energy flux density of nonthermal electrons ($5 \times 10^{11}$ \fluxunits) similar to 
that inferred from the thick target modeling of RHESSI hard X-ray observations combined with high spatial resolution areal constraints \citep{Kleint2016}.  The model comparison 
to the data is summarized in Table \ref{table:modelcheck}.  We found
that the NUV continuum intensity and Fe \textsc{ii} line profile shapes in the 5F11 model are in satisfactory agreement with the observations of the brightest
flare footpoints that were observed spectroscopically during the \marflare\ X1 solar flare.  Given the uncertainty in the duration and time-profile of the beam heating, we used a constant heating rate for a short time (4~s) and an extended time (15~s).  
The instantaneous excess continuum intensity near $\lambda\sim2826$ \AA\ in the 5F11 is achieved after only a few seconds of beam heating, and an exposure average including
the early rise phase emission and the development of the chromospheric condensation is necessary to reproduce the two chromospheric Fe \textsc{ii} emission line profile components.
 The Fe \textsc{ii} chromospheric emission lines and NUV continuum intensity originate over two flare 
layers and are complementary constraints on the model predictions of flare heating and the atmospheric conditions that produce white-light emission.

\section{Conclusions} \label{sec:conclusions}

The conclusions from our work are the following:  
\begin{itemize}

\item 
The radiative-hydrodynamic modeling of the atmospheric response to a high flux (5F11) nonthermal electron beam is an improvement over static flare modeling because the dynamic effects on the emergent spectrum can be critically examined and compared to red-shifted emission line components.  The NUV continuum 
brightness changes due to the atmosphere density evolution within a heated, chromospheric compression (condensation) that develops on short (several second) timescales in the 5F11 model.
The 5F11 produces an electron density ($\sim4-5\times10^{13}$ cm$^{-3}$) in the early phase that is consistent with the values inferred from previous static, slab modeling of the Balmer continuum and lines \citep{Donati1985},
but the density in NUV continuum emitting layers increases by another order of magnitude as the condensation develops.  

\item  In the brightest spectra of the \marflare, the excess NUV continuum excess can be explained by hydrogen Balmer recombination emission over several hundred km at chromospheric heights ($z\sim630-1020$ km) with low optical depth ($\tau_{2826} \le 0.2$).  The excess NUV continuum emission originates over two flaring layers that are heated by the nonthermal electron beam:  a chromospheric 
condensation with vertical downward velocities of $\sim20-55$ \kms\ and stationary flare layers just below the condensation.  

A variety of methods have inferred a low optical depth over the region producing the white-light continuum intensity \citep{Hudson1972, Neidig1983, Potts2010, Heinzel2014}. The high flux 5F11 RHD model
 demonstrates that a heated, downflowing compression increases the density to large values ($n_{\rm{H, max}} \sim 5\times10^{14}$ cm$^{-3}$) after 4~s when the continuum has nearly reached maximum  brightness, but this density is not enough to produce a large optical depth.  

\item The low energy electrons ($E\sim25-50$ keV) in the beam heat the chromospheric condensation and higher
energy electrons ($E\gtrsim 50$ keV initially, $E\gtrsim 80$ keV after the condensation becomes dense) heat the stationary flare layers.   In the first few seconds of the beam heating the 
NUV continuum emission in the stationary flare layers contributes to the majority of the emergent intensity enhancement.  
Therefore, we expect less continuum intensity in the first few seconds of footpoint brightening (before the CC becomes bright and dominates the 
contribution to the emergent intensity) for flares with softer (higher $\delta$) time-averaged electron beams (but for same energy flux) because fewer $E>50$ keV electrons are available to heat the stationary flare layers. 
Although there has been no significant observational relationship
established between hard X-ray spectral hardness and flare peak optical continuum intensity  \citep{Matthews2003, Kuhar2016}, higher cadence observations than currently 
available for flare white-light emission are needed to critically test the models by constraining the properties of the continuum emission when it predominantly originates from the stationary flare layers,  before the RWA line components develop.
It has been proposed that the hardness of the electron beam for very high beam fluxes could explain the interflare variation in observed continuum flux ratios in dMe flares \citep{Kowalski2016}.

\item We have developed a technique to include the Mg II $h+k$ wing opacity in the calculations of the excess continuum intensity for an accurate comparison to IRIS NUV observations.  The far wing opacities are important for an accurate treatment of (sub-)photospheric continuum dimming as well as the amount of Mg II wing emission in the upper photosphere in response to a moderate temperature change.  If these opacities are neglected,  
 errors of $15-30$\% result for the excess continuum intensity in lower beam flux models such as the F11 model (Table \ref{table:model}).

\item The \feone\  and \fetwo\ 
profiles provide new information on the dynamics in continuum-emitting layers at $T\sim8500-25,000$ K and electron densities of $\sim 5\times10^{13} - 5\times10^{14}$.  
The LTE Fe \textsc{ii} line profiles predicted by the 5F11 are
qualitatively consistent with the spectral observations of the brightest flaring footpoints:  the profiles are spectrally resolved and exhibit an emission component close to the rest wavelength that is produced in the stationary
flare layers and a bright redshifted emission 
component that is produced in the chromospheric condensation.   

\item The physical depth range parameter is an important parameter for understanding how the emergent intensity varies as a function of wavelength for emission lines and continua
because the physical depth range reflects the variation of $\tau_{\lambda}$.
 The \feone\ line is optically thin enough at line center to probe the conditions over a significant physical depth range in the flare chromosphere, but is more optically thick than the NUV continuum intensity.  Interestingly, the spectrum of BFP1 (with an exposure time of 8~s) and the spectrum of BFP2 (with an exposure time of 2.4~s)
exhibited a similar excess NUV continuum intensity and brightness in the rest-wavelength component of the chromospheric emission lines, but showed striking differences in the strength and redshift of the red-wing asymmetry (RWA) emission component in the chromospheric line profiles (Figures \ref{fig:feii_2814} and \ref{fig:feii_2832}).  

The range of relative brightnesses of the two emission components ($I_{\rm{RWA}} / I_{\lambda_{\rm{rest}}}$) for each Fe II line is adequately reproduced in the 5F11 model (using simulated exposure times of 2.4-8~s).
  The differences in $I_{\rm{RWA}} / I_{\lambda_{\rm{rest}}}$ between \feone\ and \fetwo\  are due to optical depth differences at \lamr: \fetwo\ is more optically thick at \lamr\ and has a 
  larger value of $I_{\rm{RWA}} / I_{\lambda_{\rm{rest}}}$.
Although there may be departures from LTE that are not accounted for in our models of Fe II, the LTE assumption identifies several important atmospheric parameters that lead to the line profile properties over an exposure time: 1)  the brightness at \lamr\ is determined by 
the optical depth at line center, the beam energy deposition evolution, and the extent to which the stationary flare layers have been accrued by the chromospheric condensation; 2)
the peak wavelength (and to some extent the broadening) of the red shifted emission line component is determined by the density and velocity evolution of the CC as it cools from $T=25,000$ K to $T=8500$ K. 

\item The Fe \textsc{ii} lines do not exhibit measurable thermal or pressure broadening, and a nonthermal
broadening is required to account for the width of the observed line component
at \lamr.  We find that a time-variable nonthermal broadening given by a simple physical prescription for velocity broadening is
appropriate for the profile shape, reproduces the observed bisector velocity, and accounts for some of the far red emission of the RWA emission component.

\item The coronal heating model and 5F11 beam model produce redshifts
  of the H$\alpha$ line by $\sim30$ \kms\ and a relatively bright RWA emission
  component in the LTE model of the \feone\ line, but the conductive heating flux into the chromosphere does not produce bright NUV continuum intensity as observed in the spectra of BFP1 and BFP2. 

\item At the brightest times in the 5F11, the continuum intensity exceeds the spectroscopic constraints of the continuum intensity, and the predicted Fe \textsc{ii} profiles are very bright (a factor of $1.5-3$ brighter than the spectral observations). The goal of this study is not to precisely match all possible observables to the model intensity values, given the limits on spatial resolution and our simplified prescription of the electron beam heating function such as a constant heating flux and not including return current effects.  Our goal is to quantify the range of continuum brightness values and line profile shapes achieved by the models and determine if they sweep through the regime of the observations as we change the beam flux to the highest values that are within reason.  In future work, we intend to explore the range of fluxes between F11 and 5F11, which were chosen to bracket the value of the inferred flux for the brightest kernel \citep{Kleint2016}, and the range of possible values of $E_c$.  A 2.5F11 $-$ 3.5F11 may produce a closer match to the observed intensity in the spectra.  We speculate that a soft-hard-soft variation as observed in short hard X-ray flare bursts on the Sun \citep[e.g.,][]{Fletcher2002, Grigis2004} may also help reduce the rest-wavelength intensity averaged over an exposure time because less beam energy is deposited in the stationary flare layers for softer (higher $\delta$) beams.

Nonetheless, we have tested the 5F11 against the slit jaw constraints of the brightest kernel in the flare and found that the very bright continuum and Fe II emission lines at $t\sim4$ sec are consistent with these constraints (Section \ref{sec:sjibright}).  However, spectroscopic confirmation of these extremely bright continuum and emission line intensities is necessary.  In a follow-up paper on the 5F11 model, we present optical predictions of the Balmer and Paschen jump region at $t=3.97$~s using the modeling techniques of the Balmer edge region in dMe flares (Kowalski et al. 2015b) in order to test the 5F11 model against future spectral observations of the brightest flaring kernels, such as with the Daniel K. Inoue Solar Telescope.

\end{itemize}

A high flux electron beam using the free-streaming, thick-target model reproduces several of the critical observations for the March 29th 2014 X1 flare, but the high nonthermal electron flux of 5F11
will require further modeling of the energy loss from the return current electric field as well as including the effects of beam instabilities.   
 \emph{Despite the simplifications in the thick target electron beam model employed here, we conclude that flare heating with a high flux electron beam can be used as a powerful tool to interpret spectral phenomena and to understand important radiative-hydrodynamic processes in the brightest flaring footpoints.}
The consistencies between the 5F11 model predictions and the IRIS observations will serve as a benchmark for models with additional physical 
processes that address observational challenges 
\citep{Battaglia2006, Krucker2011, Martinez2012, Dickson2013, Simoes2013} to the standard electron beam model for the brightest hard X-ray flare footpoints.  
 Measurements of the Balmer jump ratio as a constraint on optical depth, the broadening of the hydrogen lines as a constraint on electron density, and modeling lines such as Si I \citep{Judge2014} for constraints on the heating of the upper photospheric layers will complement future comparisons of IRIS flare spectra and RHD models.

\section*{Appendix A:  The Balmer Jump Ratio, Upper Photospheric Heating, and Red Optical Continuum Emission in the 5F11 Model}\label{sec:phot}
\begin{figure}[H]
\begin{center}
\includegraphics[scale=0.6]{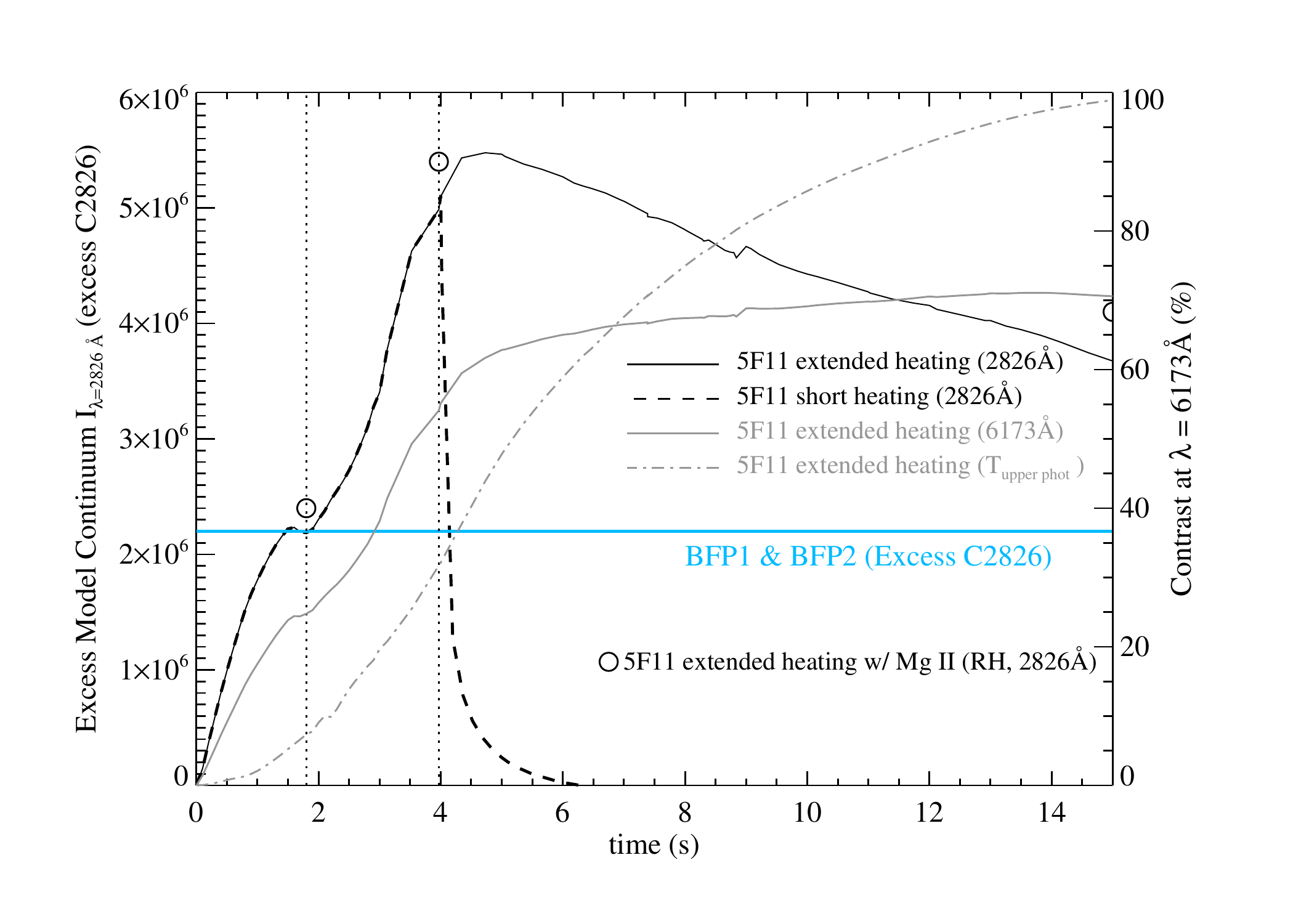}
\caption{ Light curve of the excess C2826 continuum intensity from the RADYN 5F11 simulation with short heating (dashed line) and extended heating (solid line).  The excess continuum intensity at C2826 is shown for BFP1 and BFP2 (solid light blue horizontal line). The
times analyzed in the early phase of the 5F11 run are indicated by the two vertical dotted lines at 1.8~s and 3.97~s.  The open circles are the 
 excess C2826 values calculated with the Mg II $h+k$ wing opacities at selected times (see text).  The 5F11 model attains an excess NUV continuum intensity that 
 is consistent with the spectral observations at 1.8~s and then continues to brighten as the chromospheric condensation increases in density and cools from $T\sim25,000$ K to $T\sim10,000$ K. The evolution of the continuum after 4~s is markedly different if the beam heating continues or is turned off.     The continuum contrast at $\lambda=6173$ \AA\ (gray solid line) and the temperature increase in the upper photosphere (dashed-dotted line) are discussed in Appendix A; the temperature increase in the upper photosphere ranges from 4750 K to 5680 K on the right axis.   } \label{fig:lc2826}
\end{center}
\end{figure}

\begin{figure}[H]
\begin{center}
\includegraphics[scale=0.6]{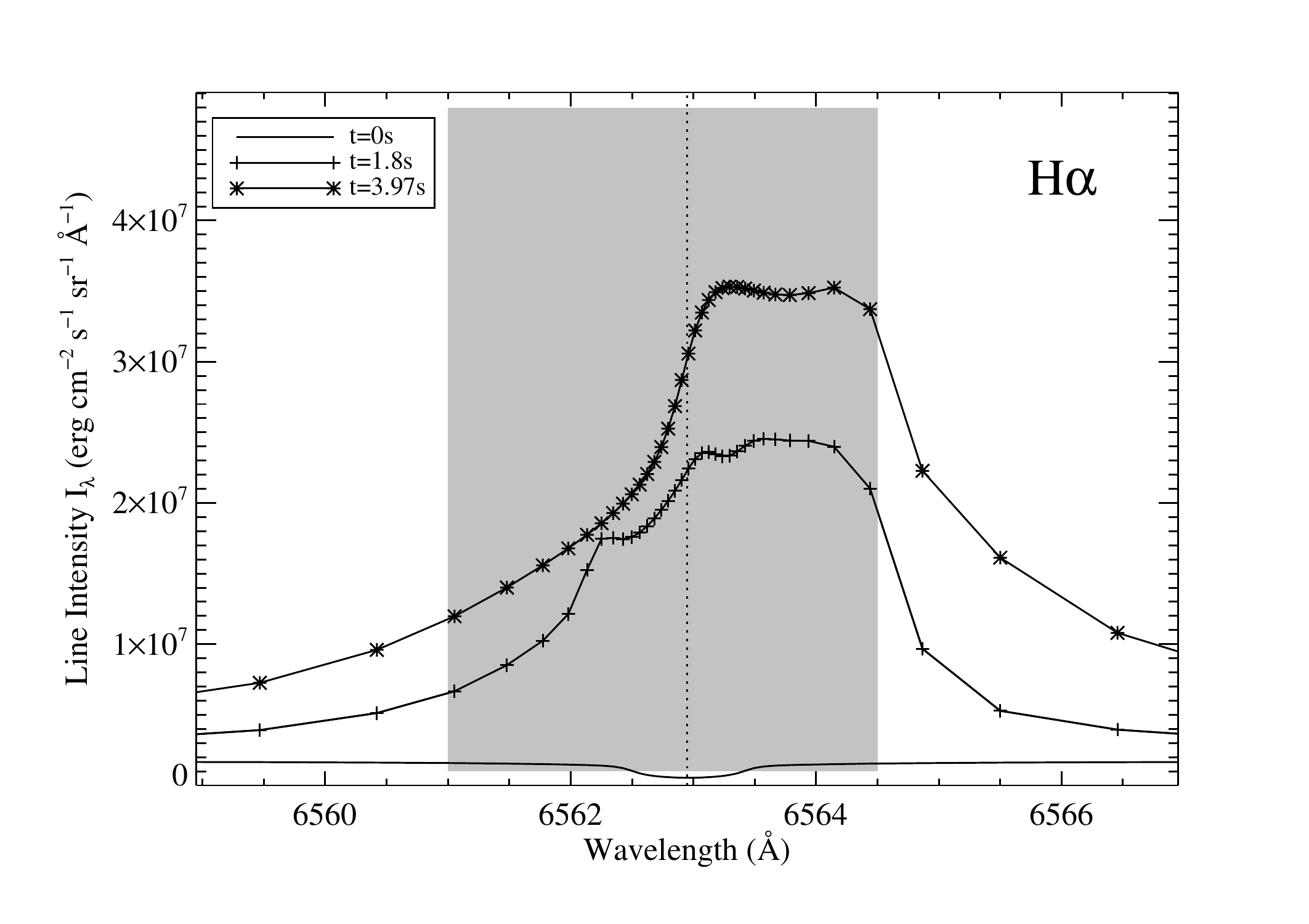}
\caption{Instantaneous H$\alpha$ line profiles from the 5F11 simulation.  The rest wavelength is indicated by a vertical dotted line.  This wavelength range is the same
as in the \cite{Canfield1987} figures, to facilitate comparison.  The wavelength range of the IBIS spectral scans of H$\alpha$ during this flare from \cite{Fatima2016} are indicated
by the gray shaded region.  Note, the default value of the microturbulence parameter in RADYN of 2 \kms\  is used for these profiles.   A bisector velocity 
of $\sim30$ \kms\ is achieved at $t=3.97$~s, which is similar to the value obtained from the Mg II lines from IRIS. } \label{fig:Halpha}
\end{center}
\end{figure}

\begin{figure}[H]
\begin{center}
\includegraphics[scale=0.6]{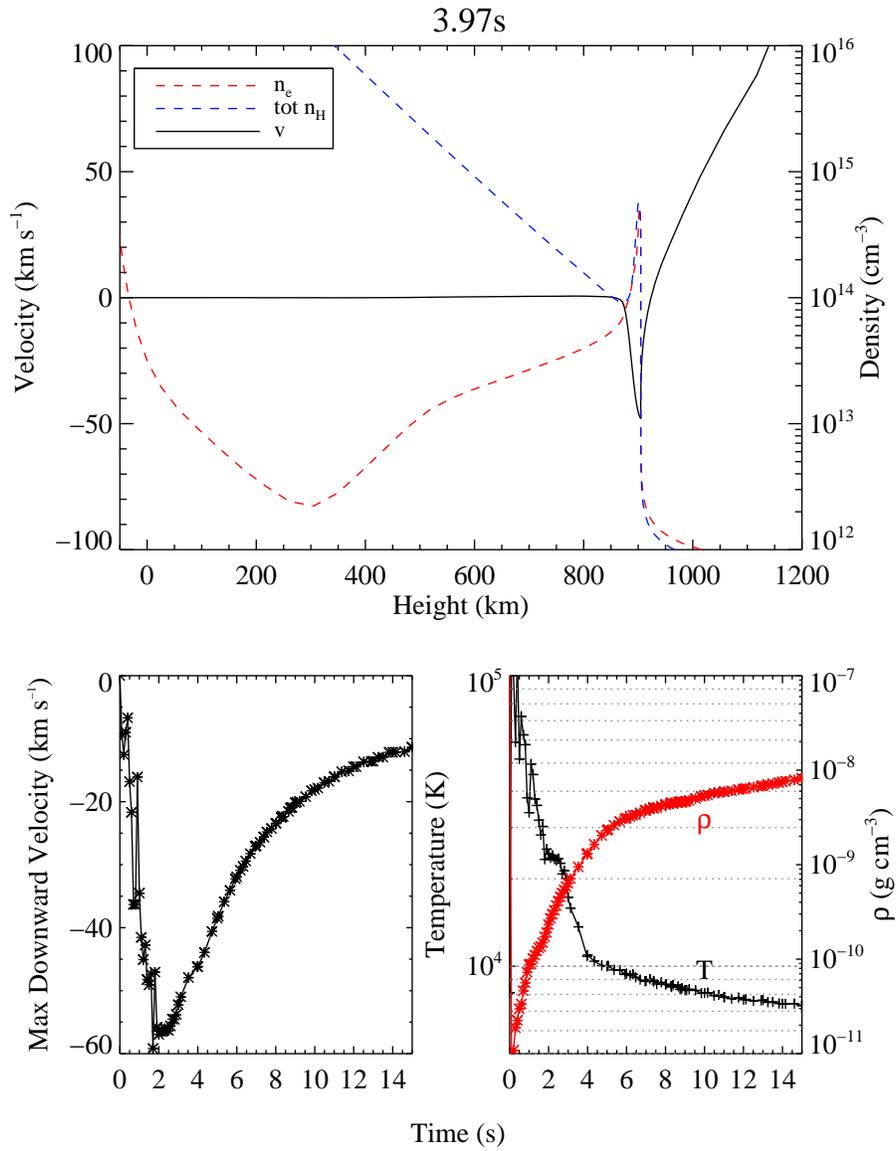}
\caption{(Top) The density and velocity profiles of the lower atmosphere at $t=3.97$~s in the 5F11 model.
(Bottom, left) Time evolution of the maximum downward vertical velocity. 
 (Bottom, right) Time-evolution of the temperature (black crosses) and density (red asterisks) at the location of the densest region of the CC for the extended 5F11 heating run.  In the bottom right panel,
 horizontal dashed lines indicate the logarithmic tick marks on the left axis.  The density increase in the top panel at $z\sim900$ km is the CC, and the temperature and density evolution are anti-correlated.   } \label{fig:vz_evol}
\end{center}
\end{figure}

\begin{figure}[H]
\begin{center}
\includegraphics[scale=0.6]{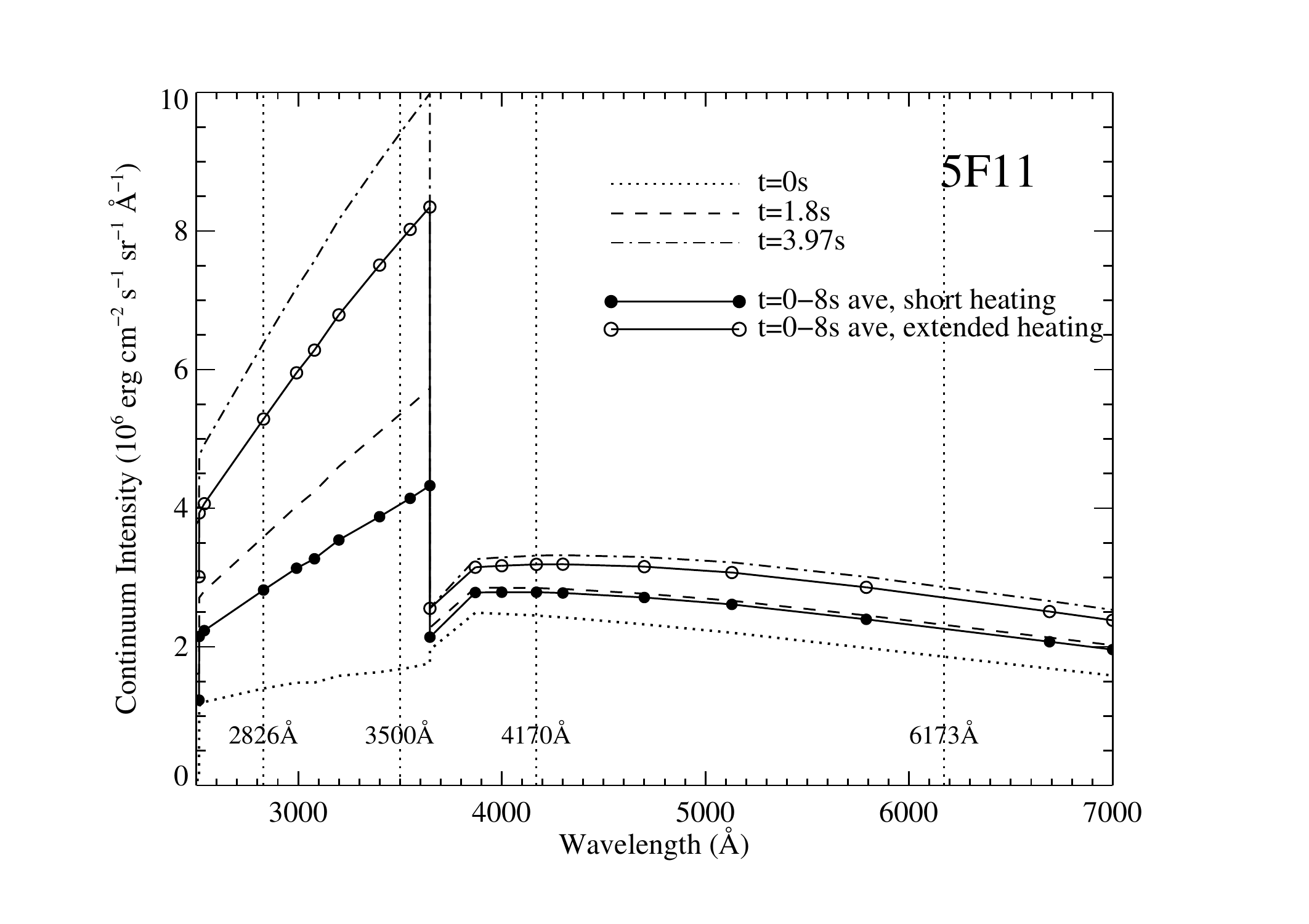}
\caption{Continuum spectral evolution at selected times (0~s, 1.8~s, and 3.97~s) in the 5F11 and averaged over the first 
8~s for the extended and short 5F11 model runs.  Specific wavelengths discussed in the text and Appendix A are indicated
by vertical dotted lines.  The excess continuum emission at $\lambda < 3646$ \AA\ (the Balmer limit) is predominantly hydrogen recombination radiation emitted
over low ($\tau_{\lambda} \lesssim 0.2$) optical depth in two flaring layers, with relatively more emergent intensity from the CC at 3.97~s than at 1.8~s.   The ratio 
of excess NUV continuum to excess optical continuum is large (see Appendix A) due to the low optical depth at $T\sim10,000$ K. } \label{fig:contspec}
\end{center}
\end{figure}

\begin{figure}[H]
\begin{center}
\includegraphics[scale=0.75]{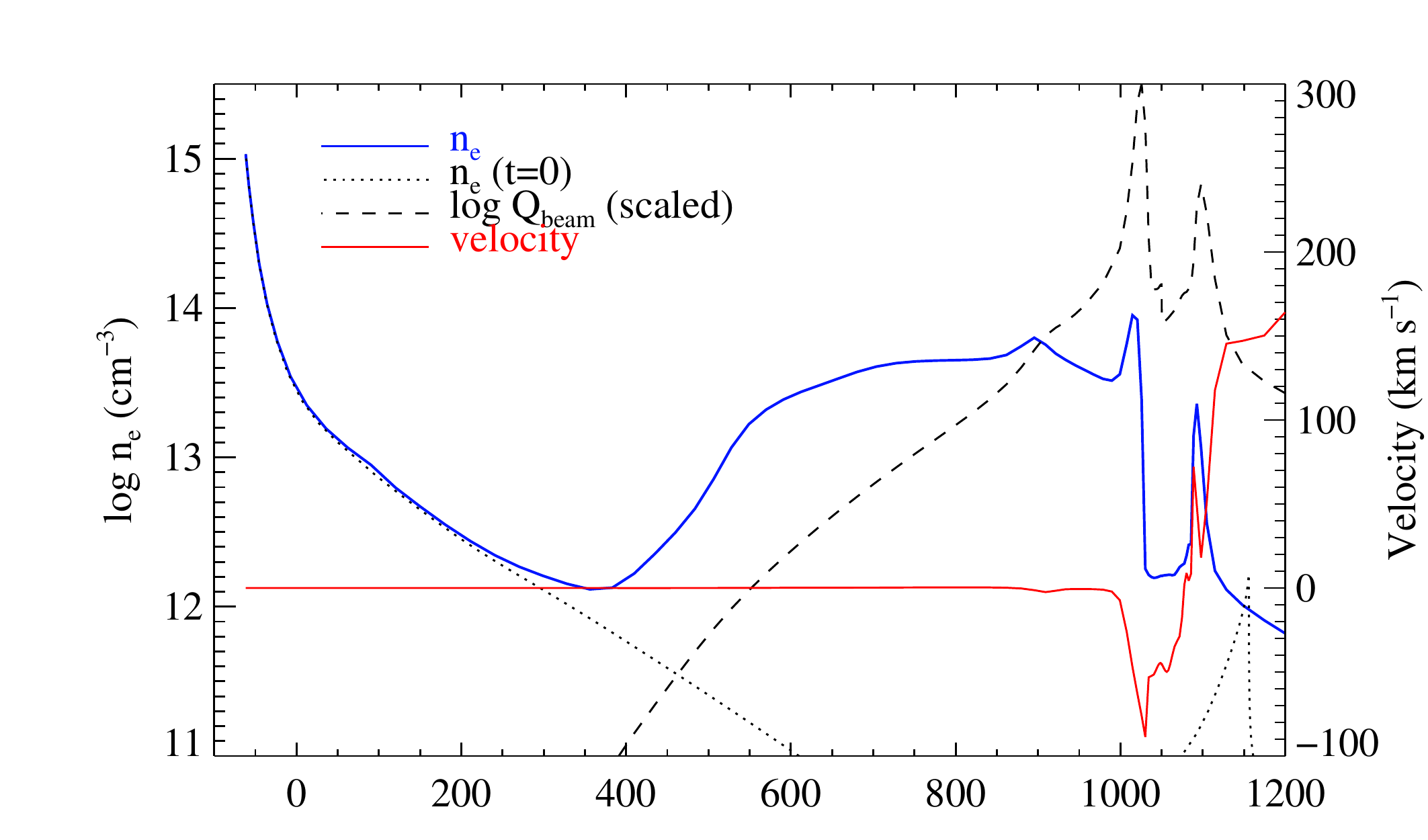} \\
\includegraphics[scale=0.75]{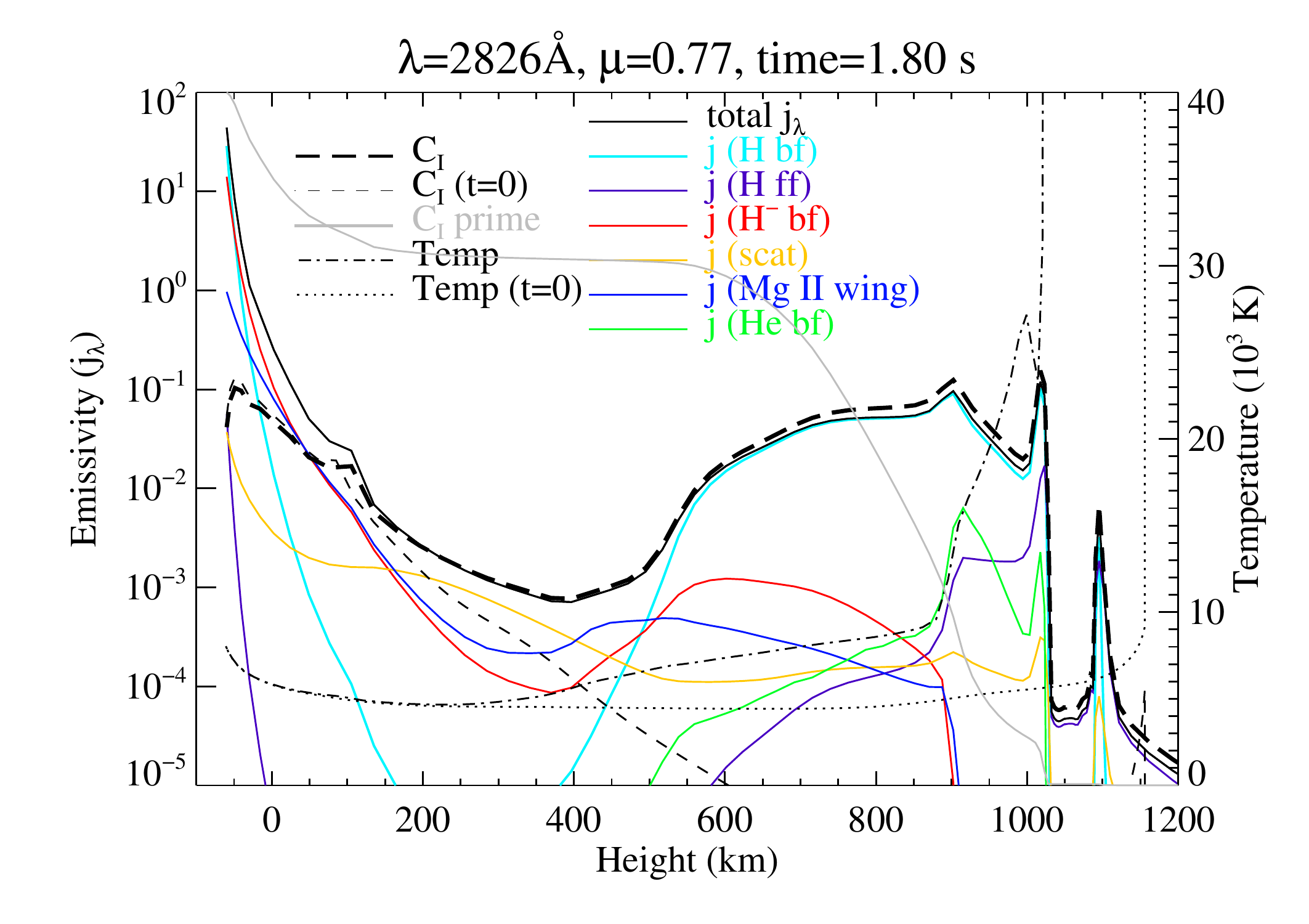}
\caption{ (Top) Atmospheric parameters at $t=1.8$~s.  The volumetric beam heating is shown over 4.5 dex on the left axis.  (Bottom) The $\lambda=2826$ \AA\ contribution function at $\mu=0.77$ at $t=1.8$~s compared to the temperature profile (right axis).   The total emissivity and the emissivity for relevant atomic processes are shown as a function of height; the solid gray line is the cumulative contribution function (C$_I^{\prime}$), which ranges from 0 to 1 and is scaled linearly to the right axis.  The pre-flare contribution function is shown as a dashed line.  $j_{\rm{scat}}$ includes Rayleigh and Thomson scattering.  Spontaneous hydrogen b-f Balmer emissivity from the stationary flare layers from $z\sim600-1000$ km dominates the emergent NUV intensity at this time when the CC is $T\sim25,000$ K.} \label{fig:jtot1}
\end{center}
\end{figure}

\begin{figure}[H]
\begin{center}
\includegraphics[scale=0.75]{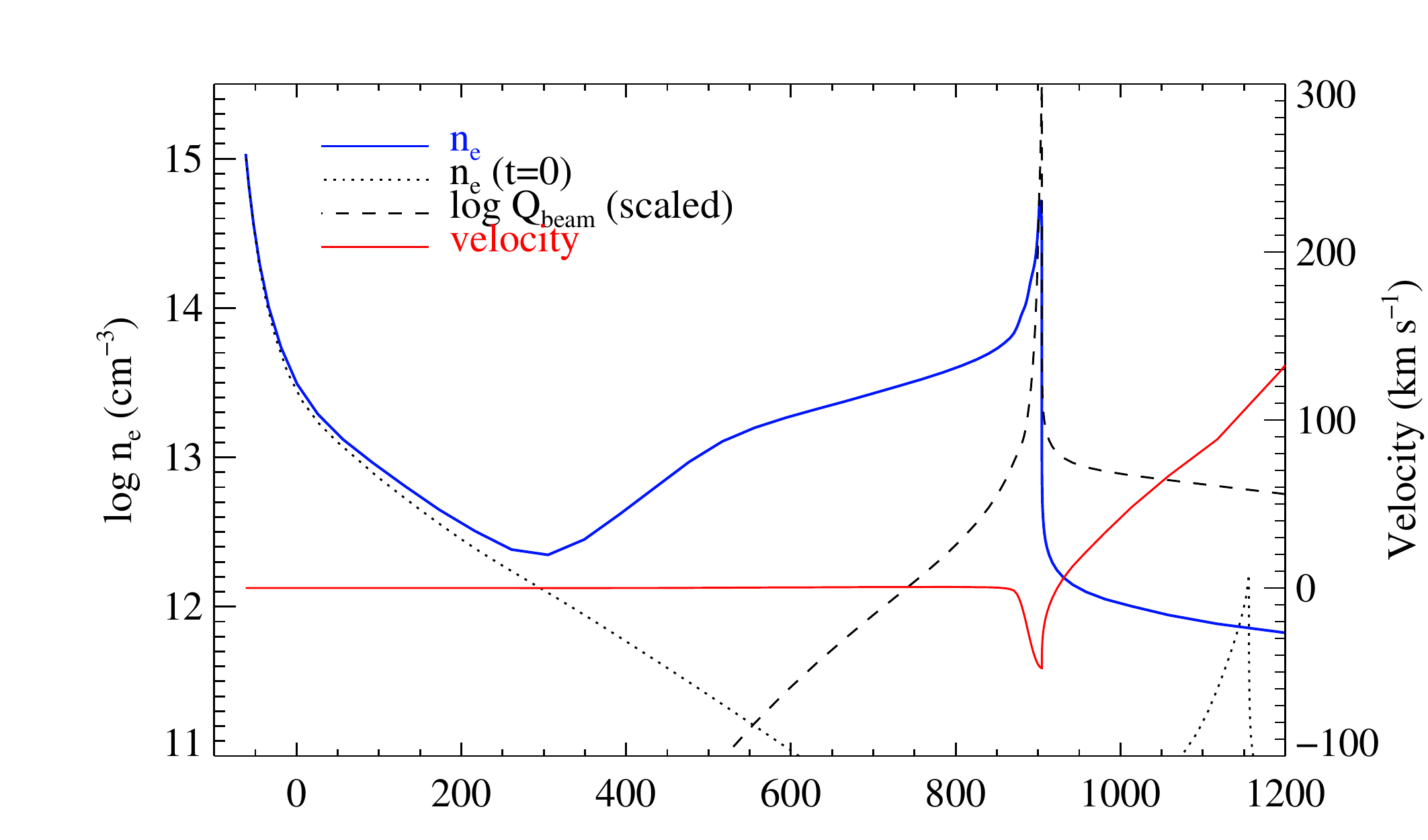} \\
\includegraphics[scale=0.75]{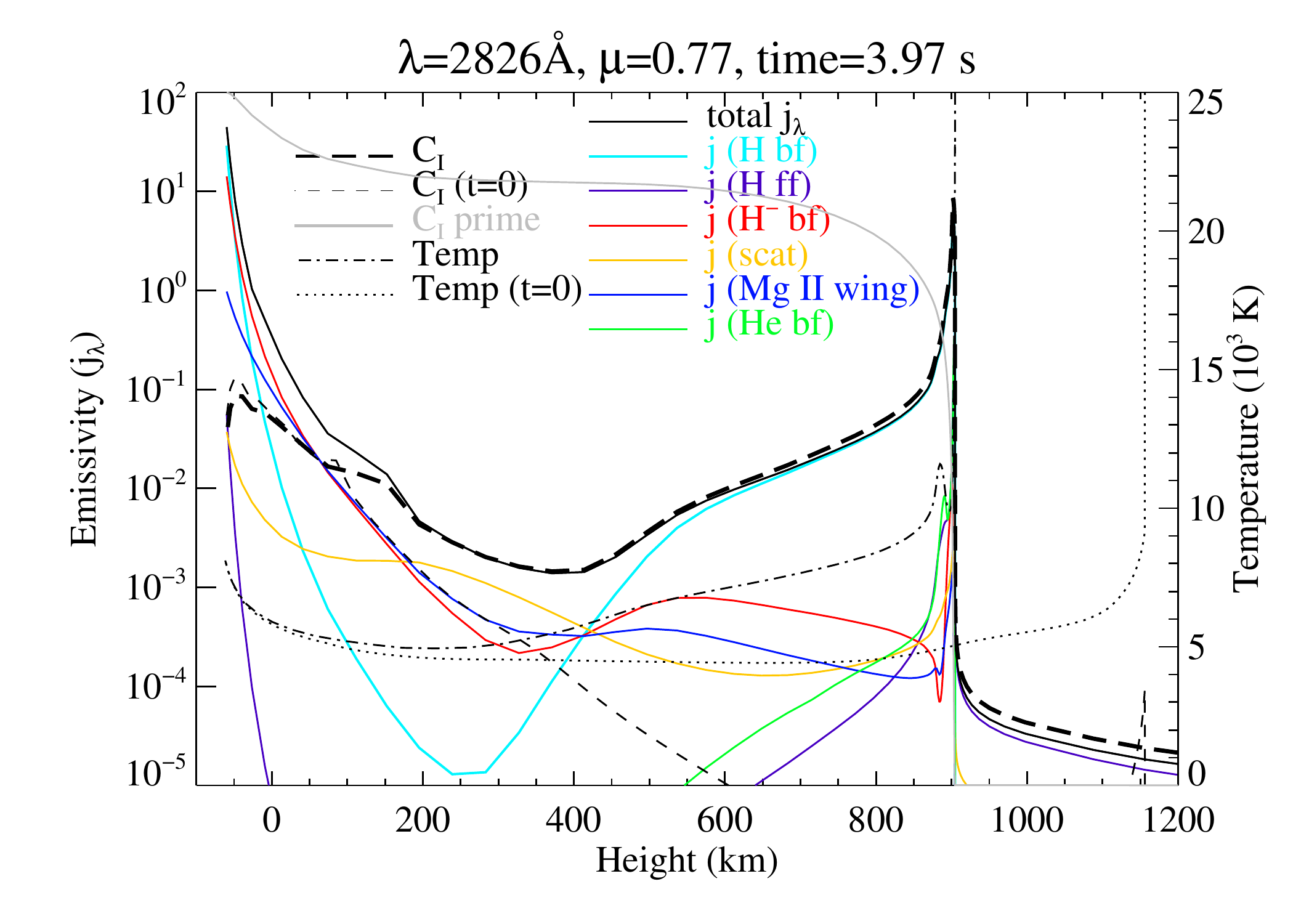}
\caption{ Same as for Figure \ref{fig:jtot1} but at $t=3.97$s.  Spontaneous hydrogen Balmer b-f emissivity from a narrow height range in the CC at $z\sim900$ km dominates the emergent NUV continuum intensity at this time. } \label{fig:jtot2}
\end{center}
\end{figure}

\begin{figure}
\plottwo{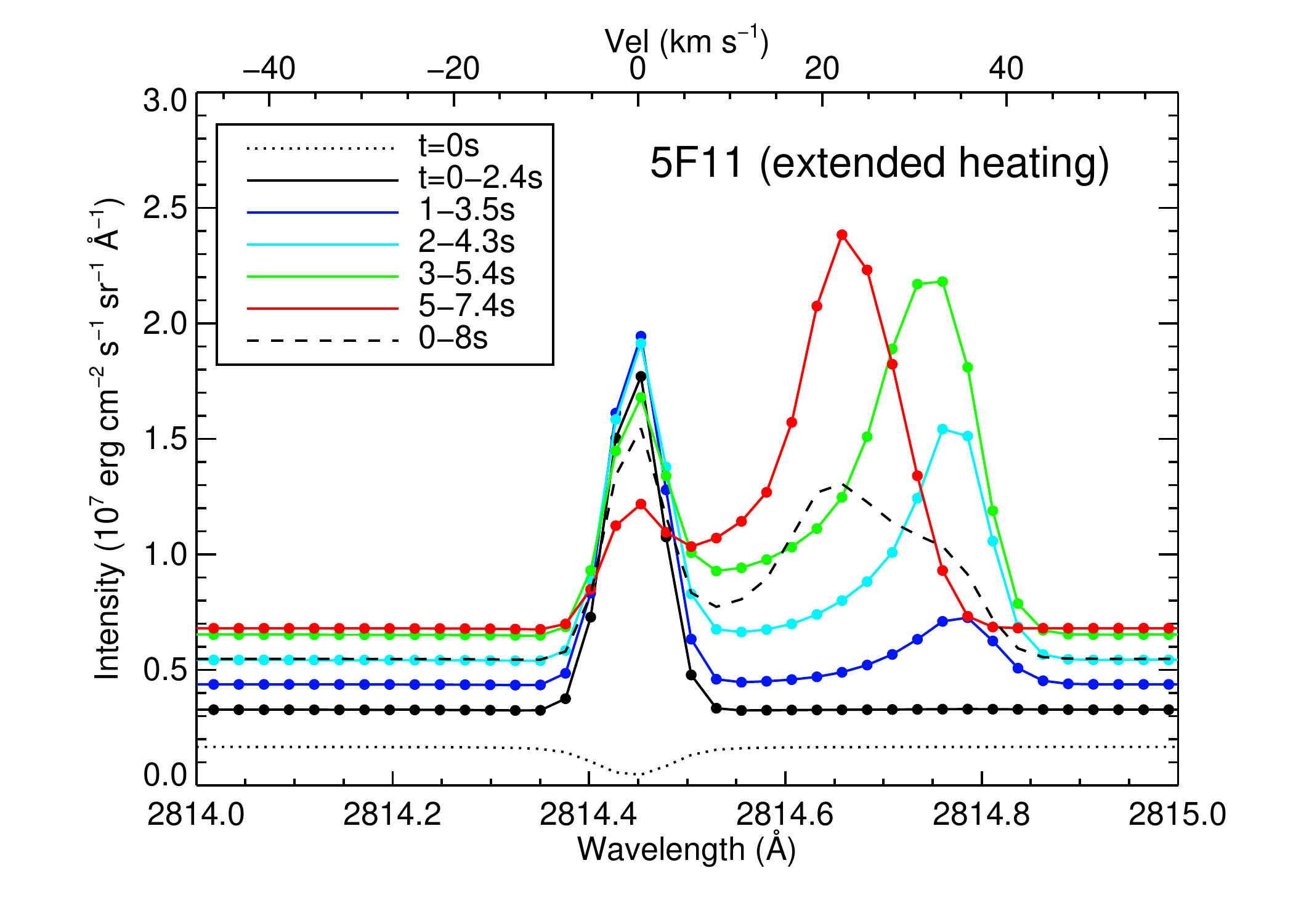}{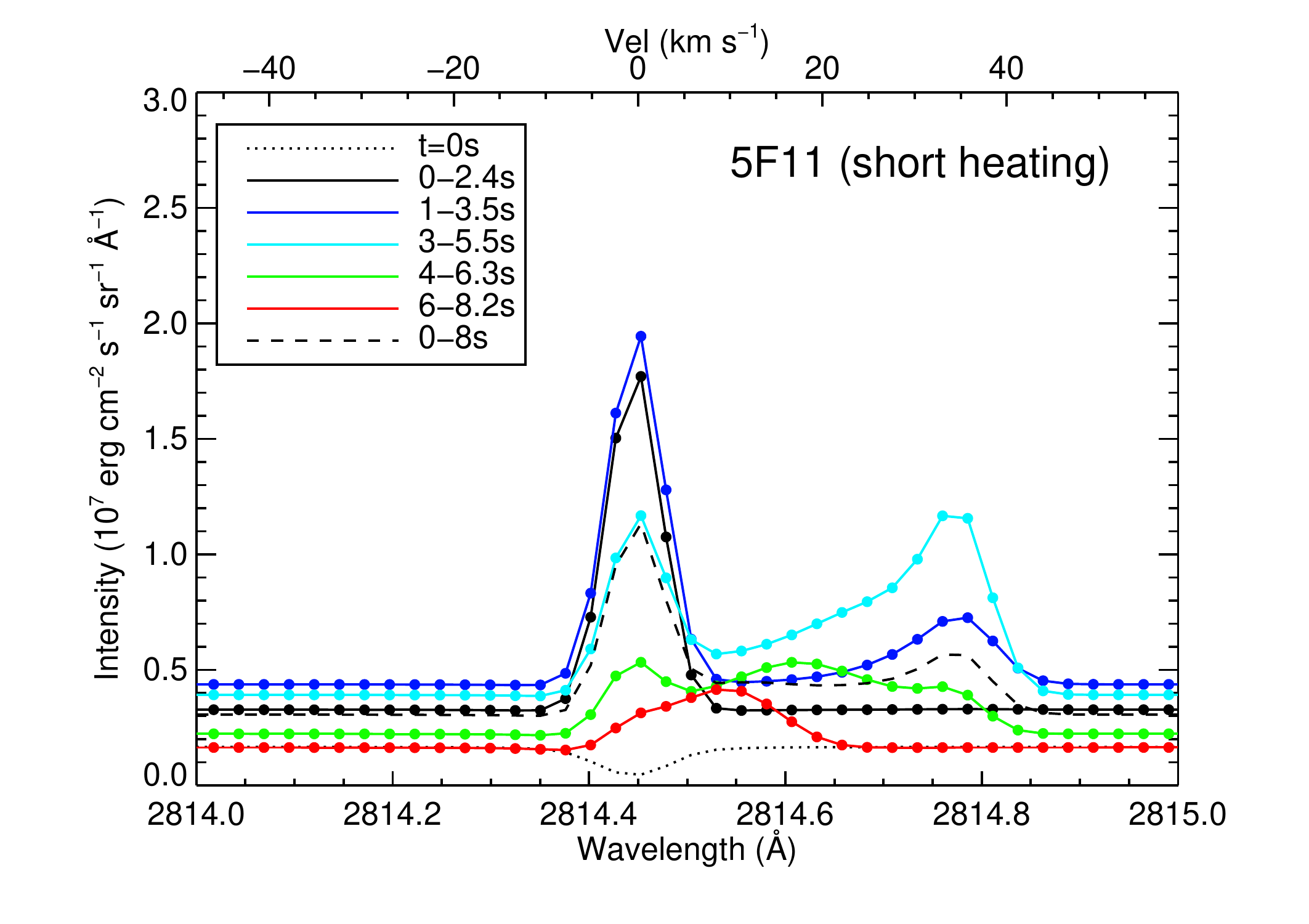}
\plottwo{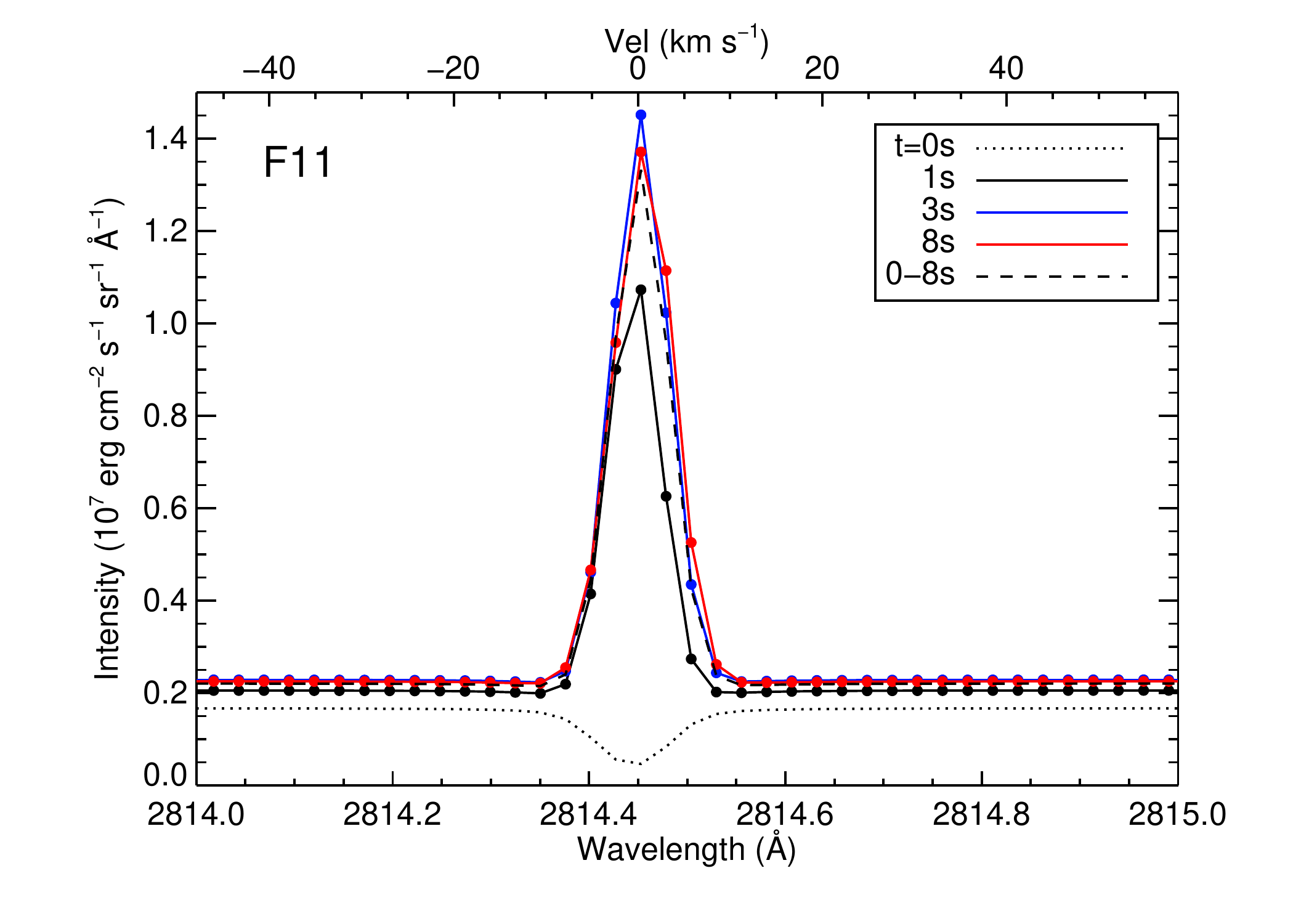}{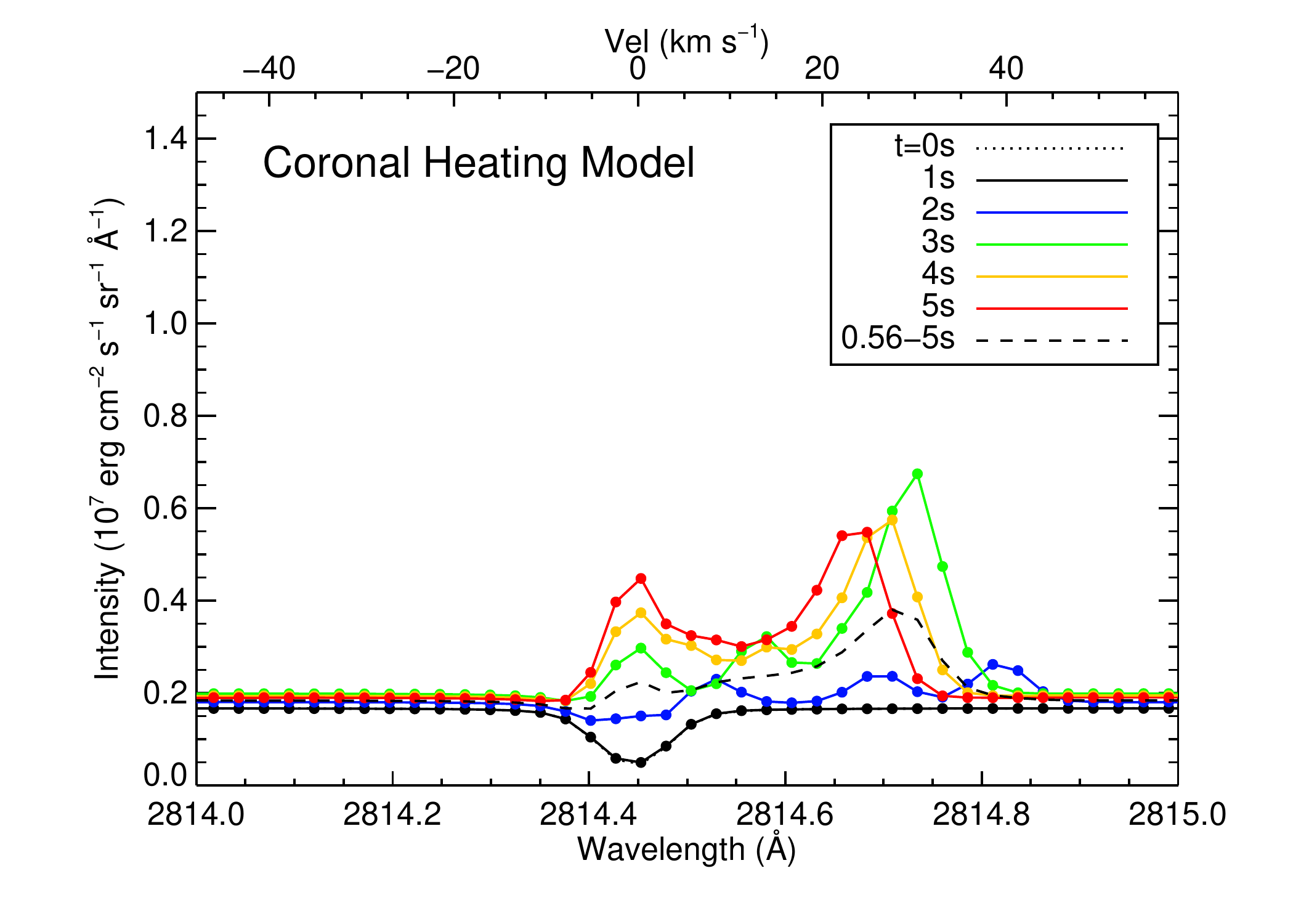} 

\caption{ (Top) The evolution of the \feone\ line for the 5F11 model with extended heating (top left) and short heating (top right) averaged over select 2.4~s intervals.  The preflare LTE calculation is shown as a dotted line, and the dashed curve is the 8~s exposure average spectrum from Figure \ref{fig:feii_2814}. (Bottom left) The evolution of instantaneous LTE \feone\ profiles in the F11 simulation at selected times; the average over the first 8~s is shown as a dashed line. (Bottom right) The evolution of the instantaneous LTE \feone\ profiles for the coronal heating model at selected times; the average over the first 5~s is shown as a dashed line. Note the small increase in the continuum emission and the smaller ranges on the y-axes in the bottom panels.  Nonthermal broadening is not used in these calculations.
Averaged over the heating, the F11 model produces no RWA intensity, the coronal heating model produces too much RWA intensity compared to intensity at the rest wavelength, and the 5F11 model produces
the observed amount of RWA intensity relative to the intensity at the rest wavelength.   The RWA intensity in the 5F11 model appears at large positive velocity and the peak shifts to bluer wavelengths as the CC decreases in velocity (Figure \ref{fig:vz_evol}).  }
\label{fig:feii_sequence}
\end{figure}

\begin{figure}[H]
\begin{center}
\includegraphics[scale=0.6]{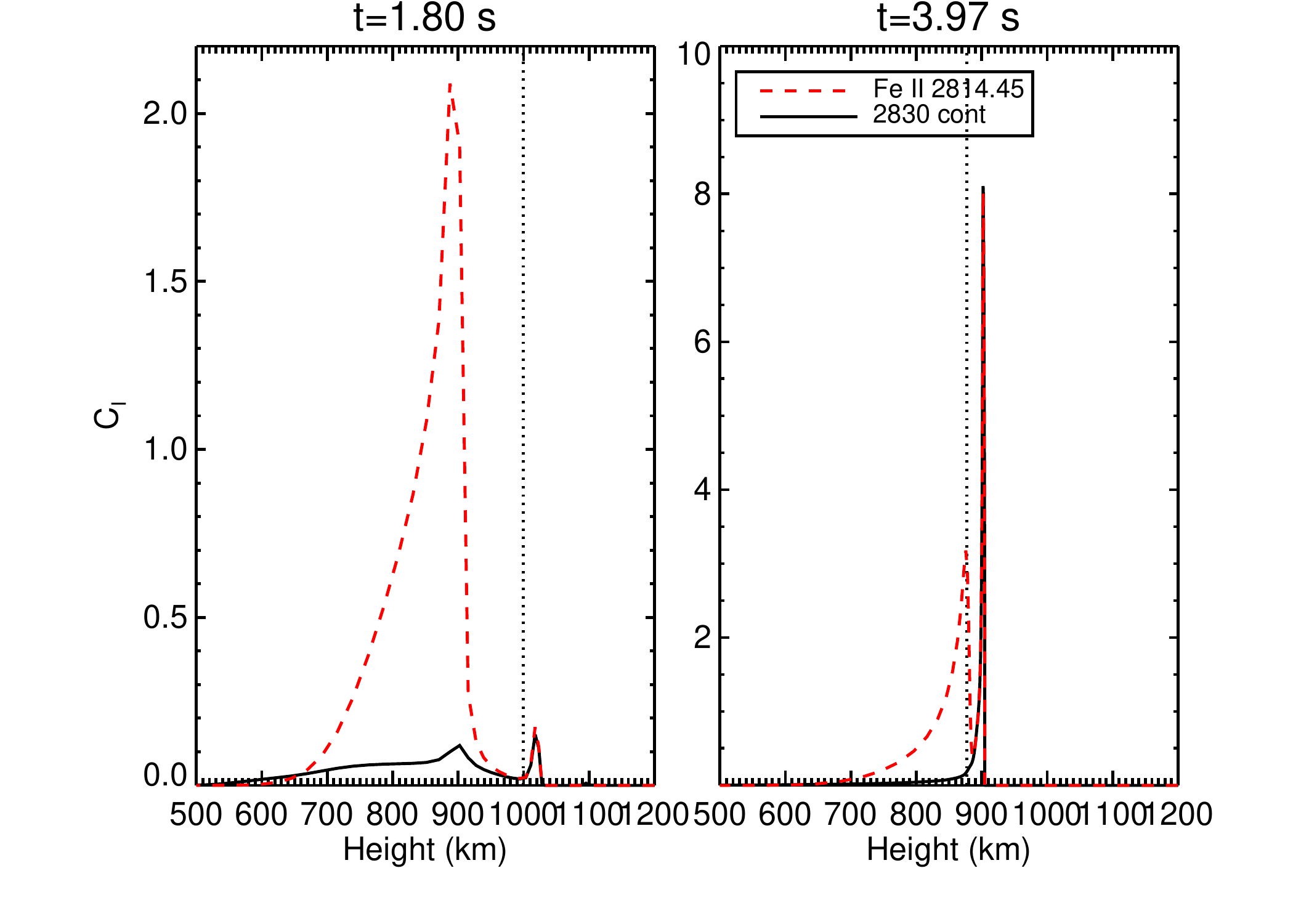}
\caption{  Contribution function for the rest wavelength of \feone\ 
and for C2826 at $t=1.8$~s and $t=3.97$~s in the 5F11 model.  The vertical dotted lines
indicate the bottom of the chromospheric condensation ($z=z_2$ in Table \ref{table:params}).  A nonthermal broadening parameter of $\xi=7$ \kms\ is used here and the value of $\tau=1$ for line-center of \feone\ is between $z=805-810$ km at both times. \feone\ is not completely optically thin and thus probes the conditions in the stationary flare layers from $800-900$ km; the intensity at \lamr\ decreases as the chromospheric condensation descends into the stationary flare layers below $z\sim900$ km.}
\label{fig:feii_ci}
\end{center}
\end{figure}

An important property of optically thin hydrogen recombination emission is a large ratio of NUV to optical excess continuum intensity \citep{Kowalski2015}, which is evident
in the continuum spectrum of the 5F11 model in Figure \ref{fig:contspec}.
The ROSA instrument \citep{Jess2008} employs two custom continuum filters at  $\lambda=3500$ \AA\ and 4170 \AA\ which can provide measurements of the 
Balmer jump ratio as done for flares on other stars with the high speed camera ULTRACAM \citep{Kowalski2016}.  For the 5F11
model the Balmer jump ratio ($F_{3500}/F_{4170}$) of the excess continuum emission is $\sim9$ in the rise phase and peak of Figure \ref{fig:lc2826}, which is 
consistent with an optically thin hydrogen recombination spectrum. These Balmer jump ratio measurements are not available for this flare, so we also calculate a ratio of excess C2826
to excess red optical continuum emission (C6173) near in wavelength ($\lambda=6173$ \AA) to SDO/HMI, which is often used
as a proxy for the optical continuum intensity in solar flares \citep[e.g.][]{Kuhar2016}.  Using the data from Table 2 of \cite{Kleint2016}, this ratio is approximately 5
for the brightest footpoints observed with both instruments (e.g., BFP1 and BFP2).  In the 5F11 model, the C2826/C6173 ratio is 5 at $t=1.8$~s and $t=3.97$~s, and is generally
consistent with these observational constraints\footnote{We have assumed that the SDO/HMI flare contrast is spatially resolved in this flare; using the IRIS data, the BFP1 and BFP2 footpoints are just resolved at the resolution of HMI (see Section \ref{sec:limitations}).}.  

The temperature evolution of the 5F11 atmosphere at a representative height in the pre-flare upper photosphere ($z=140$ km) is shown in Figure \ref{fig:lc2826} on the right axis with values ranging from $T=4750$ K to $T=5680$ K. The height of $z\sim140$ km in the 5F11 model is at the same column mass (0.25 g cm$^{-2}$) as $z\sim350$ km in the VAL3C used in the phenomenological modeling studies of this flare with the RH code \citep{Judge2014, Kleint2016}.  
  In the 5F11 model, the temperature increase  at $z\lesssim 500$ km is due to the absorption of Balmer continuum photons produced in the stationary flare layers and CC \citep[radiative backwarming;][]{Machado1989}, as in the RHD models of \cite{Allred2005, Allred2006, Cheng2010}. The 5F11 model predictions of the upper-photospheric heating from NUV backwarming are in general agreement with the increase in temperature in the upper photosphere from the phenomenological models of \cite{Kleint2016}.
  
   The evolution of the emergent intensity contrast ($\frac{I-I_{\rm{pre}}}{I_{\rm{pre}}}$) at $\lambda=6173$ \AA\ is also shown in Figure \ref{fig:lc2826}.  A maximum of $\sim65-70$\% contrast is achieved at $\sim5$~s while slightly increasing after this time for the extended heating run.  The maximum contrast is several times larger than the peak values in this region of the \marflare\ flare \citep[15\%;][]{Kleint2016}, but the contrast is only $\sim25$\% at 1.8~s in the 5F11.   
   
  Following the NUV continuum emissivity analysis in Section \ref{sec:continuum}, we calculate that $\sim75$\% of the excess red optical (C6173) continuum intensity in the first four seconds of the 5F11 originates from the 
 the CC (and stationary flare layers) at $z > 744$ km (Table \ref{table:params}).  
   Most of this emission is optically thin Paschen recombination radiation.
 The remaining $\sim25$\% of the 
  excess C6173 intensity originates from H$^-$ emission below the stationary flare layers at $z\sim 200-744$ km; the 
  upper photosphere at $z\lesssim200$ km does not produce large H$^-$ emissivity until it heats more than the $\Delta T \sim 300$ K that occurs within the first four seconds.
  In contrast to the NUV, the total optical continuum emissivity at $z>200$ km has 
  non-negligible amounts of hydrogen free-free emission\footnote{At $t=1.8$~s there are equal amounts of hydrogen free-free and hydrogen bound-free red optical emissivity in the CC.} at higher temperatures near $T\sim25,000$ K and H$^-$ bound-free emission from
  the cooler layers, which are heated by the beam and backwarming at $z<744$ km.

  We extend the red optical continuum analysis to $t=15$~s in the extended 5F11 heating model which gives insight into the difference in the time evolution of the excess C2826 and the C6173 contrast
  in Figure \ref{fig:lc2826}.  The temperature of the CC at 15~s ($T=7500$ K; Figure \ref{fig:vz_evol}) produces
 increasingly strong H$^{-}$ emission at $\lambda=6173$\AA\ which is comparable to the Paschen recombination emission in the CC at this time.  
Due to the temperature increase in the upper photosphere, $\Delta T_{\rm{upper phot}} \sim 900$ K, the C6713 emissivity from lower heights ($z<500$ km) is nearly comparable to the emission from the CC at $t=15$~s. 
 
 The excess red optical light curve (Figure \ref{fig:lc2826}) is rather flat at later times in the extended 5F11 simulation and the NUV light curve decreases.  Therefore, the 
 Balmer jump ratio decreases over time.  By 15~s, the amount of height-integrated hydrogen NUV (Balmer) b-f emissivity in the CC and stationary flare layers has decreased.  In the NUV, the increased amount of H$^{-}$ emission from the upper photosphere is insignificant compared to the (bright) Balmer continuum emission and thus the light curve is dominated by the evolution of the 
 NUV continuum emission in the CC.  The NUV emergent continuum intensity decreases after 5~s due to a small physical depth range ($\Delta z \sim 25$ km at 15~s, compared to $160$ km at $t=3.97$~s).
  The Balmer jump ratio of the excess spectrum decreases from 9 at $t=3.97$~s to 
$\sim5-6$ at 15~s in the extended 5F11 heating simulation; the C2826/C6173 continuum ratio decreases from 5 to $\sim3$.  The decrease in the ratio of excess NUV to optical continuum intensity is due to the smaller physical depth range of Balmer continuum intensity in the NUV 
and the increasing H$^-$ optical emissivity from the upper photosphere which heats to $T\sim5700$ K and from the CC which cools below $T\sim8000$ K.  
   
In summary, upper photospheric heating of $\Delta T \sim$900 K occurs in the 5F11 model but is delayed with respect to the NUV continuum emission.  The heating in the upper photosphere is caused by radiative backwarming by photons in the Balmer continuum range as in previous RHD models with lower beam fluxes \citep{Allred2005}.  The temperature increase inferred from the phenomenological modeling of the IRIS NUV data during this flare in \cite{Kleint2016, Judge2014} is generally consistent with the 5F11 model backwarming effects.  Upper photospheric ($z\sim140$ km) heating by $\Delta T \sim 300$ K that occurs within the first four seconds in the 5F11 but the majority of the excess red optical continuum intensity in the early times of the heating is from optically thin Paschen recombination emission in the CC and stationary flare layers.  An increase in the H$^-$ emission occurs in all layers from the upper photosphere to the chromospheric condensation as the upper photosphere heats up and the condensation cools.  A predominantly optically thin hydrogen recombination spectrum in the early phase predicts a large Balmer jump ratio, and the increasing H$^-$ emission decreases the ratio 
for extended beam heating.

\section*{Appendix B: Constraints on hot ($T\sim9000$ K) blackbody-like emission at blue optical wavelengths}
\cite{Kretzschmar2011} measured an optical color temperature of T$\sim9000$ K in Sun-as-a-Star
narrowband continuum data in blue (4020 \AA), green (5000 \AA), and red (8620 \AA) filters from the VIRGO/SPM instrument (on the SOHO spacecraft) for superposed flares of GOES class C, M, and X-class in Solar Cycle 23.  

\cite{Kleint2016} combined the IRIS NUV intensity values with HMI ($\lambda\sim6173$\AA)  and IR ($\lambda\sim10832$\AA) intensity measurements to constrain the coarse properties of the white-light spectral energy distribution (for the footpoints observed with spectra, including BFP1 and BFP2) in the \marflare\ flare.  A blackbody with $T\sim6300$ K could fit the optical and IR data,
and a Balmer continuum emission component was necessary to account for the IRIS NUV continuum enhancement \citep[see][for a similar conclusion using spectra of dMe flares]{Kowalski2010, Kowalski2013}.
Moreover, \cite{Kleint2016} noted that subtracting a pre-flare blackbody ($T=5770$ K) from the flare blackbody ($T=6300$ K) results in 
an optical color temperature of $T\sim9000$ K for the \marflare\ flare, putting the results for this flare in line 
with the results from the Sun-as-a-star analysis of \cite{Kretzschmar2011}.

We repeat the calculations (L. Kleint, priv communication)
using the data from \cite{Kretzschmar2011} while adjusting for the fraction of the solar surface no longer emitting at pre-flare values during the flare
\citep[Equation 3 of][]{Kowalski2016} for a direct comparison to a blackbody flare model.  The high-thresh area in the SJI2832 images for the \marflare\ flare is $\sim10$ arcsec$^2$ (Section \ref{sec:area}), which is also the 
 typical area of white-light emission sources in TRACE/WL data \citep[L. Fletcher priv. communication;][]{Hudson2006, Fletcher2007}.  Assuming this 
area for the optical flare emission observed with VIRGO/SPM, the
color temperature range that characterizes the peaks of the average X-class and average M-class flares becomes $T\sim8000-8500$ K, or about 1000 K less than the values\footnote{For the 2003 Oct 28 flare considered in \cite{Kretzschmar2011} with an area that has been directly measured to be 130 arcsec$^2$, we obtain a refined estimate of 7900 K.} in \cite{Kretzschmar2011}.  
However, the high-thresh (where $I_{\lambda,\rm{excess}}\ge3\times10^6$ \cgs) excess specific luminosity accounts for only $\sim$1/3 of the low-thresh excess specific luminosity in the impulsive phase
and thus the emission from the high-thresh area is not likely to dominate a Sun-as-a-star measurement as would be observed with VIRGO/SPM; a larger area may be closer to the relevant area to use to interpret the measurements in \cite{Kretzschmar2011}.  
 If we use the area of the low-thresh region for the SJI 2832 images ($\sim100$ arcsec$^2$),
then the color temperature range that characterizes the peaks of the VIRGO/SPM flares becomes $\sim7000$ K (or less for larger areas), which is closer to the observed color temperature in this flare \citep{Kleint2016}. The assumptions for these
calculations include that the area emits at the standard irradiance value in each filter before the
flare occurs. If this area were brighter 
than the standard value before the flare, then the inferred color temperature decreases; if this area were dimmer the inferred color temperature increases.

The spectra of BFP1 and BFP2 at 17:46:08$+$75~s and 17:46:24$+$75s, respectively, in Figure \ref{fig:brightest} are in the decaying phase of these sources and
show NUV continuum emission with a C2826 excess of $\sim0.25-0.5\times10^6$ \cgs.  This range is less than the low thresh criterion for SJI 2832 emission ($I_{\lambda,\rm{excess}}>0.6\times10^6$ \cgs; Section \ref{sec:area}).   Accounting for some contribution from emission lines in SJI 2832,
faint NUV continuum intensity would be expected to contribute to the areas corresponding to the low-thresh value.  However, the Balmer continuum-emitting (NUV) areas
 have been inferred to be an order of magnitude larger than optical-continuum emitting areas in dMe flares \citep{Kowalski2010}, and the optical and IR continuum source size
 in an X-class solar flare has been found to vary as a function of wavelength \citep{Xu2012}.   High spatial resolution measurements of solar flare optical and NUV areal measurements would clarify the source area to use when interpreting the data from \cite{Kretzschmar2011}.

We measure the color temperature of the blue-to-red optical continuum in the 5F11 model in Figure \ref{fig:contspec} for comparison to the observational constraints in \cite{Kleint2016}.  In the first 4~s, the color temperature from $\lambda=4020$ \AA\ to 5000 \AA\ ranges between $T=6600-6900$ K, which is comparable to the observed color temperature (6300 K) in the \marflare\ flare.
 The color temperature during the 5F11 flare heating is lower than the color temperature ($T\sim7100$ K) at $t=0$~s 
because the RADYN model\footnote{At $t=0$~s, $T_{\rm{rad}}=5740$ K and  $T_{\rm{rad}}=6040$ K at $\lambda=5000$ \AA\ thus giving a higher color temperature.} does not include line haze opacity at blue and violet wavelengths \citep{Vernazza1976}.  In the 5F11 model, the lower color temperature of the blue optical wavelength range during the flare results because the flare emission is dominated by 
 optically thin Paschen recombination radiation (with a smaller contribution from hydrogen free-free emission) as found for the red optical continuum intensity in Appendix A.

We use the high spatial resolution of the IRIS SJI 2832 data to characterize the largest possible radiation temperature for the brightest pixels of BK2830 in the \marflare\ flare, which corresponds to $T_{\mathrm{rad}} \approx 7560$ K.  
The blackbody radiation temperature of $\sim7560$ K is close to the refined range of $T=8000-8500$ K for the VIRGO/SPM data of \cite{Kretzschmar2011} using the smaller estimate for the area.
However, the 5F11 model provides an alternative explanation (Section \ref{sec:sjibright}) for the brightest NUV flare pixels in the SJI 2832 data of the \marflare\ flare as a combination of Fe \textsc{ii} emission lines and 
 hydrogen Balmer recombination radiation from low optical depth.
   For the brightest pixels with spectra and slit jaw data in the impulsive phase, a moderately hot ($T\gtrsim8000-9000$ K) blackbody-like component is thus not necessary to explain the IRIS data.
However, due to the relatively low time sampling of the NUV spectral and SJI data relative to the 
 impulsive phase duration ($\sim120$~s), the constraints on the formation of a hot blackbody-like spectrum in a given pixel for this flare are limited.   
We note that the nonthermal power for the \marflare\ flare is low \citep[$8\times10^{27}$ erg s$^{-1}$][]{Kleint2016} compared to the nonthermal power inferred for some other flares \citep[$\sim10^{29}$ erg s$^{-1}$;][]{Matthews2003, Fletcher2007, Milligan2014}.  The super-posed epoch analysis of data from \cite{Kretzschmar2011} may be biased towards the brighter white-light solar flares \citep{Kerr2014}, and brighter white-light emission generally occurs in flares with higher electron power above $50$ keV \citep{Kuhar2016}.

\section*{Appendix B.1: Comparison of 5F11 and F13 RHD Models}
In the impulsive phase of impulsive-type flares from active M dwarf stars, the NUV and blue optical continuum distribution often exhibits a color temperature 
of $T \sim 10,000 - 12,000$ K, with a small Balmer jump in emission \citep{HawleyPettersen1991, Kowalski2013, Kowalski2016}. Due to a low surface flux of an M dwarf in quiescence,  subtracting an M dwarf pre-flare spectrum from a flare observation (to infer a color temperature) does not 
artificially produce a hot continuum in the blue and NUV as is possible for the Sun as discussed in \cite{Kleint2016}.

A hot blackbody-like continuum spectrum in the NUV and optical has been produced recently in an RHD simulation of the response of an M dwarf atmosphere to an F13 nonthermal electron beam \citep{Kowalski2015, Kowalski2016}, which also produces white-light emitting CC and stationary flare layers.
 In the 5F11 electron beam model, the Balmer continuum emission originates over a large physical depth range ($\Delta z = 50 - 390$ km; Table \ref{table:params}), which is not strongly wavelength dependent,
due to the low optical depth in the CC.  In the F13 simulation, the Balmer continuum emission originates over a much smaller physical depth range of $\sim$1 km \citep{Kowalski2015IAU}\footnote{Using $\Delta z$ defined as we have in this paper;  \cite{Kowalski2015} used the FHWM of the contribution function as an indication of the physical depth range.  Using the FWHM of the contribution function at $t=3.97$~s for the 5F11 model, the physical depth range of C2826 continuum intensity is 3.5 km.} due to the larger density and optical depth in the CC.  In the F13 model, the physical depth range is strongly wavelength dependent, and only the blue optical photons ($\lambda \sim 4300$ \AA) have an optical depth  that is low enough to escape from the stationary flare layers.  Thus, we predict that optically thick lines should not exhibit an emission component at \lamr\ for flare atmospheres with CC's that exhibit large values of the optical depth at continuum wavelengths, such as a hot ($T\ge9000$ K) blackbody-like spectrum.

\acknowledgments
We thank an anonymous referee for improvements to the manuscript.  
AFK thanks Dr. L. Kleint for helpful discussions about the 2014 Mar 29 flare, Dr. J. Klimchuk for discussions on non-thermal broadening and coronal heating, Dr. F. Reale for the suggestion to compare to a coronal heating model, Dr. A. Inglis and Dr. B. Dennis for discussions about RHESSI imaging and spectroscopy, Dr. S. Jaeggli for discussions on IRIS data, Dr. J. Drake for discussions about 
beam instabilities and double-layers, Dr. H. Uitenbroek for 
the use of the RH code, M. Alaoui for helpful discussions on the return current effects during solar flares, Dr. G. Del Zanna for helpful conversations about non-equilibrium ionization, and Dr. H. Hudson, Dr. L. Fletcher for helpful discussions about this work.  
AFK also acknowledges helpful discussions at Dr. P. Testa's workshop on 
microflare heating at the International Space Science Institute and discussions at Dr. L. Harra's workshop on energy transformation in solar and stellar flares at the International Space Science Institute,
AFK acknowledges funding that supported this work from  the NASA Heliophysics Guest Investigator Grant NNX15AF49G and funding from the Goddard Planetary Heliophysics Institute (GPHI) Task 132,
and support from the AAS Solar Physics Division 2015 Metcalf Travel Award to present results from this work at the Hinode 9 workshop at Queen's University Belfast.  
The research leading to these results has received funding from the European Community's Seventh Framework Programme (FP7/2007-2013) under grant agreement no. 606862 (F-CHROMA)and ERC grant agreement no. 291058 (CHROMPHYS).  IRIS is a NASA small explorer mission developed and operated by LMSAL 
with mission operations executed at NASA Ames Research center and major 
contributions to downlink communications funded by ESA and the Norwegian 
Space Centre.

\clearpage
\bibliographystyle{apj}
\bibliography{mar29flare}
\clearpage

\clearpage

\clearpage

\clearpage

\global\pdfpageattr\expandafter{\the\pdfpageattr/Rotate 90}

\begin{turnpage}
\begin{deluxetable}{ccccccc}
\tabletypesize{\scriptsize}
\tablewidth{0pt}
\tablecaption{IRIS Continuum Observations}
\tablehead{
\colhead{label} &
\colhead{mid-time} &
\colhead{raster \#} &
\colhead{x [\arcsec] (raster position)} &
\colhead{y [\arcsec] (pixel)} &
\colhead{C2826 (pre) [$10^6$ \cgs]} &
\colhead{C1349 (pre) [$10^6$ \cgs]} }
\startdata
BFP1 & 17:46:08 & 173 & 518.2 (4) & 262.7 (437) &  2.84 (0.54) & 0.45 (\nodata) \\
BFP2 & 17:46:24 & 173 & 522.2 (6) & 261.5 (430) &  3.10 (0.96) & 0.27 (\nodata) \\
\enddata 
\tablecomments{ The y-pixel corresponds to the y-pixel location after
  a $+$1 pixel and $-2$ pixel shift has been applied to obtain the
  level 3 FUV and
  NUV spectral datacubes, respectively. For the $x$ and $y$ positions, IDL indices of level 3 datacubes
  correspond to the values in the parentheses after subtracting 1.}
\label{table:bfpvals}
\end{deluxetable}
\end{turnpage}
\clearpage

\begin{turnpage}
\begin{deluxetable}{cccc}
\tabletypesize{\scriptsize}
\tablewidth{0pt}
\tablecaption{Model Observables}
\tablehead{
\colhead{time} &
\colhead{excess C2826 (RADYN)} &
\colhead{excess C2826 (RH)} &
\colhead{H$\alpha$ 30\% bisector velocity [km s$^{-1}$]} }
\startdata
5F11 & & & \\
\hline
$t=1.8$~s & 2.2   &  2.4 &  $\sim15$ \\
$t=3.97$~s & 5.1  & 5.4 &  $\sim30$ \\
$t=15$~s (extended heating) & 3.7 & 4.1 &  $\sim10$ \\
$t=0-8$~s (short heating)  &   1.2       &   \nodata     &  \nodata    \\
$t=0-8$~s (extended heating)  &   3.9    &   \nodata     &   \nodata   \\
\hline
\hline
F11 & & & \\
\hline
$t=3$~s &  0.64   &  0.78 &  $\sim0$ \\
$t=18$~s &  0.7   &  1  &  $<5$ \\
\hline
\hline
coronal heating model & & & \\
\hline
$t=3$~s & 0.4 & 0.4 &  $\sim30$ \\
\enddata 
\tablecomments{The excess C2826 values are the excess intensity values in units of $10^6$ \cgs.  The pre-flare intensity from the RADYN calculation at 2826 \AA\ is 1.4$\times10^6$ \cgs. The pre-flare intensity from the RH calculation including the Mg II $h+k$ wing opacity is 8.4$\times10^5$ \cgs.  A value of ``\nodata'' means that these values were not necessary to calculate in this work.  }
\label{table:model}
\end{deluxetable}
\end{turnpage}
\clearpage

\begin{turnpage}
\begin{deluxetable}{cccccccccccc}
\tabletypesize{\scriptsize}
\tablewidth{8.5in}
\tablecaption{Essential Parameters of the Lower Flare Atmosphere}
\tablehead{
\colhead{Time} &
\colhead{$\Delta z$} & 
\colhead{$z_1$} &
\colhead{$z_2$} &
\colhead{$z_3$} &
\colhead{$T(z_1-z_2)$} &
\colhead{$T(z_2)$ (K) } &
\colhead{$T(z_2 - z_3)$ } &
\colhead{$n_e (z_1-z_2)$ } &
\colhead{$n_e (z_2 -z_3)$} &
\colhead{$Q_b (z>z_2)$} &
\colhead{$Q_b (z>z_3)$} \\ 
\colhead{[s]} & \colhead{[km]} & \colhead{[km]} & \colhead{[km]} & \colhead{[km]} &\colhead{[K]} & \colhead{[K]} & \colhead{[K]} & \colhead{[cm$^{-3}$]} & \colhead{[cm$^{-3}$]} & \colhead{} & \colhead{} }

\startdata
5F11 extended heating model & & && & & & & & & &  \\
\hline
1.8   & 390   & 632 &    1000    &  1020    &  10570   & 27170    &  24250  & 4.6e+13  & 7.4e+13  &   0.85  & 0.30  \\

3.97 & 160  & 744  &   878   &   904    &  8790    &  10330 & 12000  & 4.8e+13  & 3.6e+14  &   0.96  &   0.47  \\

7.0  &  50 & 741     & 766   &   794     & 8190    &  8520  & 10670  & 4.3e+13 &  4.6e+14  &   0.99   & 0.65 \\
\hline
\enddata
\tablecomments{ $\Delta z$ is the physical depth range at $\lambda=2826$ \AA; we used 500 km as the minimum height ($z_{\rm{lim}}$ in Equation \ref{eq:ciprime}).
 $z_1$ is the bottom of the stationary flare layer where $C_I^{\prime}=0.95$, $z_2$ is the bottom of the chromospheric condensation where the downward gas speed falls below 5 \kms, $z_3$ is the top of the chromospheric condensation where $C_I^{\prime}=0.05$.  The temperature and density values are weighted by the continuum contribution function in each region of the atmosphere, and $Q_b$ is the fraction of beam energy flux deposited
  at heights greater than $z$.   }
\label{table:params}
\end{deluxetable}
\end{turnpage}
\clearpage

\begin{turnpage}
\begin{deluxetable}{cc}
\tabletypesize{\scriptsize}
\tablewidth{0pt}
\tablecaption{Time-variable nonthermal broadening in the CC}
\tablehead{
\colhead{Time [s]} &
\colhead{$\xi$ [km s$^{-1}$]} } 
\startdata

 1 & 68 \\
 1.5  & 55 \\
 2  & 42 \\
 2.5  & 31 \\
 3  & 25 \\
 4 & 18 \\
 5 & 13 \\
 6 & 11 \\
 7 & 10 \\
 8 & 10 \\
 15 & 7 \\
 
 \enddata
\tablecomments{  Values of the nonthermal broadening (microturbulence) parameter $\xi$ assumed at heights corresponding to the chromospheric condensation; see Section \ref{sec:nonthermal} for details.    }
\label{table:xi}
\end{deluxetable}
\end{turnpage}
\clearpage

\begin{turnpage}
\begin{deluxetable}{cccccc}
\tabletypesize{\scriptsize}
\tablewidth{0pt}
\tablecaption{Model Comparison Checklist}
\tablehead{
\colhead{time} &
\colhead{NUV continuum intensity} &
\colhead{\lamr\ intensity} &
\colhead{$I_{\rm{RWA}} / I_{\lambda_{\rm{rest}}}$} & \colhead{H$\alpha$ bisector} &
\colhead{SJI 2832 Intensity} \\
\colhead{} & \colhead{excess C2826} & \colhead{\feone, \fetwo} & \colhead{\feone, \fetwo} & \colhead{$\ge 30$\kms} &  \colhead{ }  }
\startdata
5F11 & & & \\
\hline
$t=1.8$~s & \checkmark  & $>$, -- & X, -- & $<$ & -- \\
$t=3.97$~s & $>$  &  $>$, -- & \checkmark, -- & \checkmark & \checkmark \\
$t=0-8$~s ave (short heating)  &  $<$    &  \checkmark, $>$  & \checkmark, \checkmark & -- & -- \\
$t=0-8$~s ave (extended heating)  &   $>$    &  $>$, $>$  &   \checkmark, \checkmark & -- & --   \\
\hline
\hline
F11 & & & \\
\hline
$t=3$~s &  $<$  &  $>$, -- & X, -- & X  & -- \\
$t=18$~s & $<$   & $>$, --  & X, -- & X & -- \\
\hline
\hline
coronal heating model & & & \\
\hline
$t=3$~s & $<$ & $<$, -- & $>$, -- & \checkmark & -- \\
\enddata 
\tablecomments{For each of the models at selected times, we indicate where they are consistent with the data constraints (\checkmark),
 where they exceed the data constraints ($>$), where they are too low ($<$), where they fail altogether to produce the indicated
 feature (''X''), and where they are not compared to the data in this work (--).  Except for the last column, the data 
 that we compare to are the quantities obtained from BFP1 at 17:46:08 and BFP2 at 17:46:24 (Table \ref{table:bfpvals}), and the H$\alpha$ bisector is used as a proxy for the bisector of an
 optically thick chromospheric line.   }
\label{table:modelcheck}
\end{deluxetable}
\end{turnpage}
\clearpage

\end{document}